\documentclass[acmsmall]{acmart}

\usepackage[strict]{changepage}
\usepackage{cleveref}
\usepackage{framed}
\usepackage{multirow}
\usepackage{rotating}
\usepackage{tcolorbox}
\usepackage{xspace}

\graphicspath{{figures/}}
\tcbset{arc=1.5pt, boxrule=\heavyrulewidth, colback=gray!1, colframe=black}

\definecolor{darkblue}{rgb}{0.0, 0.0, 0.55}
\newenvironment{quotebox}{%
    \vspace{-.1cm}
	\MakeFramed{\advance\hsize-\width\FrameRestore}%
	\noindent\hspace{-4.55pt}
	\begin{adjustwidth}{}{7pt}%
		\vspace{-1pt}\vspace{1pt}%
	}
	{%
		\vspace{1pt}\end{adjustwidth}\endMakeFramed%
}

\newcommand{\rqi}{What are the significant features of contributor-abandoned PRs in the studied projects?\xspace}
\newcommand{\rqii}{How do different features impact the probability of PR abandonment in the studied projects?\xspace}
\newcommand{\rqiii}{What are the probable reasons why contributors abandon their PRs in the studied projects?\xspace}

\newcommand{\dataframe}{435,048\xspace}
\newcommand{\dataset}{265,325\xspace}
\newcommand{\abandoned}{4,450\xspace}
\newcommand{\sample}{354\xspace}

\begin{document}

\acmJournal{TOSEM}

\title{On Wasted Contributions: Understanding the Dynamics of Contributor-Abandoned Pull Requests}
\subtitle{A Mixed-Methods Study of 10 Large Open-Source Projects}

\author{SayedHassan Khatoonabadi}
\orcid{0000-0003-0615-9242}
\affiliation{%
    \department[0]{Data-driven Analysis of Software (DAS) Lab}
    \department[1]{Department of Computer Science \& Software Engineering}
    \institution{Concordia University}
    \city{Montreal}
    \state{QC}
    \country{Canada}
}
\email{sayedhassan.khatoonabadi@concordia.ca}

\author{Diego Elias Costa}
\orcid{0000-0001-7084-2594}
\affiliation{%
    \department[0]{Data-driven Analysis of Software (DAS) Lab}
    \department[1]{Department of Computer Science \& Software Engineering}
    \institution{Concordia University}
    \city{Montreal}
    \state{QC}
    \country{Canada}
}
\email{diego.costa@concordia.ca}

\author{Rabe Abdalkareem}
\orcid{0000-0001-9914-5434}
\affiliation{%
    \department{School of Computer Science}
    \institution{Carleton University}
    \city{Ottawa}
    \state{ON}
    \country{Canada}
}    
\email{rabe.abdalkareem@carleton.ca}

\author{Emad Shihab}
\orcid{0000-0003-1285-9878}
\affiliation{%
    \department[0]{Data-driven Analysis of Software (DAS) Lab}
    \department[1]{Department of Computer Science \& Software Engineering}
    \institution{Concordia University}
    \city{Montreal}
    \state{QC}
    \country{Canada}
} 
\email{emad.shihab@concordia.ca}



\begin{abstract}
    Pull-based development has enabled numerous volunteers to contribute to open-source projects with fewer barriers. Nevertheless, a considerable amount of pull requests (PRs) with valid contributions are abandoned by their \emph{contributors}, wasting the effort and time put in by both the contributors and maintainers. To better understand the underlying dynamics of contributor-abandoned PRs, we conduct a mixed-methods study using both quantitative and qualitative methods. We curate a dataset consisting of \dataset PRs including \abandoned abandoned ones from ten popular and mature GitHub projects and measure 16 features characterizing PRs, contributors, review processes, and projects. Using statistical and machine learning techniques, we find that complex PRs, novice contributors, and lengthy reviews have a higher probability of abandonment and the rate of PR abandonment fluctuates alongside the projects' maturity or workload. To identify why contributors abandon their PRs, we also manually examine a random sample of \sample abandoned PRs. We observe that the most frequent abandonment reasons are related to the obstacles faced by contributors, followed by the hurdles imposed by maintainers during the review process. Finally, we survey the top core maintainers of the studied projects to understand their perspectives on dealing with PR abandonment and on our findings.
\end{abstract}

\maketitle

\renewcommand{\shortauthors}{Khatoonabadi et al.}

\section{Introduction}
Pull-based development has been popularized by social coding platforms such as GitHub and is widely adopted by distributed software teams, especially within the open-source community \citep{gousios_exploratory_2014}. In this development model, developers fork a project (i.e., create a personal copy of the project) before making their changes. Whenever ready, the developers request their changes to get merged into the project by submitting a Pull Request (PR). The maintainers then review the PR and decide whether to merge it into their project. Compared to traditional methods, pull-based development reduces the time taken to review and merge the contributions \citep{gousios_exploratory_2014, zhu_effectiveness_2016}.

The streamlined contribution mechanism enabled by PRs has encouraged numerous external developers to contribute to open-source projects with fewer barriers \citep{gousios_exploratory_2014, pinto_more_2016, zhu_effectiveness_2016}. However, an industrial report \citep{davis_8_2018} estimates that 8\% of PRs are wasted and never merged. Such PRs are either rejected by the maintainers or abandoned by their contributors. In contrast to rejected PRs, abandoned PRs are valid contributions that are not finalized because their contributors have left the review process unfinished. Unfortunately, abandoned PRs waste a considerable amount of time and effort that is often put in both by the contributors to prepare and submit such PRs and by the maintainers to manage and review them.

The literature has extensively studied how various technical, social, and personal factors influence the acceptance and review process of PRs. However, PR abandonment as a challenge that results in a great opportunity cost for the open-source community, especially for the contributors and the reviewers of abandoned PRs, has only recently received attention from \citet{li_are_2021}. Based on a survey of open-source developers, they explained how abandoned PRs impact project maintainers and discussed why developers abandon their PRs. While their findings shed light on PR abandonment from the perspective of developers, the influence of the factors related to PRs, contributors, review processes, and projects on PR abandonment is still unknown.

To gain a better and more comprehensive understanding of the underlying dynamics of contributor-abandoned PRs, we conduct a mixed-methods study using both quantitative and qualitative methods \citep{creswell_research_2017}. \emph{For the sake of brevity, we refer to external contributors as contributors throughout the paper.} First, we curate a dataset consisting of \dataset contributor PRs from ten popular and mature GitHub projects (namely, Homebrew Cask, Kubernetes, Kibana, Ansible, DefinitelyTyped, Rust, Odoo, Legacy Homebrew, Elasticsearch, and Swift). Then, we devise heuristics to identify \abandoned candidate PRs with a high chance of being truly abandoned by their contributors. Next, we measure 16 features to characterize the PRs, their contributors, their review processes, and their projects for our quantitative analyses. We aim to answer the following three research questions in this paper:

\begin{itemize}
    \item[\textbf{RQ\textsubscript{1}:}] \textbf{\rqi} We find that contributor-abandoned PRs are usually more complex, their contributors are usually less experienced, and their review process is usually lengthier than nonabandoned PRs. Furthermore, as the projects mature, contributor-abandoned PRs have become more frequent in three projects (i.e., Kubernetes, Swift, and DefinitelyTyped) and less frequent in five other projects (i.e., Kibana, Ansible, Elasticsearch, Odoo, and Homebrew Cask).
    \smallskip
    \item[\textbf{RQ\textsubscript{2}:}] \textbf{\rqii} We find that the features of the review process, contributor, and project are more important in predicting PR abandonment than the features of PRs themselves. Specifically, PRs with more than three responses from the participants or the contributors, and those submitted by novice contributors are more likely to get abandoned. Also, the abandonment probability changes as the projects evolve, with half of the projects showing a decrease in abandonment in their mature stages and the other half showing an increase in abandonment.
    \smallskip
    \item[\textbf{RQ\textsubscript{3}:}] \textbf{\rqiii} We find that the most frequent abandonment reasons are related to the obstacles faced by contributors followed by the hurdles imposed by maintainers during the review process. Specifically, difficulty addressing the maintainers' comments, lack of review from the maintainers, difficulty resolving the CI failures, and difficulty resolving the merge issues are the most common reasons why contributors abandon their PRs.
\end{itemize}

\noindentparagraph{\emph{\textbf{Our Contributions.}}} In summary, we make the following key contributions in this paper:

\begin{itemize}
    \item We identify the features of PRs, their contributors, their review processes, and their projects that significantly differ between abandoned and nonabandoned PRs.
    \item We rank the features based on their relative importance for predicting PR abandonment and describe how different values of these features vary the predicted probability of abandonment.
    \item We identify the probable reasons why contributors abandon their PRs and survey the core developers of studied projects to understand their perspectives on dealing with PR abandonment and our findings.
    \item To promote the reproducibility of our study and facilitate future research, we also share our dataset at \url{https://doi.org/10.5281/zenodo.4892277}.
\end{itemize}

\noindentparagraph{\emph{\textbf{Paper Organization.}}} The remainder of this paper is organized as follows. Section~\ref{sec:methodology} presents our research methodology and Section~\ref{sec:rqi}~to~\ref{sec:rqiii} present our findings for each research question. Then, Section~\ref{sec:perspectives} reports the perspectives of maintainers on dealing with PR abandonment and our findings. Next, Section~\ref{sec:discussion} further discusses our findings and Section~\ref{sec:related_work} reviews the related work. Finally, Section~\ref{sec:limitations} discusses the limitations of our study and Section~\ref{sec:conclusion} concludes the paper.

\section{Methodology}
\label{sec:methodology}
In the following, we explain how we design our study (\Cref{sec:design}), select the study projects (\Cref{sec:projects}), collect the required data (\Cref{sec:data}), identify abandoned PRs (\Cref{sec:abandoned}), and extract features from PRs (\Cref{sec:features}).

\subsection{Study Design}
\label{sec:design}
To gain a more comprehensive understanding of the underlying dynamics of contributor-abandoned PRs, we conduct a mixed-methods study using both quantitative and qualitative methods \citep{creswell_research_2017}. First, we perform statistical analysis to identify the significant features of abandoned PRs in RQ\textsubscript{1} (\Cref{sec:rqi}). Then, we use machine learning techniques to determine the relative importance of the features and describe how each feature varies the predicted probability of abandonment in RQ\textsubscript{2} (\Cref{sec:rqii}). Finally, we manually examine a random sample of \abandoned abandoned PRs to identify the reasons why contributors abandon their PRs in RQ\textsubscript{3} (\Cref{sec:rqiii}).

\subsection{Studied Projects}
\label{sec:projects}
For our study, we need open-source projects that are popular among the community and have a rich history of adopting pull-based development. For this purpose, we rely on GitHub as a pioneer in supporting the pull request model and the largest open-source ecosystem \citep{github_state_2020}, which have also been the subject of many software engineering studies \citep{kalliamvakou_-depth_2016}. To focus on the most popular projects, we use the number of stars as a proxy \citep{borges_understanding_2016, borges_whats_2018} and retrieve the list of the top 1,000 most-starred projects. Among these projects, we focus on the top ten with the most number of PRs to ensure that each project has enough historical data for our study. As shown in \Cref{tab:projects}, the studied projects cover multiple application domains and programming languages, with each project having at least 14 thousand stars, 31 thousand PRs, 800 external contributors, and 4 years of PR history.

\begin{table}
    \caption{Overview of the projects selected to study contributor-abandoned PRs.}
    \label{tab:projects}
    \resizebox{\textwidth}{!}{%
        \begin{tabular}{@{}l|ccrr|ll@{}}
            \toprule
            \textbf{Project} & \textbf{PRs} & \textbf{Stars} & \textbf{Contributors} & \textbf{Months} & \textbf{Domain}         & \textbf{Language(s)} \\
            \midrule
            Homebrew Cask    & 78,446       & 17,077         & 7,246                 & 98              & Package Manager         & Ruby                 \\
            Kubernetes       & 56,721       & 66,644         & 3,628                 & 71              & Container Orchestration & Go                   \\
            Kibana           & 43,324       & 14,313         & 896                   & 87              & Analytics Dashboard     & TypeScript           \\
            Ansible          & 42,338       & 43,333         & 7,168                 & 98              & Automation Platform     & Python               \\
            DefinitelyTyped  & 38,645       & 28,316         & 13,866                & 91              & Type Definitions        & TypeScript           \\
            Rust             & 38,361       & 45,261         & 3,181                 & 116             & Programming Language    & Rust                 \\
            Odoo             & 38,241       & 17,636         & 1,822                 & 72              & Business Apps           & JavaScript, Python   \\
            Legacy Homebrew  & 33,577       & 27,786         & 7,904                 & 75              & Package Manager         & Ruby                 \\
            Elasticsearch    & 33,411       & 49,134         & 2,350                 & 116             & Analytics Engine        & Java                 \\
            Swift            & 31,984       & 51,831         & 974                   & 54              & Programming Language    & C++, Swift           \\
            \bottomrule
        \end{tabular}
    }
\end{table}

\subsection{Data Collection}
\label{sec:data}
To identify abandoned PRs, we require the timeline activity of PRs, which records all the events during the lifecycle of a PR. For this purpose, we use the \texttt{PyGithub} package \citep{jacques_pygithub_2021} to retrieve the required data from GitHub. On May 30th, 2020, we collected the timeline, commits, and changed files metadata \citep{github_issues_2021, github_pulls_2021} for the \dataframe PRs of the studied projects.

\subsection{Abandoned PRs Identification}
\label{sec:abandoned}
After collecting the PRs data, we need to identify those abandoned by their contributors. Each PR in GitHub has one of the following three states: (i) \textit{open} indicates that the PR is not finalized and might be in progress, (ii) \textit{closed} indicates that the PR is either rejected by the maintainers or abandoned by its contributor, and (iii) \textit{merged} indicates that the PR is merged into the project. Abandoned PRs are a subset of the \textit{open} or \textit{closed} PRs that are wasted because their contributors have left them unfinished. The contributors of such PRs may either explicitly declare their abandonment decision or implicitly stop addressing the maintainers' comments. The maintainers often employ bots like Stale \citep{github_stale_2021} to close abandoned PRs after a period of inactivity \citep{wessel_should_2019}, or they manually find and close the abandoned PRs.

However, GitHub does not assign a specific status for abandoned PRs to explicitly distinguish them from nonabandoned ones. Therefore, we cannot simply retrieve the list of abandoned PRs neither directly through the GitHub API \citep{github_rest_2021} nor using existing archives such as GHTorrent \citep{gousios_ghtorent_2013} and GH Archive \citep{grigorik_gh_2021}. Therefore, we resort to heuristics to identify abandoned PRs based on the collected metadata. Heuristics are not guaranteed to be optimal and are subject to an inherent trade-off between their accuracy and completeness \citep{khatoonabadi_gap2wss_2021}. To determine the best balance between the precision and recall of our dataset, we experimented with different heuristics before finalizing the following:

\noindentparagraph{\emph{\textbf{Step 1: Exclude PRs from core developers.}}} Our study focuses on contributions from external developers, which are more prone to get abandoned. Therefore, we exclude the PRs from core developers to focus on external contributions. GitHub defines different roles for the authors of PRs within a project \citep{github_enums_2021}. Among these roles, \textit{owner} refers to the owners of the project, \textit{member} refers to the members of the organization owning the project, and \textit{collaborator} refers to those invited to collaborate on the project. Since these three roles typically have push/merge permissions within a GitHub repository, we consider them as core developers and exclude their PRs from our dataset.

\noindentparagraph{\emph{\textbf{Step 2: Exclude PRs from deleted accounts.}}} GitHub allows its users to remove their accounts permanently and afterward refers to the contributors of PRs from such accounts as \textit{ghost}. Since there is no straightforward way to distinguish between the contributors of such PRs \citep{kalliamvakou_-depth_2016}, we exclude them from our study.

\noindentparagraph{\emph{\textbf{Step 3: Exclude recently updated PRs.}}} To minimize the chance of marking PRs that are still in progress as abandoned, we exclude the PRs that their contributors have recently updated. To be conservative, we exclude the PRs that their contributors have updated (i.e., new comments or commits) within the last six months of the data collection date (i.e., May 30th, 2020).

\noindentparagraph{\emph{\textbf{Step 4: Exclude merged PRs.}}} We consider merged PRs as not wasted and thus not abandoned. However, the maintainers may not always use the merging methods provided through the GitHub interface to merge PRs. To account for such PRs, we resort to heuristics similar to \citet{kalliamvakou_-depth_2016}. Specifically, we exclude the PRs with a \textit{merged} status (i.e., merged using the GitHub interface) and those PRs that are closed without a \textit{merged} status, but have a merged commit inside the project that references them (e.g., ``Close \#123'').

\noindentparagraph{\emph{\textbf{Step 5: Search keywords in discussion comments.}}} As the last step for identifying abandoned PRs, we rely on keyword searching within all the discussion comments of PRs similar to \citet{li_are_2021}. First, we remove code snippets and reply quotes from these comments and then search for keywords representing the unresponsiveness of contributors. To determine such keywords, we consider the keywords used in \citet{li_are_2021} as our initial set. Then, we manually examine a sample of known abandoned PRs from our studied projects and iteratively refine our keywords. Finally, we find the following keywords are commonly used to refer to abandoned PRs:

\texttt{\{abandon, stale, any update, lack of update, no update, inactive, inactivity, lack of activity, no activity, not active, lack of reply, no reply, lack of response, no response\}}.\\

Using our heuristics, we identified \abandoned abandoned PRs among the curated \dataset PRs. As with any heuristic, ours may return some nonabandoned PRs (i.e., false positives), given that the review process of PRs often involves social interactions between the contributors and the reviewers before getting finalized. To validate the quality of our dataset, we manually analyze 100 PRs (10 PRs from each project) to verify if they have been truly marked as abandoned. We find seven false positives out of the 100 examined PRs as these PRs were rejected while including the keywords representing abandonment (e.g., \citep{macrae_added_2014}). Still, we believe that a false-positive rate of 7\% gives us enough confidence to rely on this dataset for our study.

\subsection{Feature Extraction}
\label{sec:features}
To identify the features that are possibly associated with PR abandonment, we consult the literature on pull-based development \citep{gousios_dataset_2014, zhang_shoulders_2020, zhang_pull_2021, zhang_pull_2022}. As shown in \Cref{tab:features}, we extract 16 features covering four different dimensions: (i) PR features, (ii) contributor features, (iii) review process features, and (iv) project features. In the following, we describe the extracted features for each dimension in more detail.

\begin{table}
    \caption{Overview of the features extracted to characterize PRs, their contributors, their review process, and their projects.}
    \label{tab:features}
    \resizebox{\textwidth}{!}{%
        \begin{tabular}{@{}l|l|l@{}}
            \toprule

            \textbf{Dimension}                       &\textbf{Feature}                   & \textbf{Description}                                                   \\
            \midrule
            \multirow{4}{*}{\textbf{Pull Request}}   & pr\_description                   & Number of words in the title and description of the PR                 \\
                                                     & pr\_commits                       & Number of commits during the lifecycle of the PR                       \\
                                                     & pr\_changed\_lines                & Number of changed lines during the lifecycle of the PR                 \\
                                                     & pr\_changed\_files                & Number of changed files during the lifecycle of the PR                 \\
            \midrule
            \multirow{4}{*}{\textbf{Contributor}}    & contributor\_contribution\_period & Number of months since the first PR of the contributor in the project  \\
                                                     & contributor\_pulls                & Number of prior PRs by the contributor in the project                  \\
                                                     & contributor\_acceptance\_rate     & Ratio of previously merged PRs by the contributor in the project       \\
                                                     & contributor\_abandonment\_rate    & Ratio of previously abandoned PRs by the contributor in the project    \\
            \midrule
            \multirow{4}{*}{\textbf{Review Process}} & review\_response\_latency         & Number of days till the first response in the PR                       \\
                                                     & review\_participants              & Number of participants during the lifecycle of the PR                  \\
                                                     & review\_participants\_responses   & Number of responses by the participants during the lifecycle of the PR \\
                                                     & review\_contributor\_responses    & Number of responses by the contributor during the lifecycle of the PR  \\
            \midrule
            \multirow{4}{*}{\textbf{Project}}        & project\_age                      & Number of months since the starting date of the project                \\
                                                     & project\_pulls                    & Number of prior PRs in the project                                     \\
                                                     & project\_contributors             & Number of prior contributors in the project                            \\
                                                     & project\_open\_pulls              & Number of open PRs in the project at the submission time of the PR     \\
            \bottomrule
        \end{tabular}
    }
\end{table}

\subsection*{PR Features:}
\noindentparagraph{\emph{\textbf{Description Length.}}} The description length of PRs is found to negatively impact their acceptance probability and review time \citep{yu_determinants_2016}. We aim to understand whether PRs with shorter descriptions are more frequently abandoned than verbosely described PRs. To characterize a PR's description length, we measure the number of words that have been used in its title and description (denoted by \textit{pr\_description}).

\noindentparagraph{\emph{\textbf{Change Complexity.}}} The complexity of changes has been extensively shown to negatively impact the acceptance probability and the review time of PRs \citep{tsay_influence_2014, soares_acceptance_2015, yu_wait_2015, yu_determinants_2016, kononenko_studying_2018}. We aim to understand whether complex PRs are more prone to get abandoned. To characterize a PR's change complexity, we measure the number of commits that have been submitted during the PR's lifecycle (denoted by \textit{pr\_commits}); the number of lines (denoted by \textit{pr\_changed\_lines}); and the number of files (denoted by \textit{pr\_changed\_files}) that have been changed (i.e., additions or deletions) as part of the submitted commits.

\subsection*{Contributor Features:}
\noindentparagraph{\emph{\textbf{Experience Level.}}} The experience of contributors has been extensively shown to positively impact the acceptance probability and the review time of PRs \citep{gousios_exploratory_2014, soares_acceptance_2015, yu_determinants_2016, kononenko_studying_2018}. We aim to understand whether more experienced contributors are less likely to abandon their PRs. To characterize a contributor's experience within a project, we measure the number of months that have been elapsed since the first submitted PR of the contributor to the project (denoted by \textit{pr\_contribution\_period}); the number of PRs that the contributor has previously submitted to the project (denoted by \textit{contributor\_pulls}); and the ratio of the previously submitted PRs by the contributor that had been merged into the project (denoted by \textit{contributor\_acceptance\_rate}).

\noindentparagraph{\emph{\textbf{Abandonment History.}}} To the best of our knowledge, the abandonment history of contributors has not been previously studied. We aim to understand whether contributors who have a long history of abandonment are more likely to abandon their PRs. To characterize a contributor's abandonment history within a project, we measure the ratio of the previously submitted PRs by the contributor that we have marked as abandoned in the project (denoted by \textit{contributor\_abandonment\_rate}).

\subsection*{Review Process Features:}
\noindentparagraph{\emph{\textbf{Response Latency.}}} The response latency is found to negatively impact the acceptance probability and the review time of PRs \citep{yu_wait_2015, yu_determinants_2016}. We aim to understand whether PRs that take longer to receive a first response from the reviewers are more likely to get abandoned. To characterize a PR's response latency, we measure the number of days that have been taken to receive their first response (i.e., comment or review) from the participants (denoted by \textit{pr\_response\_latency}).

\noindentparagraph{\emph{\textbf{Participants Activity.}}} The activity of participants (i.e., anyone participating in the review process except the contributor) is found to negatively impact the acceptance probability of PRs \citep{tsay_influence_2014, kononenko_studying_2018}. We aim to understand whether PRs with a higher activity from their participants are more likely to get abandoned. To characterize the participants' activity in a PR, we measure the number of participants in its review process (denoted by \textit{review\_participants}); and the number of responses (i.e., comments or reviews) that have been submitted by the participants (denoted by \textit{review\_participants\_responses}) during the review process.

\noindentparagraph{\emph{\textbf{Contributor Activity.}}} Similar to the participants' activity, we aim to understand whether PRs with a higher activity from their contributors are also more likely to get abandoned. To characterize the contributor's activity in a PR, we measure the number of responses (i.e., comments or self-reviews) that the contributor has submitted during the review process of the PR (denoted by \textit{review\_contributor\_responses}).

\subsection*{Project Features:}
\noindentparagraph{\emph{\textbf{Maturity Level.}}} The maturity of projects is found to have a mixed impact on the acceptance probability and the review time of their PRs \citep{tsay_influence_2014, yu_determinants_2016}. We aim to understand whether the rate of abandoned PRs changes as projects become more mature. To characterize a project's maturity, we measure the number of months that have been elapsed since the creation date of the project until the submission date of the PR (denoted by \textit{project\_age}); the number of PRs that have been previously submitted to the project (denoted by \textit{project\_pulls}); and the number of developers who have previously contributed to the project (denoted by \textit{project\_contributors}) at the submission time of the PR.

\noindentparagraph{\emph{\textbf{Maintainers Workload.}}} The workload of maintainers is found to negatively impact the acceptance probability and the review time of PRs \citep{yu_wait_2015, yu_determinants_2016}. We aim to understand whether the high workload of maintainers increases the rate of abandoned PRs. To characterize a project's workload, we measure the number of submitted PRs that were still open at the submission time of the PR (denoted by \textit{project\_open\_pulls}).

\section{RQ\texorpdfstring{\textsubscript{1}}{1}: \rqi}
\label{sec:rqi}
PR abandonment is a challenge that results in a significant opportunity cost for the open-source community, especially for the contributors and the reviewers of abandoned PRs. A recent study by \citet{li_are_2021} has surveyed open-source developers to explain why PRs become abandoned. However, the influence of different factors on PR abandonment has not been studied yet. As our first research question, we aim to understand which features of PRs, their contributors, their review processes, and their projects are associated with PR abandonment. Specifically, we want to investigate how significantly abandoned PRs differ from nonabandoned ones.

\subsection{Approach}
We perform statistical analyses to identify the significant features of abandoned PRs compared with nonabandoned PRs. First, we compare the distribution of the extracted features between abandoned and nonabandoned PRs and then test their statistical and practical significance. In the following, we explain each step in more detail:

\noindentparagraph{\emph{\textbf{Step 1: Compare distribution of features.}}} To compare the distribution of features between abandoned and nonabandoned PRs, we generate violin plots \citep{hintze_violin_1998} for each project using the \texttt{ggstatsplot} package \citep{patil_visualizations_2021}. The generated plots for each feature are presented in \Cref{appendix:stats}, specifying their median values (denoted by $M$), interquartile ranges (the box inside the violin), and probability densities (the width of the violin at each value).

\noindentparagraph{\emph{\textbf{Step 2: Test statistical significance of features.}}} To test the statistical difference between the features of abandoned and nonabandoned PRs, we apply the Mann–Whitney $U$ test \citep{mann_test_1947} with a 95\% confidence level (i.e., $\alpha = 0.05$). We use this nonparametric test because we cannot assume the distribution of our features to be normal. To calculate this statistic, we use the \texttt{stats} package \citep{r_core_team_r_2021} and add the results to the plots generated in Step~1. For easier comparison, we denote $p < 0.05$ with *, $p < 0.01$ with **, and $p < 0.001$ with ***.

\noindentparagraph{\emph{\textbf{Step 3: Test practical significance of features.}}} While statistical significance verifies whether a difference exists between the features of abandoned and nonabandoned PRs, we also need to test their practical difference \citep{kirk_practical_1996}. For this purpose, we use Cliff's delta \citep{cliff_dominance_1993} to estimate their magnitude of difference (i.e., effect size). The value of Cliff's delta (denoted by $d$) ranges from $-1$ to $+1$: a positive $d$ implies that the values of the feature in abandoned PRs are often greater than those of nonabandoned PRs, while a negative $d$ implies the opposite. To calculate this statistic, we use the \texttt{effectsize} package \citep{ben-shachar_effectsize_2020} and add the results to the plots generated in Step~1. For easier comparison, we convert the $d$ values to qualitative magnitudes based on the following thresholds as suggested by \citet{hess_robust_2004}:

\begin{equation*}
    \text{Effect size}=
    \begin{cases}
        \text{Negligible}, & \text{if         $\lvert d \rvert \leq 0.147$} \\
        \text{Small},      & \text{if $0.147 < \lvert d \rvert \leq 0.33$}  \\
        \text{Medium},     & \text{if $0.33  < \lvert d \rvert \leq 0.474$} \\
        \text{Large},      & \text{if $0.474 < \lvert d \rvert \leq 1$}     \\
    \end{cases}
\end{equation*}

\subsection{Findings}
\label{sec:rqi_findings}
\Cref{tab:significance} summarizes the significance of different features across the studied projects. We consider a feature significant if its difference between abandoned and nonabandoned PRs is both statistically significant (i.e., $p < 0.05$) and practically significant (i.e., the effect size is small, medium, or large) in at least one project. Overall, we observe that the most significant features are related to the review process and contributors of PRs. We also find that four features (characterizing the review process and contributor) are significant across all the projects, and eight other features (encompassing all the dimensions) are significant in at least half the projects. In the following, we discuss the significance of each dimension in more detail.

\begin{table}
    \caption{Significance of different features across the studied projects. $\uparrow$ shows that abandoned PRs have values greater than nonabandoned ones, $\downarrow$ shows that abandoned PRs have values smaller than nonabandoned ones, and $\uparrow\downarrow$ shows a mixed relationship.}
    \label{tab:significance}
    \resizebox{\textwidth}{!}{%
        \begin{tabular}{@{}l|l|c|ccc@{}}
            \toprule
            \textbf{Dimension}                       & \textbf{Feature}                  & \textbf{Significant} & \textbf{Small}           & \textbf{Medium}  & \textbf{Large}           \\
            \midrule
            \multirow{4}{*}{\textbf{Pull Request}}   & pr\_description                   & 8                    & 7 ($\uparrow$)           & 1 ($\uparrow$)   & --                       \\
                                                     & pr\_commits                       & 6                    & 6 ($\uparrow$)           & --               & --                       \\
                                                     & pr\_changed\_lines                & 3                    & 1 ($\uparrow$)           & --               & 2 ($\uparrow$)           \\
                                                     & pr\_changed\_files                & --                   & --                       & --               & --                       \\
            \midrule
            \multirow{4}{*}{\textbf{Contributor}}    & contributor\_pulls                & \textbf{10}          & 4 ($\downarrow$)         & 3 ($\downarrow$) & 3 ($\downarrow$)         \\
                                                     & contributor\_acceptance\_rate     & \textbf{10}          & 5 ($\downarrow$)         & 4 ($\downarrow$) & 1 ($\downarrow$)         \\
                                                     & contributor\_contribution\_period & 5                    & 2 ($\downarrow$)         & 1 ($\downarrow$) & 2 ($\downarrow$)         \\
                                                     & contributor\_abandonment\_rate    & 2                    & 2 ($\uparrow$)           & --               & --                       \\
            \midrule
            \multirow{4}{*}{\textbf{Review Process}} & review\_participants\_responses   & \textbf{10}          & 1 ($\uparrow$)           & 1 ($\uparrow$)   & 8 ($\uparrow$)           \\
                                                     & review\_participants              & \textbf{10}          & 2 ($\uparrow$)           & 1 ($\uparrow$)   & 7 ($\uparrow$)           \\
                                                     & review\_contributor\_responses    & 7                    & 3 ($\uparrow$)           & 3 ($\uparrow$)   & 1 ($\uparrow$)           \\
                                                     & review\_response\_latency         & 4                    & 4 ($\uparrow$)           & --               & --                       \\
            \midrule
            \multirow{4}{*}{\textbf{Project}}        & project\_age                      & 8                    & 6 ($\uparrow\downarrow$) & --               & 2 ($\uparrow\downarrow$) \\
                                                     & project\_pulls                    & 8                    & 6 ($\uparrow\downarrow$) & --               & 2 ($\uparrow\downarrow$) \\
                                                     & project\_contributors             & 8                    & 6 ($\uparrow\downarrow$) & --               & 2 ($\uparrow\downarrow$) \\
                                                     & project\_open\_pulls              & 6                    & 4 ($\uparrow\downarrow$) & --               & 2 ($\uparrow\downarrow$) \\
            \bottomrule
        \end{tabular}
    }
\end{table}

\noindentparagraph{\emph{\textbf{Abandoned PRs are usually more complex than nonabandoned PRs.}}} As shown in the PR dimension of \Cref{tab:significance}, abandoned PRs tend to have lengthier descriptions (8 projects), contain more commits (6 projects), and involve more changed lines (3 projects). However, abandoned and nonabandoned PRs tend to be similar in their number of changed files across all the projects. The results suggest that abandoned PRs receive even more effort from their contributors, highlighting the waste resulting from the abandonment.

\noindentparagraph{\emph{\textbf{The contributors of abandoned PRs usually have less experience than the contributors of nonabandoned PRs.}}} As shown in the contributor dimension of \Cref{tab:significance}, the contributors of abandoned PRs tend to have previously submitted fewer PRs (all the projects), have a lower acceptance rate (all the projects), have a lower contribution period (5 projects), and have a higher abandonment rate (2 projects). However, the results cannot be attributed to the expected higher familiarity and expertise of the maintainers because we only consider external contributors in our study (\Cref{sec:abandoned}).

\noindentparagraph{\emph{\textbf{The review process of abandoned PRs is usually lengthier than the review process of nonabandoned PRs.}}} As shown in the review process dimension of \Cref{tab:significance}, the review process of abandoned PRs tends to receive more responses from its participants (all the projects), involve more participants (all the projects), receive more responses from the contributors (7 projects), and have a higher latency to receive the first response from the participants (4 projects). The results suggest that abandoned PRs are not just abandoned after the PR was submitted but have received even more effort from both their contributors and reviewer, again highlighting the waste resulting from the abandonment.

\noindentparagraph{\emph{\textbf{The project features play both a positive and negative role in PR abandonment.}}} As shown in the project dimension of \Cref{tab:significance}, we observe contrasting patterns in how the rate of abandoned PRs change alongside the project maturity (8 projects) or workload (6 projects). For easier comparison, we group these projects based on their similarities (i.e., positive or negative) in \Cref{tab:project_features}. In the first group (i.e., Kubernetes, Swift, and DefinitelyTyped), abandoned PRs tend to become more frequent as the projects become more mature (i.e., an increase in project\_age, project\_pulls, or project\_contributors). In two of these projects (i.e., Kubernetes and Swift), abandoned PRs also become more frequent as they experience a higher workload (i.e., an increase in project\_open\_pulls). In contrast to the first group, the second group (i.e., Kibana, Ansible, Elasticsearch, Odoo, and Homebrew Cask) experienced fewer abandoned PRs as the projects become more mature (i.e., an increase in project\_age, project\_pulls, or project\_contributors). Surprisingly, in four of these projects (i.e., Kibana, Ansible, Elasticsearch, and Odoo), abandoned PRs are more frequent when the projects have a lower workload (i.e., a decrease in project\_open\_pulls). The results may be associated with the change in the team structure, policies, or processes. For example, DefinitelyTyped has refined its review process since 2016 by rotatively assigning a TypeScript employee each week to focus on merging PRs \citep{therox_changes_2020}.

\begin{table}
    \caption{Difference of the project features between abandoned and nonabandoned PRs.}
    \label{tab:project_features}
    \resizebox{\textwidth}{!}{%
        \begin{tabular}{@{}cl|ccc|c@{}}
            \toprule
                                &                  & \multicolumn{3}{c|}{\textbf{Maturity}}                                           & \textbf{Workload}             \\
            \textbf{Group}      & \textbf{Project} & \textbf{project\_age} & \textbf{project\_pulls} & \textbf{project\_contributors} & \textbf{project\_open\_pulls} \\
            \midrule
            \multirow{3}{*}{I}  & Kubernetes       & Large ($\uparrow$)    & Large ($\uparrow$)      & Large ($\uparrow$)             & Large ($\uparrow$)            \\
                                & Swift            & Small ($\uparrow$)    & Small ($\uparrow$)      & Small ($\uparrow$)             & Small ($\uparrow$)            \\
                                & DefinitelyTyped  & Small ($\uparrow$)    & Small ($\uparrow$)      & Small ($\uparrow$)             & --                            \\
            \midrule
            \multirow{5}{*}{II} & Kibana           & Large ($\downarrow$)  & Large ($\downarrow$)    & Large ($\downarrow$)           & Large ($\downarrow$)          \\
                                & Ansible          & Small ($\downarrow$)  & Small ($\downarrow$)    & Small ($\downarrow$)           & Small ($\downarrow$)          \\
                                & Elasticsearch    & Small ($\downarrow$)  & Small ($\downarrow$)    & Small ($\downarrow$)           & Small ($\downarrow$)          \\
                                & Odoo             & Small ($\downarrow$)  & Small ($\downarrow$)    & Small ($\downarrow$)           & Small ($\downarrow$)          \\
                                & Homebrew Cask    & Small ($\downarrow$)  & Small ($\downarrow$)    & Small ($\downarrow$)           & --                            \\
            \bottomrule
        \end{tabular}
    }
\end{table}

\bigskip
\begin{tcolorbox}
    \paragraph{\emph{\textbf{Answer to RQ\textsubscript{1}.}}} Our findings suggest that contributor-abandoned PRs are usually more complex, their contributors are usually less experienced, and their review process is usually lengthier than nonabandoned PRs. Furthermore, as the projects mature, contributor-abandoned PRs have become more frequent in three projects (i.e., Kubernetes, Swift, and DefinitelyTyped) and less frequent in five other projects (i.e., Kibana, Ansible, Elasticsearch, Odoo, and Homebrew Cask).
\end{tcolorbox}
\medskip

\section{RQ\texorpdfstring{\textsubscript{2}}{2}: \rqii}
\label{sec:rqii}
In RQ\textsubscript{1}, we investigated what features of PRs, their contributors, their review processes, and their projects are associated with PR abandonment. As our second research question, we aim to better understand which PRs have a higher probability of getting abandoned by their contributors. Specifically, we want to identify which features are the most important for predicting PR abandonment and describe how each feature can influence the abandonment probability of PRs.

\subsection{Approach}
We use machine learning techniques to understand how each feature varies the predicted probability of PRs getting abandoned. First, we consider the features that we found to be significant in abandoned PRs and remove correlated and redundant features to ensure the quality of our models. Then, we build and evaluate the classifier models that we later use to analyze the relative importance and impact of each feature on the abandonment probability. In the following, we explain each step in more detail:

\noindentparagraph{\emph{\textbf{Step 1: Remove insignificant features.}}} In RQ\textsubscript{1}, we found that the number of changed files in a PR (i.e., \textit{pr\_changed\_files}) does not significantly differ between abandoned and nonabandoned PRs in any of the studied projects. Therefore, we exclude this feature because it is not valuable for analyzing the abandonment probability of PRs and consider the remaining 15 features for our analysis.

\begin{figure}
    \includegraphics[width=0.75\textwidth]{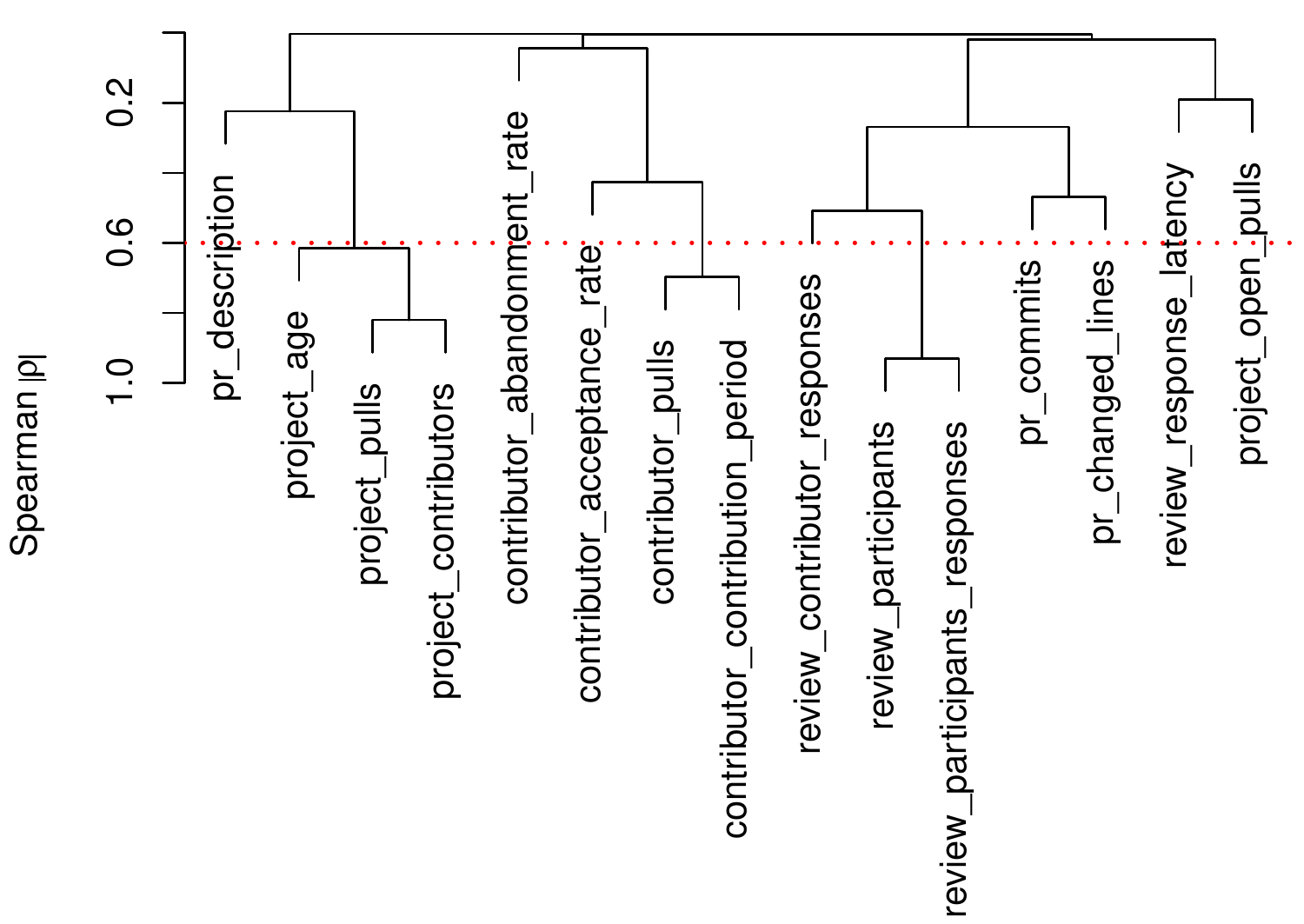}
    \caption{Spearman's $\rho$ correlation among different pairs of features across the combined data of the studied projects.}
    \label{fig:correlations}
\end{figure}

\noindentparagraph{\emph{\textbf{Step 2: Remove correlated features.}}} To focus on the most important features, we eliminate highly correlated features, which negatively affect the interpretation of models \citep{dormann_collinearity_2013}. To check the monotonic relationship between each pair of features, we use Spearman's $\rho$ \citep{spearman_proof_2010} as a nonparametric test because we cannot assume the distribution of our features to be normal. We measure correlations on the combined data of the studied projects to ensure that any correlation exists across all of them. \Cref{fig:correlations} presents a hierarchical cluster of the correlations generated using the \texttt{Hmisc} package \citep{harrell_hmisc_2021}. For each group of strongly correlated features (i.e., $\lvert \rho \rvert \geq 0.6$ as suggested by \citet{evans_straightforward_1996}), we keep the feature that is easier to interpret for our study and remove the rest. Accordingly, we drop the following four features from our analysis: \textit{project\_pulls}, \textit{project\_contributors}, \textit{contributor\_contribution\_period}, and \textit{review\_participants}.

\noindentparagraph{\emph{\textbf{Step 3: Remove redundant features.}}} While we remove highly correlated features in Step~1, we also need to eliminate redundant features to focus on the most important ones. To identify redundant features, we use the \texttt{Hmisc} package \citep{harrell_hmisc_2021}, which applies flexible parametric additive models to measure how well each feature can be predicted from other features. Similar to our correlation analysis, we measure redundancy on the combined data of the studied projects to ensure that any redundancy exists across all of them. Accordingly, we did not find any redundant features.

\noindentparagraph{\emph{\textbf{Step 4: Build classifier models.}}} To gain deeper insights on PR abandonment, we build a random forest classifier for each project using the \texttt{ranger} package \citep{wright_ranger_2017}. To model the abandonment probability, we consider the type of PR (i.e., abandoned or nonabandoned) as the dependent variable and the selected 11 features as the independent variables. Random forests \citep{breiman_random_2001} are commonly used in various domains and outperform linear models in both the predictive power and the ability to learn complex relations. To boost the predictive power of each model, we use the \texttt{tuneRanger} package \citep{probst_hyperparameters_2019}. This package automatically tunes the following three hyperparameters of random forests using sequential model-based optimization \citep{jones_efficient_1998}: (i) the number of variables randomly drawn for each split, (ii) the fraction of instances randomly drawn for training each tree, and (iii) the minimum number of samples that a node must have to split.

\noindentparagraph{\emph{\textbf{Step 5: Evaluate performance of models.}}} To ensure that the models are reliable for our analysis, we evaluate their predictive power using the following two recommended metrics for binary classifiers \citep{haibo_he_learning_2009}:

\begin{itemize}
    \item \textbf{AUC-ROC:} which measures the area under the Receiver Operating Characteristic (ROC) curve \citep{bradley_use_1997}. The ROC curve plots the true positive rate (i.e., the ratio of correctly classified abandoned PRs to truly abandoned PRs) against the false positive rate (i.e., the ratio of incorrectly classified abandoned PRs to nonabandoned PRs) across different thresholds. The value of AUC-ROC ranges from 0 to 1, with values more than 0.5 indicating better performance than a no-skill classifier (i.e., baseline). Note that the value of AUC-ROC is the same for both positive (i.e., abandoned PRs) and negative (i.e., nonabandoned PRs) classes.
    \smallskip
    \item \textbf{AUC-PR:} which measures the area under the Precision-Recall (PR) curve \citep{fawcett_introduction_2006}. The PR curve plots the precision (i.e., the ratio of correctly classified abandoned PRs to all classified abandoned PRs) against recall (i.e., the ratio of correctly classified abandoned PRs to truly abandoned PRs) across different thresholds. The value of AUC-PR also ranges from 0 to 1, but the performance of a no-skill classifier (i.e., baseline) is determined by the distribution of classes in a dataset (i.e., distribution of abandoned and nonabandoned PRs). Note that, unlike AUC-ROC, the value of AUC-PR is different between positive (i.e., abandoned PRs) and negative (i.e., nonabandoned PRs) classes.
\end{itemize}

To reduce bias in our performance evaluations, we perform a stratified 10-fold cross-validation with ten repeats (a total of 100 iterations) for each model using the \texttt{mlr} package \citep{bischl_mlr_2016}. \Cref{tab:performance} presents the results of our performance evaluation for each model, where the baseline column shows the ratio of the minority class (i.e., abandoned PRs). We observe that our models have a good performance with an average AUC-ROC of 0.87 and perform at least four times better than the baseline in terms of AUC-PR.

\begin{table}
    \caption{Performance scores of our model for each studied project.} 
    \label{tab:performance}
    \begin{tabular}{@{}l|c|ccc@{}}
        \toprule
        \textbf{Project} & \textbf{AUC-ROC} & \textbf{AUC-PR} & \textbf{Baseline} & \textbf{AUC-PR / Baseline} \\
        \midrule
        Ansible          & 0.88             & 0.042           & 0.006             & 7.08x                      \\
        DefinitelyTyped  & 0.86             & 0.262           & 0.054             & 4.84x                      \\
        Elasticsearch    & 0.86             & 0.021           & 0.003             & 6.14x                      \\
        Homebrew Cask    & 0.96             & 0.063           & 0.001             & 42.78x                     \\
        Kibana           & 0.96             & 0.095           & 0.002             & 50.27x                     \\
        Kubernetes       & 0.89             & 0.375           & 0.060             & 6.23x                      \\
        Legacy Homebrew  & 0.92             & 0.116           & 0.013             & 8.85x                      \\
        Odoo             & 0.77             & 0.029           & 0.002             & 17.25x                     \\
        Rust             & 0.81             & 0.109           & 0.020             & 5.50x                      \\
        Swift            & 0.82             & 0.059           & 0.003             & 19.36x                     \\
        \bottomrule
    \end{tabular}
\end{table}

\noindentparagraph{\emph{\textbf{Step 6: Analyze importance of features.}}} So far, we have built classifiers that can aptly model the abandonment probability of PRs. To compare the relative importance of different features, we perform permutation feature importance analysis \citep{fisher_all_2019} for each model using the \texttt{iml} package \citep{molnar_iml_2018}. This approach permutes a feature to break the association between the feature and the outcome (i.e., the abandonment probability in our case). The importance of the feature is then measured by how much error the permutated data introduces compared to the original error (i.e., loss in AUC in our case) after 100 iterations. Therefore, the most important features have the largest impact on the performance of our models and thus are more valuable for predicting PR abandonment.

\noindentparagraph{\emph{\textbf{Step 7: Analyze impact of features.}}} After measuring the relative importance of the features in Step~6, we aim to describe how each feature varies the abandonment probability. For this purpose, we generate Accumulated Local Effects (ALE) plots \citep{apley_visualizing_2020} using the \texttt{iml} package \citep{molnar_iml_2018}. ALE shows the effect of a feature at a certain value compared to the average prediction of the data. In other words, a downward trend implies a reduced probability of abandonment, an upward trend implies an increased probability of abandonment, and a stable ALE implies no changed probability of abandonment. To focus on the most common values of features, we filter out the values over the 99th percentile for each feature in each project. The plots are then created by dividing a feature into ten intervals selected based on its quantiles. For each interval, the PRs that fall into that interval are considered for calculating the difference in their prediction when replacing the value of the feature with the upper and lower limits of the interval. We model the abandonment probability as a function of the selected features, and thus, any relationship is causal for the model and may not hold in the real world \citep{molnar_interpretable_2022}.

\subsection{Findings}
\label{sec:rqii_findings}
\Cref{tab:importance} summarizes the importance of different features in each project. We find that the features of the review process, contributors, and projects play a more prominent role in PR abandonment than the features of PRs themselves. Specifically, the number of responses from the participants is the most important feature by a large margin, indicating that the activity of reviewers is essential in classifying abandoned PRs. The second and third most important features are the acceptance rate and the number of previously submitted PRs of the contributor, respectively, highlighting the impact of the contributor experience on PR abandonment. The fourth and fifth most important features are the number of responses from the contributor and the age of the project. Other features, except for the abandonment rate of the contributor, are also among the top five in at least one project, showing that different features have a different impact on PR abandonment due to the inherent differences between the projects. Unexpectedly, we also observe that the number of commits in the PR, the abandonment rate of the contributor, and the latency to the first response from the participants are overall the least important features, respectively. In the following, we describe how the top five features impact the predicted abandonment probability of PRs. The ALE plots for the rest of the features can be found in \Cref{appendix:ale}.

\begin{table}
    \caption{Importance of different features across the studied projects.}
    \label{tab:importance}
    \resizebox{\textwidth}{!}{%
        \begin{tabular}{@{}l|l|cccccccccc|cc@{}}
            \toprule
            \textbf{Dimension}                       & \textbf{Feature}                         & \begin{sideways}\textbf{Ansible}\end{sideways} & \begin{sideways}\textbf{DefenitelyTyped}\end{sideways} & \begin{sideways}\textbf{Elasticsearch}\end{sideways} & \begin{sideways}\textbf{Homebrew Cask}\end{sideways} & \begin{sideways}\textbf{Kibana}\end{sideways} & \begin{sideways}\textbf{Kubernetes}\end{sideways} & \begin{sideways}\textbf{Legacy Homebrew}\end{sideways} & \begin{sideways}\textbf{Odoo}\end{sideways} & \begin{sideways}\textbf{Rust}\end{sideways} & \begin{sideways}\textbf{Swift}\end{sideways} & \begin{sideways}\textbf{Overall Rank}\end{sideways} & \begin{sideways}\textbf{Average Loss}\end{sideways} \\
            \midrule
            \multirow{3}{*}{\textbf{Pull Request}}   & pr\_changed\_lines                       & 7                                              & 8                                                      & 7                                                    & \textbf{3}                                           & 6                                             & 9                                                 & \textbf{2}                                             & 7                                           & 6                                           & 9                                            & 6                                                   & 1.55                                                \\
                                                     & pr\_description                          & 8                                              & \textbf{3}                                             & \textbf{4}                                           & \textbf{5}                                           & 8                                             & 6                                                 & 9                                                      & 9                                           & 7                                           & 10                                           & 8                                                   & 1.48                                                \\
                                                     & pr\_commits                              & 10                                             & \textbf{5}                                             & 10                                                   & 11                                                   & 10                                            & 10                                                & 10                                                     & 10                                          & 10                                          & 8                                            & 11                                                  & 1.21                                                \\
            \midrule
            \multirow{3}{*}{\textbf{Contributor}}    & \textit{contributor\_acceptance\_rate}   & \textbf{5}                                     & \textbf{4}                                             & \textbf{1}                                           & 6                                                    & \textbf{5}                                    & \textbf{5}                                        & 8                                                      & \textbf{3}                                  & \textbf{1}                                  & \textbf{3}                                   & \textbf{2}                                          & 1.95                                                \\
                                                     & \textit{contributor\_pulls}              & \textbf{3}                                     & 9                                                      & \textbf{3}                                           & \textbf{4}                                           & \textbf{3}                                    & 7                                                 & 6                                                      & \textbf{4}                                  & \textbf{4}                                  & \textbf{1}                                   & \textbf{3}                                          & 1.81                                                \\
                                                     & contributor\_abandonment\_rate           & 11                                             & 11                                                     & 11                                                   & 8                                                    & 11                                            & 8                                                 & 11                                                     & 11                                          & 9                                           & 6                                            & 10                                                  & 1.24                                                \\
            \midrule
            \multirow{3}{*}{\textbf{Review Process}} & \textit{review\_participants\_responses} & \textbf{1}                                     & \textbf{1}                                             & \textbf{2}                                           & \textbf{1}                                           & \textbf{1}                                    & \textbf{1}                                        & \textbf{1}                                             & \textbf{1}                                  & \textbf{5}                                  & \textbf{2}                                   & \textbf{1}                                          & 4.45                                                \\
                                                     & \textit{review\_contributor\_responses}  & 9                                              & 6                                                      & 6                                                    & \textbf{2}                                           & \textbf{2}                                    & \textbf{4}                                        & \textbf{3}                                             & \textbf{5}                                  & \textbf{2}                                  & 11                                           & \textbf{4}                                          & 1.74                                                \\
                                                     & review\_response\_latency                & \textbf{2}                                     & 10                                                     & 9                                                    & 9                                                    & 9                                             & 11                                                & \textbf{5}                                             & \textbf{2}                                  & 11                                          & \textbf{5}                                   & 9                                                   & 1.33                                                \\
            \midrule
            \multirow{2}{*}{\textbf{Project}}        & \textit{project\_age}                    & 6                                              & \textbf{2}                                             & \textbf{5}                                           & 7                                                    & \textbf{4}                                    & \textbf{2}                                        & \textbf{4}                                             & 8                                           & \textbf{3}                                  & 7                                            & \textbf{5}                                          & 1.73                                                \\
                                                     & project\_open\_pulls                     & \textbf{4}                                     & 7                                                      & 8                                                    & 10                                                   & 7                                             & \textbf{3}                                        & 7                                                      & 6                                           & 8                                           & \textbf{4}                                   & 7                                                   & 1.49                                                \\
            \bottomrule
        \end{tabular}
    }
\end{table}

\begin{figure}
    \includegraphics[width=\textwidth]{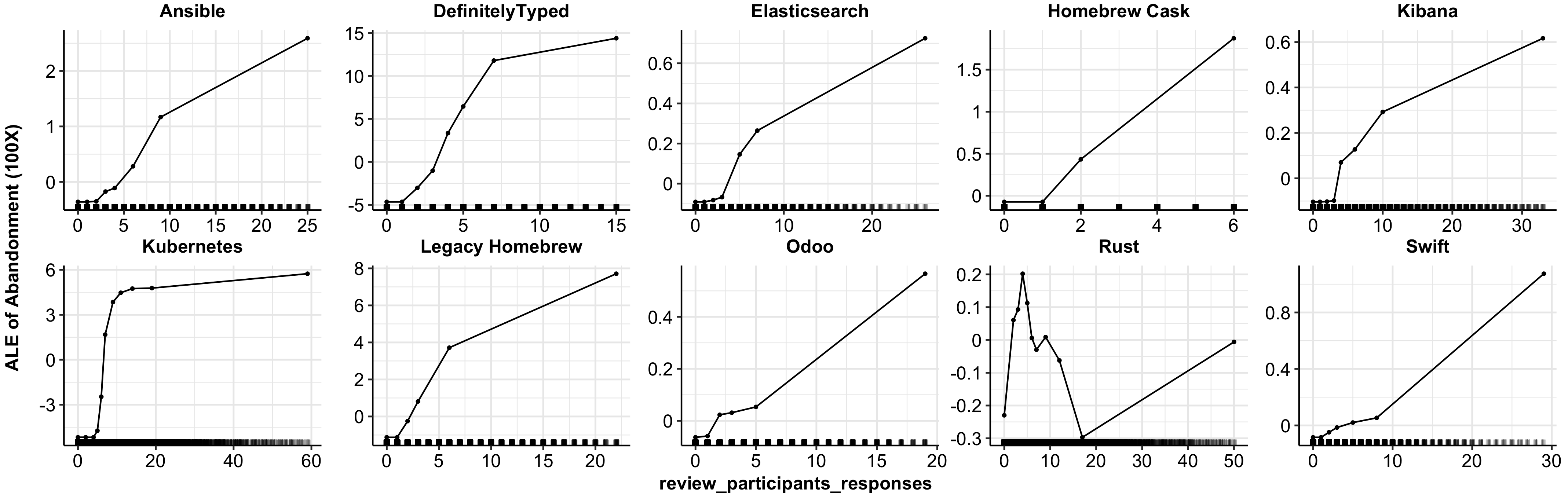}
    \caption{ALE plots showing how \textit{review\_participants\_responses} varies the abandonment probability of PRs across the studied projects.}
    \label{fig:review_participants_responses_ale}
\end{figure}

\begin{figure}
    \includegraphics[width=\textwidth]{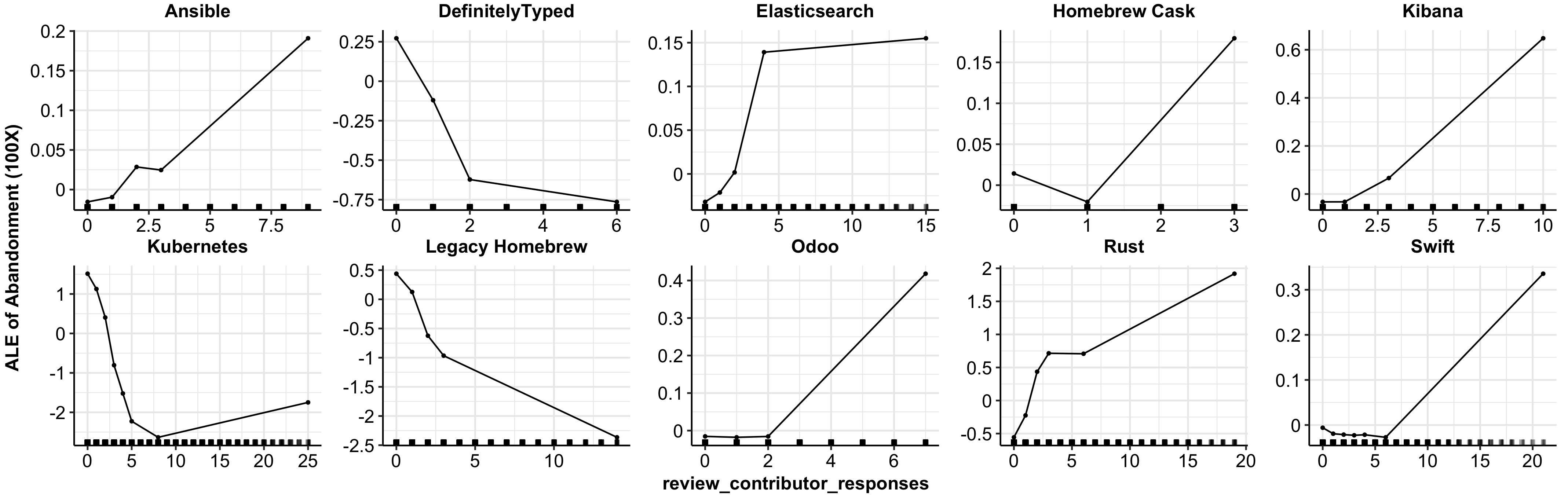}
    \caption{ALE plots showing how \textit{review\_contributor\_responses} varies the abandonment probability of PRs across the studied projects.}
    \label{fig:review_contributor_responses_ale}
\end{figure}

\noindentparagraph{\emph{\textbf{PRs with long discussions are more likely to get abandoned.}}} \Cref{fig:review_participants_responses_ale,fig:review_contributor_responses_ale} show how the number of responses from the participants and from the contributor varies the abandonment probability of a PR across the studied projects, respectively. We find that the probability of abandonment increases in most of the projects as the number of responses from the participants or the contributor increase (i.e., upwards trend). We also observe that PRs that receive more than three responses from either the participants or the contributor have an increased probability of abandonment in most of the projects. The results provide further evidence that abandoned PRs often demand more time and effort from both their reviewers and their contributors (see \Cref{sec:rqi_findings}).

\begin{figure}
    \includegraphics[width=\textwidth]{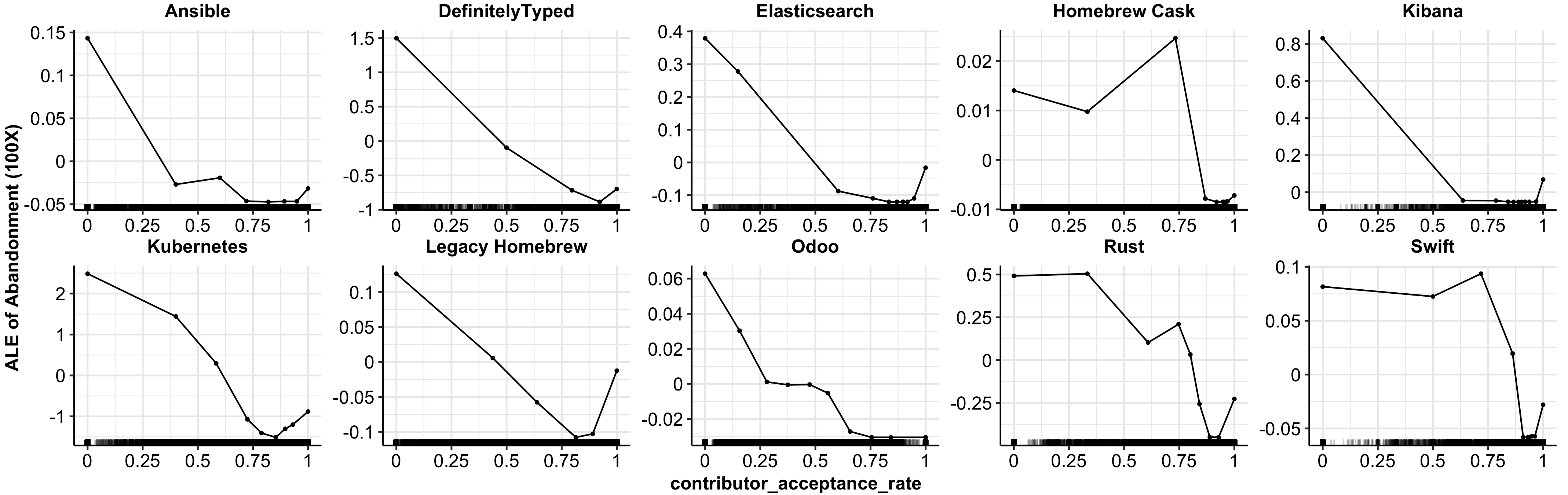}
    \caption{ALE plots showing how \textit{contributor\_acceptance\_rate} varies the abandonment probability of PRs across the studied projects.}
    \label{fig:contributor_acceptance_rate_ale}
\end{figure}

\begin{figure}
    \includegraphics[width=\textwidth]{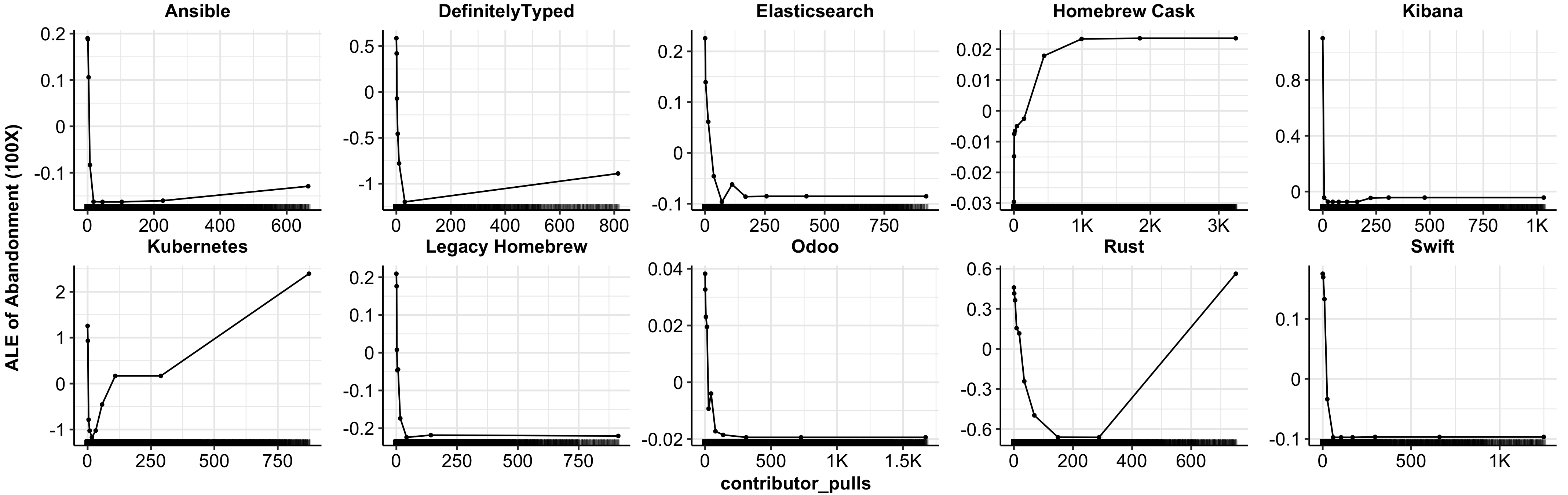}
    \caption{ALE plots showing how \textit{contributor\_pulls} varies the abandonment probability of PRs across the studied projects.}
    \label{fig:contributor_pulls_ale}
\end{figure}

\noindentparagraph{\emph{\textbf{Novice contributors are more likely to abandon their PRs.}}} \Cref{fig:contributor_acceptance_rate_ale,fig:contributor_pulls_ale} show how the acceptance rate and the number of previously submitted PRs by the contributor vary the abandonment probability of a PR across the studied projects, respectively. We observe that contributors with zero experience have the highest probability of abandonment in almost all the projects. Surprisingly, highly experienced contributors also have an increased probability of abandonment in a few projects. The results suggest that the contributors of abandoned PRs may need more guidance and attention from the maintainers.

\begin{figure}
    \includegraphics[width=\textwidth]{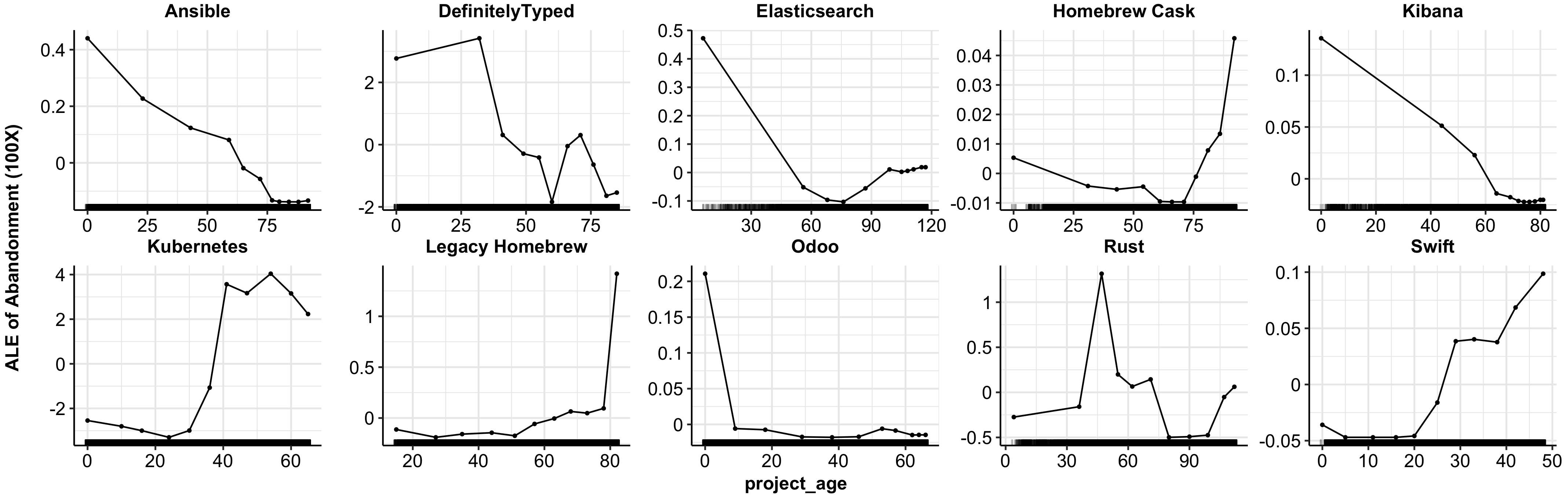}
    \caption{ALE plots showing how \textit{project\_age} varies the abandonment probability of PRs across the studied projects.}
    \label{fig:project_age_ale}
\end{figure}

\noindentparagraph{\emph{\textbf{PR abandonment has changed throughout the history of projects.}}} \Cref{fig:project_age_ale} shows how the age of projects varies the abandonment probability of a PR across the studied projects. Similar to RQ\textsubscript{1} (\Cref{sec:rqi_findings}), we observe two contrasting patterns: PR abandonment improves throughout the time in some projects and worsens in some other projects. In the first group (i.e., Ansible, DefinitelyTyped, Elasticsearch, Kibana, and Odoo), the abandonment probability is highest when the project age is low (i.e., downwards trend). However, in the second group (i.e., Homebrew Cask, Kubernetes, Legacy Homebrew, and Swift), the abandonment probability increases when the project age increases (i.e., upwards trend). While this fluctuation may be associated with the expected change in the workload of projects as they grow, we found that the number of open PRs is less important to our models than the age of the project.

\bigskip
\begin{tcolorbox}
    \paragraph{\emph{\textbf{Answer to RQ\textsubscript{2}.}}} Our findings suggest that the features of the review process, contributor, and project are more important in predicting PR abandonment than the features of PRs themselves. Specifically, PRs with more than three responses from the participants or the contributors, and those submitted by novice contributors are more likely to get abandoned. Also, the abandonment probability changes as the projects evolve, with half of the projects showing a decrease in abandonment in their mature stages and the other half showing an increase in abandonment.
\end{tcolorbox}
\medskip

\section{RQ\texorpdfstring{\textsubscript{3}}{3}: \rqiii}
\label{sec:rqiii}
In RQ\textsubscript{1} and RQ\textsubscript{2}, we quantitatively analyzed contributor-abandoned PRs to understand how different factors influence their abandonment probability. As our last research question, we aim to complement our previous findings and gain a deeper understanding of the underlying dynamics of abandoned PRs. Specifically, we want to look for clues in the review discussions of abandoned PRs to better understand why contributors abandon their PRs.

\subsection{Approach}
We perform a manual examination to establish a taxonomy of the abandonment reasons by following the coding guidelines presented by \citet{seaman_qualitative_1999}. First, we label a sample of abandoned PRs to identify the probable reasons why contributors abandon their PRs and then calculate our interrater agreement. In the following, we explain each step in more detail:

\noindentparagraph{\emph{\textbf{Step 1: Identify abandonment reasons.}}} First, we randomly select \sample PRs from \abandoned abandoned PRs of the studied projects (confidence level of 95\% with a $\pm$5\% confidence interval). Then, the first three authors are required to manually examine the discussion comments of each PR and try to pinpoint the primary reason(s) why its contributor has abandoned the PR. In most cases, the contributor has abandoned the PR without any explanation, and thus we look for clues in the interactions between the contributor and the reviewers to identify the most probable reasons for the abandonment. In cases where the contributor provided a reason for their abandonment decision, we also investigate other major reasons that might have led to their abandonment.

We perform the labeling in two rounds. In the first round, three annotators independently label a random sample of 60 PRs from the selected \sample PRs to establish the classification scheme. In the second round, we divide the sample \sample abandoned PRs (including the 60 PRs from the first round) into two sets to be labeled using the classification scheme from the previous round: the first set was labeled independently by the first and second authors, and the second set was labeled independently by the first and third authors. Finally, the annotators merged the labels and further refined the labels. In each round, when the annotators had different opinions, they discussed until they reached an agreement and then retroactively updated all the previously labeled PRs to ensure a coherent classification.

\noindentparagraph{\emph{\textbf{Step 2: Calculate interrater agreement.}}} To ensure the quality of our taxonomy, we calculate the Cohen’s Kappa coefficient \citep{cohen_coefficient_1960} using the \texttt{scikit-learn} package \citep{pedregosa_scikit-learn_2011}. This statistic is commonly used to evaluate the interrater agreement in different domains. The value of Cohen’s Kappa ranges from $-1.0$ to $+1.0$, with values more than 0 indicating an agreement better than chance. We obtained a Kappa score of 0.73, which is considered a substantial agreement as suggested by \citet{landis_measurement_1977}.

\subsection{Findings}
As shown in \Cref{tab:reasons}, we identified ten major reasons why contributors abandon their PRs, grouped into three categories: (i) contributor-related reasons, (ii) maintainer-related reasons, and (iii) PR-related reasons. Note that the total frequency of the identified reasons is greater than 100\% because we observe multiple reasons for some abandoned PRs. We find that the most frequent abandonment reasons are related to the obstacles faced by contributors followed by the hurdles imposed by maintainers during the review process. Particularly, difficulty addressing the maintainers' comments, lack of review from the maintainers, and difficulty resolving the CI failures are the most frequent reasons (observed in more than 20\% of abandoned PRs). In the following, we discuss the identified reasons in the order of their frequency.

\begin{table}
    \caption{Probable reasons why contributors abandon their PRs.}
    \label{tab:reasons}
    \begin{tabular}{@{}l|l|c@{}}
        \toprule
        \textbf{Category}                             & \textbf{Reason}                                    & \textbf{Frequency (\%)} \\
        \midrule
        \multirow{4}{*}{\textbf{Contributor-related}} & Difficulty addressing the maintainers' comments    & 45.8                    \\
                                                      & Difficulty resolving the CI failures               & 20.9                    \\
                                                      & Difficulty resolving the merge issues              & 14.1                    \\
                                                      & Difficulty complying with the project requirements & 1.4                     \\
        \midrule
        \multirow{4}{*}{\textbf{Maintainer-related}}  & Lack of review from the maintainers                & 22.6                    \\
                                                      & Lack of answer from the maintainers                & 9.3                     \\
                                                      & Lack of integration by the maintainers             & 6.5                     \\
                                                      & Lack of consensus among the maintainers            & 4.5                     \\
        \midrule
        \multirow{2}{*}{\textbf{PR-related}}.         & Existence of duplicated work                       & 3.1                     \\
                                                      & Dependency on upcoming changes                     & 1.4                     \\
        \bottomrule
    \end{tabular}
\end{table}

\noindentparagraph{\emph{\textbf{Difficulty addressing the maintainers' comments (45.8\%).}}} In almost half of the abandoned PRs, their contributors found it difficult to address the maintainers' comments, questions, or change requests. The contributors did not have the required technical knowledge or enough time to continue the work (e.g., [P198] said \textit{``Nope. Had no time and will to take on this.''}). Interestingly, contributors may ask others to continue the work (e.g., [P340] said \textit{``If you don't mind taking it over, that would be fantastic. Happy to provide whatever help I can!''}).

\noindentparagraph{\emph{\textbf{Lack of review from the maintainers (22.6\%).}}} The second common reason why contributors abandon their PRs is that they did not receive a timely (if any) review from the maintainers. Sometimes, a PR gets reviewed, and the contributor addresses the maintainers' comments, but the maintainers do not follow it up. For example, [P7] was closed after multiple rounds of discussions, even though the contributor addressed the maintainers' comments (\textit{``Oh dang, I guess it doesn't warn you after the first time. I'll get this re-made soon and try to be a little more proactive about getting input.''}). Not receiving timely reviews from the maintainers may also send a signal to the contributors that their work is not treated seriously (e.g., [P182] said \textit{``I'm happy to resolve these merge conflicts now, but before I do, I am curious if I have messed something else up in my contribution? Since this didn't get a comment since July of last year, I am afraid I've missed something crucial.''}). If PRs are not reviewed in a timely manner, they may become outdated and require extra work from their contributors to rebase and update them.

\noindentparagraph{\emph{\textbf{Difficulty resolving the CI failures (20.9\%).}}} The third reason includes cases where the contributors found it difficult to resolve the Continuous Integration (CI) failures arisen during the review process. Such failures are often brought up by the project bots even before the PR gets a review from the maintainers. Sometimes, the contributors do not even know how to fix the CI failures and ask the maintainers for help (e.g., [P346] said \textit{``I've looked at the errors in Travis. Most of them seem unrelated to the PR. Could you please help me out with this?''}).

\noindentparagraph{\emph{\textbf{Difficulty resolving the merge issues (14.1\%).}}} The fourth reason includes cases where the contributors found it difficult to resolve the merge issues arisen during the review process. If the project codebase has been updated, the contributors are asked to rebase their local branch, resolve any merge conflicts, and then push their changes again. Such issues typically arise when the maintainers take a long time to review the PR, and the PR becomes outdated (e.g., [P181] said \textit{``I am willing to keep rebasing (and certainly willing to continue responding to comments), but not without some indication that it will eventually be merged.''}). We also observe that contributors are sometimes asked to squash their pushed commits into a single commit to reduce the noise in the revision history, which also requires additional effort and time from them.

\noindentparagraph{\emph{\textbf{Lack of answer from the maintainers (9.3\%).}}} In some cases, we observe the reason for abandonment is because the contributors had not received a timely (if any) answer from the maintainers when they asked for help to complete a task (e.g., [P300] said \textit{``CI is timing out. Everything is fine until the timeout. Anything I can do to get this merged?''}) or asked for clarification (e.g., [P313] said \textit{``Modifying the passed in proxyTransport cannot be considered `safe'?''}) or asked for confirmation (e.g., [P273] said \textit{``Is that the right way?''}). Note that this reason is different from ``lack of review from the maintainers'' as such PRs are blocked because the contributor is awaiting an answer to a question from the maintainer and not awaiting a review for the applied changes.

\noindentparagraph{\emph{\textbf{Lack of integration by the maintainers (6.5\%).}}} This reason includes cases where a PR has been already approved by the reviewers but has been pending integration. In some projects, such as Kubernetes, a PR should undergo a two-phase review process. In the first round, reviewers approve the changes, and then project integrators need to merge the changes. However, contributors may abandon their PRs if the integrators do not attempt to merge the PR in a timely (if any) manner after the reviewers have approved the PR. For example, in [P1], the PR has been approved for integration multiple times. However, since the integrators were not responsive, the PR needs to be rebased, requiring additional work from the contributor. In [P307], a reviewer suggested \textit{``Might be worth pinging the reviewers too if that is all that is stopping progress.''}

\noindentparagraph{\emph{\textbf{Lack of consensus among the maintainers (4.5\%).}}} This reason includes cases where the reviewers could not reach a consensus on how to continue with the PR. This disagreement typically arises when there is no straightforward solution to resolve the PR issues, and each alternative has its own advantages and disadvantages. Such PRs often undergo a long discussion and demand lots of time and effort from both reviewers and contributors (e.g., [P354]). However, the PR eventually gets abandoned by the contributor due to inconsistent feedback and often overlong discussions.

\noindentparagraph{\emph{\textbf{Existence of duplicated work (3.1\%).}}} This reason includes cases where the work is either a duplicate of an existing PR (e.g., [P294] said \textit{``I am abandoning this PR in favor of npm-ramda, but I wont close it in case someone want to make use of it.''}), another contributor has submitted a more comprehensive PR, or the issue addressed in the PR is no longer applicable (e.g., [P340] said \textit{``pkg-config should no longer be able to pick up non-deps under superenv. Hopefully that means this is resolved.''}).

\noindentparagraph{\emph{\textbf{Difficulty complying with the project requirements (1.4\%).}}} This reason includes rare cases where the contributors found it difficult to comply with the project-specific requirements. In projects such as Kubernetes, the contributors are asked to sign the contribution level agreements before their PR even gets reviewed by the maintainers (e.g., [P149]). Also, many projects provide templates for the PR description and ask the contributors to update the description according to the templates (e.g., [P134]).

\noindentparagraph{\emph{\textbf{Dependency on upcoming changes (1.4\%).}}} In rare cases, the abandoned PRs were not valuable on their own and depended on other changes that must be merged first before the proposed changes can be considered (e.g., [P11] said \textit{``Hopefully, once mappable and partial types land in TS, I will fix this.''}).

\bigskip
\begin{tcolorbox}
    \paragraph{\emph{\textbf{Answer to RQ\textsubscript{3}.}}} Our findings suggest that the most frequent abandonment reasons are related to the obstacles faced by contributors followed by the hurdles imposed by maintainers during the review process. Specifically, difficulty addressing the maintainers' comments, lack of review from the maintainers, difficulty resolving the CI failures, and difficulty resolving the merge issues are the most common reasons why contributors abandon their PRs.
\end{tcolorbox}
\medskip

\section{Perspectives of the Project Maintainers}
\label{sec:perspectives}
To gain deeper insights on PR abandonment, we design a survey asking the core maintainers of the studied projects about their perspectives on our findings and how to tackle PR abandonment. After explaining the goal of this research and presenting a summary of our findings, we ask the following three open-ended questions in the survey:

\begin{itemize}
    \item Does your team implement any approaches to deal with abandoned PRs? If so, what approaches does your team use?
    \item Do you suggest any approaches that can minimize the risk of a PR being abandoned by its contributor? (these can be approaches that your team does or does not use)
    \item Do you have any feedback about the four findings of our study? (we would love to hear it, positive or negative)
\end{itemize}

We send e-mails to the top 25 core developers of each studied project (a total of 250 e-mails) to invite them to participate in our survey. From these invitations, we receive a total of 16 responses (6.5\%) to our survey. Our response rate is similar to the 5\% response rate, commonly observed in software engineering studies \citep{lethbridge_studying_2005}. From the ten studied projects, we received responses from all the projects except Elasticsearch. As our sample is small, we refrain from discussing particular projects and instead present the overarching themes that appeared in the participant responses.

\subsection{How do projects \emph{deal} with abandoned PRs?} In this question, we aim to understand the processes and practices that the projects have already put in place to deal with abandoned PRs. Out of the 16 participants, six responded that their team had implemented approaches to deal with abandoned PRs. These approaches are:

\noindentparagraph{\emph{\textbf{Holding triage meetings (4x).}}} The most commonly mentioned approach to deal with abandoned PRs is holding (recurrent) triage meetings, where developers review the status of PRs and support PRs that need attention. According to the survey participants, triage meetings streamline the communication and help find problems before it causes the PRs to become abandoned.

\begin{quotebox}
    \textit{``We have regular triage meetings to review the status of PRs that abandoned or about to be abandon. We also help new comers and new maintainers for reviving the old PRs and merge them.''} [S1]
    \medskip\\
    \textit{``Better upfront planning and communication help eliminate abandoned PRs. Many times, a quick zoom with a teammate to talk through the proposed change goes a long way into finding problems before a large effort is needed.''} [S6]
\end{quotebox}

\noindentparagraph{\emph{\textbf{Using bots to auto-close abandoned PRs (2x).}}} While all the projects use Stale bot \citep{github_stale_2021} or a similar in-house implementation (e.g., fejta-bot in Kubernetes \citep{kubernetes_fejta-bot_2021}) to follow up with PRs that are about to get abandoned and auto-close already abandoned PRs, two participants explicitly mentioned their use of such bots. As a respondent explained:

\begin{quotebox}
    \textit{``A bot first pings owners of the code after a week. Then it pings the submitter a couple of weeks later, telling them that it will be closed soon.''} [S2]
\end{quotebox}

Most of the participants (10 responses) reported that their project does not implement any particular approach to deal with abandoned PRs. Of these participants, four explained why their team had not adopted any specific approach to prevent PR abandonment:

\noindentparagraph{\emph{\textbf{The maintainers are overwhelmed with work (2x).}}} Two maintainers reported that they are overwhelmed with work and thus have decided to reduce their efforts on retaining PRs from external contributors. As a respondent explained:

\begin{quotebox}
    \textit{``We are overwhelmed with work, and most community PRs have a low ration value/effort needed to merge it, so we essentially gave up, except for rare cases.''} [S13]
\end{quotebox}

\noindentparagraph{\emph{\textbf{The project does not rely on open source contributions (2x).}}} Two participants mentioned that their project does not rely on external contributions for the implementation of new features or improvements. Therefore, there is little incentive to spend special time and effort in retaining PRs from external contributors. As a respondent explained:

\begin{quotebox}
    \textit{``We appreciate community PRs, but the vast majority of the contributors to our project are employees and so we don't rely on open source contributions for features or improvements. Community PRs are usually applicable to a specific niche usecase which we'd be happy to accept if the contributor is willing to go through the process with us.''} [S5]
\end{quotebox}

\subsection{How do maintainers \emph{recommend} to mitigate PR abandonment?} In this question, we aim to understand the approaches that the maintainers recommend to minimize the risk of PRs getting abandoned by their contributors, whether their team has adopted them yet or not. Out of the 16 participants, 15 responded with suggestions that they believe could mitigate PR abandonment. In the following, we summarize these recommendations into suggestions for contributors and suggestions for maintainers.

\noindentparagraph{\emph{\textbf{Maintainers should strive to make the contributor experience as smooth as possible (6x).}}} It stands to reason that the more obstacles contributors face, the higher the chances of abandonment. One participant (S5) mentioned that having helpful and understandable error messages can help contributors fix issues in their PRs. Another participant (S3) mentioned that maintainers should ``put kid gloves on'' when dealing with newcomers and make community contributions as painless as possible. One participant (S15) even suggested that maintainers merge PRs with minor issues and then either make the required changes themselves or open an issue in the project. As one respondent aptly summarizes: 

\begin{quotebox}
    \textit{``Responding quickly and encouragingly, and making your PR contributor experience as seamless as possible (automatic CI, helpful and understandable error messages) are likely all you can do. It's a lot of work even to submit a PR, the extra work necessary when updates are needed isn't something most people will be willing to give.''} [S5]
\end{quotebox}

The participants also mentioned that improving the project's testing documentation (S1) and contribution guidelines (S2), and making bot instructions more understandable (S2) could help to mitigate PR abandonment. One participant (S13) also cited that projects need to increase their available resources, with another participant (S1) mentioning that increasing community reach and having maintainers from the community may help maintainers better handle the required workload.

\noindentparagraph{\emph{\textbf{Maintainers should establish a triage process for external contributions (2x).}}} Once again, the participants mentioned the importance of a triage process in mitigating PR abandonment. A triage process helps assign reviewers to a PR based on their expertise and experience and may lead to timely responses to contributors. One participant (S14) suggested that PRs should preferably be assigned to one reviewer instead of an entire team to compel reviewers to act and prevent idleness. Another respondent described that the triage process should also monitor the status of PRs and act if reviewers have not responded to the contributors yet:

\begin{quotebox}
    \textit{``[Projects should have] a human rotation that triages pull requests and checks if they are progressing, or if someone has dropped the ball or is waiting on some event that will never happen.``} [S16]
\end{quotebox}

\noindentparagraph{\emph{\textbf{Contributors should create PRs that are clear and concise (3x).}}} The participants emphasized the importance of PRs to focus only on a single use case and provide a clear description of changes. PRs that include multiple unrelated changes create a burden for reviewers that need to ensure all changes are correct. As two participants responded:

\begin{quotebox}
    \textit{``Make the proposal clear and concise. The reviewer might take 10 or 15 seconds to figure out the problem being solved and the solution. If it takes longer, the reviewer will probably give up.''} [S11]
    \medskip\\
    \textit{``Smaller PRs are obviously better. We try not to nit-pick though it is human nature that a 100 line diff gets no nitpicking and a 4 line diff gets plenty. This can hardly make people want to contribute and is unfortunate.''} [S9]
\end{quotebox}

\noindentparagraph{\emph{\textbf{Contributors should assess the project's interest in the proposed changes before submitting PRs (3x).}}} Managing expectations is important in contributing to open source projects. As different projects have different philosophies and needs regarding community contributions, contributors should assess whether maintainers value their PRs. Typically, new features are harder to integrate than bug fixes and performance improvement patches.

\begin{quotebox}
    \textit{``In large projects, a performance improvement or a bug fix proposed by an external contributor is more likely to be merged than a new feature. Not only because it's usually simpler, but because the feature might not be in line with the project's interests. A bug fix or a performance improvement is always in line with the project's interest.''} [S10]
    \medskip\\
    \textit{``Managing expectations could be a factor. We won't merge entirely new features just like that, simply because it is really important that Odoo remains and improves on being kept simple. E.g. PRs could come from a client project that needs a certain feature, but merging it like this, might do harm to a lot of other clients or cause more further problems where we need some distance to really think about the best solution.''} [S12]
\end{quotebox}

One way to assess the validity of the contribution is to open an issue in the project. The contribution guidelines of the majority of the projects explicitly state that contributors should first open an issue to discuss their contributions and defer the implementation until when the maintainers have agreed on the usefulness of the proposed changes. As a participant responded:

\begin{quotebox}
    \textit{``Opening issues to discuss changes prior to posting PRs helps reduce abandoned PRs. Discussing the change ahead of time gives developers and the community time to explore the change and ensure that the 1) proposed change is one that the project wants to maintain, 2) proposed change is scalable, 3) proposed change is a feature that is needed by many use cases and not a one off for specific use case, 4) proposed implementation is maintainable and fits with the architecture and future of the project.''} [S6]
\end{quotebox}

\noindentparagraph{\emph{\textbf{Both contributors and maintainers should be more upfront about their intentions (2x).}}} Two participants stated that communication should be improved from both sides. Maintainers need to be upfront about their intentions on merging (or not) the contribution from contributors to avoid wasting effort and time. As a participant responded:

\begin{quotebox}
    \textit{``There needs to be a clear signal from the project to the PR contributor if there is no interest at all in the PR, or if there is interest, what needs to be fixed to be accepted. Then participate or help, or signal when the effort can no longer be sustained. If there is no clear signal, the contributor has no idea what's going on. Automated "closer-bots" cannot solve this problem (or make it worse).''} [S8]
\end{quotebox}

Similarly, contributors should mention their willingness to make the requested changes. One participant mentioned that anxiety could play an important factor leading to PR abandonment, particularly when miscommunication happens to newcomers to the project:

\begin{quotebox}
    \textit{``Anxiety about contribution probably leads to some abandonment. I suggest to all newish contributors that they remember the people in charge of these projects where just like them once.''} [S9]
\end{quotebox}

\subsection{How do developers \emph{interpret} our findings?} We provided participants with a pre-print of this manuscript and encouraged participants to report any negative or positive remarks they had about our findings. Out of the 16 participants, 14 participants commented on our findings (~87.5\%). \Cref{tab:explanation} overviews the explanations provided by survey participants regarding each of our main findings. In the following, we discuss the survey responses for each of our findings in more detail.

\begin{table}
    \caption{Overview of the explanation of our findings based on our survey responses.}
    \label{tab:explanation}
    \resizebox{\textwidth}{!}{%
        \begin{tabular}{@{}l|l@{}}
            \toprule
            \textbf{Study Findings}                                                    & \textbf{Survey Explanations}                                                                                                                                                                                      \\
            \midrule
            Complex PRs are more likely to get abandoned.                              & \begin{tabular}{@{\labelitemi\hspace{\tabcolsep}}l@{}}Complex PRs are less likely to get a timely review.\\Contributor becomes frustrated by frequent change requests.\end{tabular}                               \\
            \midrule
            Novice contributors are more likely to abandon their PRs.                  & \begin{tabular}{@{\labelitemi\hspace{\tabcolsep}}l@{}}Maintainers expect quality changes regardless of the contributor experience.\\Novice contributors may find the contribution process difficult.\end{tabular} \\
            \midrule
            PRs with long discussions are more likely to get abandoned.                & \begin{tabular}{@{\labelitemi\hspace{\tabcolsep}}l@{}}Long discussions often indicate a controversial PR.\\Lack of unanimous decisions lengthens the review process.\end{tabular}                                 \\
            \midrule
            Projects have a significant influence on the likelihood of PR abandonment. & \begin{tabular}{@{\labelitemi\hspace{\tabcolsep}}l@{}}Maintainer team structure, attitude, and workload influence PRs.\\Project scope, architecture, and ownership influence PRs.\end{tabular}                    \\
            \bottomrule
        \end{tabular}
    }
\end{table}

\noindentparagraph{\emph{\textbf{Complex PRs are more likely to get abandoned.}}} Our findings suggest that PRs with lengthier descriptions, more commits, or more changed lines are more likely to get abandoned (RQ\textsubscript{1}--RQ\textsubscript{2}). The participants argued that such complex PRs require extra efforts from both their contributors and reviewers. Therefore, such PRs might linger for a while before getting reviewed, and also might require more changes from the contributor to become satisfactory:

\begin{quotebox}
    \textit{``The more complex a PR is, the less likely a reviewer is going to spend valuable time on it, in particular if the contributor is not well known.''} [S11]
    \medskip\\
    \textit{``Either the maintainers leave them open for months or the contributor is frustrated by the numerous requested changes.''} [S9]
\end{quotebox}

\noindentparagraph{\emph{\textbf{Novice contributors are more likely to abandon their PRs.}}} Our findings suggest that contributors who submitted fewer PRs, have a lower acceptance rate, have a lower contribution period, or have a higher abandonment rate within a project are more likely to abandon their PRs (RQ\textsubscript{1}--RQ\textsubscript{2}). The participants suggested that this can be due to projects expecting high-quality changes (with proper formatting, documentation, and description of changes) that often require access to experienced maintainers. However, projects can lower their expectations from external contributors:

\begin{quotebox}
    \textit{``Like any wall in life, it is scary and something to throw yourself against. The project can help—making contributions feel more welcome. Unfortunately if you are too welcoming to contribution you get rapidly overwhelmed by contribution and cannot accept it all.''} [S9]
    \medskip\\
    \textit{``The main issue I see with abandoned PRs is that the bar for a commit is too high. To give you some perspective, new developers in my team take a few weeks to land their first commit. And that is with constant access to experienced developers/mentors. This is because we expect our commit to have: proper tests, linting, inline documentation, references to relevant commits, documentation in some cases, good commit message, and targetting the correct branch. I think we should lower the bar for external contributors (so they can quickly land a fix), but eventually, still add an additional commit with a test or something if needed. Clearly, doing that requires some resources. I tried doing that a few years ago, and was quickly slammed with pings everywhere to ask me to work on those PRs!''} [S13]
\end{quotebox}

\noindentparagraph{\emph{\textbf{PRs with long discussions are more likely to get abandoned.}}} Our findings suggest that PRs involving more participants, more responses from the participants or from the contributor, or with higher latency to get a response from a reviewer in a project are more likely to get abandoned (RQ\textsubscript{1}--RQ\textsubscript{2}). The participants argued that long discussions could indicate a controversial PR lacking a unanimous solution to address the PR issues:

\begin{quotebox}
    \textit{``Long discussions themselves are not a problem. However, they are often indicative of disagreement, a complex topic, or an absence of a solution (the solution in the PR is not correct, but commenters on the PR don't have a good suggestion either).''} [S16]
    \medskip\\
    \textit{``[long discussions] seems like a proxy for controversial PRs, which I suspect is also a factor at play. In my personal experience, open source projects are often lead by passionate individuals with strong opinions about the way things should be done, which often conflict with the opinions of new people to the project. I think this could be addressed by discussing changes beforehand but this is another hurdle which makes contributing more challenging.''} [S5]
\end{quotebox}

However, a strong leadership team could prevent such controversial PRs from becoming extra lengthy:

\begin{quotebox}
    \textit{``Good projects have strong leaders that make decisions and don’t deliberate too long. Bitcoin suffered here greatly after Satoshi left.''} [S9]
\end{quotebox}

\noindentparagraph{\emph{\textbf{Projects have a significant influence on the likelihood of PR abandonment.}}} Our findings suggest that projects have a significant influence over PR abandonment and that throughout the history of projects, the rate of PR abandonment has significantly fluctuated as the projects evolved. The participants suggested that such fluctuations might be related to changes in the attitude of the team, scope and architecture of the software, and popularity and ownership of the project:

\begin{quotebox}
    \textit{``It's a multitude of factors. Attitudes of maintainers. Software architecture of the projects (if well designed you can expect well designed PRs). Scope of the projects (ie. documentation is clear on the scope to prevent PRs that feature creep). Fame: too big a project will get too much contribution and likely there will be insufficient people to manage it.''} [S9]
    \medskip\\
    \textit{``You might also take into account the ownership of the project: is it a community project? A company project? Who is writing the roadmap of the project? A company project has a roadmap in line with its business development, which is not the case for a community project.''}
\end{quotebox}

\section{Discussion}
\label{sec:discussion}
Combining the results from our quantitative (RQ\textsubscript{1}--RQ\textsubscript{2}) and qualitative (RQ\textsubscript{3}) investigation provides evidence that contributors and the review process play a more prominent role in PR abandonment than projects and PRs themselves. In the following, we integrate the findings from our three research questions and our survey with core developers and further discuss the implications of our findings.

\noindentparagraph{\emph{\textbf{The Role of Contributors in PR Abandonment.}}} Our findings indicate that the contributors of abandoned PRs usually have less experience than the contributors of nonabandoned PRs. Specifically, we observed that novice contributors who have submitted fewer PRs, have a lower acceptance rate, or have a lower contribution period within a project are more likely to abandon their PRs in most of our studied projects (RQ\textsubscript{1}--RQ\textsubscript{2}). Our survey results suggest that novice contributors often find the contribution process more difficult as maintainers typically expect high-quality changes (with proper tests, formatting, documentation, and description of changes) regardless of contributor experience before approving the changes to get merged (\Cref{sec:perspectives}). We also observed that the contributors of abandoned PRs have frequently faced many obstacles (due to lack of enough knowledge, time, or even interest) to continue and complete the review process (RQ\textsubscript{3}). Indeed, inexperienced contributors face various barriers in making their contributions accepted \citep{steinmacher_why_2013, steinmacher_preliminary_2014, steinmacher_almost_2018}. Prior studies have also reported the positive impact of contributor experience in acceptance and review time of PRs \citep{gousios_exploratory_2014, soares_acceptance_2015, yu_determinants_2016, kononenko_studying_2018}. Our survey respondents recommend maintainers either lower their expectations or be more attentive and supportive towards external contributors (especially casual contributors or newcomers) throughout the review process. Also, contributors can discuss their proposed changes before submitting a PR to facilitate the review process, especially if the change introduces new features or involves large changes. This discussion helps contributors to ensure that their proposed changes align with the project roadmap and design. Contributors are also expected to adhere to contribution guidelines and project conventions as it helps them have a better grasp of the review process.

\noindentparagraph{\emph{\textbf{The Role of Review Processes in PR Abandonment.}}} Our findings indicate that the review process of abandoned PRs is usually lengthier than the review process of nonabandoned PRs. Specifically, we observed that lengthy PRs which involve more participants, or more responses from the participants or from the contributor are more likely to get abandoned in most of our studied projects (RQ\textsubscript{1}--RQ\textsubscript{2}). Our survey results suggest that long discussions are often indicative of a controversial PR that is addressing a complex issue or does not have a unanimously accepted solution (\Cref{sec:perspectives}). We also observed that abandoned PRs frequently lack a (timely) review, response, or even action from the maintainers which also unnecessarily lengthens the review process of a PR (RQ\textsubscript{3}). Prior studies have also reported that high response latencies and lengthy discussions negatively impact the acceptance and review time of PRs \citep{tsay_influence_2014, yu_wait_2015, yu_determinants_2016, kononenko_studying_2018}. Our survey respondents recommend maintainers be more responsive and support external contributors (especially casual contributors or newcomers) till the completion of their PRs. In cases that a PR needs only trivial changes, maintainers can merge the PR as is and either implement the changes themselves or open a new issue for the required changes. Also, maintainers can hold recurrent triage meetings to review the status of PRs and support PRs that need attention to mitigate PR abandonment. In cases where a PR has become lengthy, lead maintainers should involve and decide on the outcome of the PR using a voting process.

\noindentparagraph{\emph{\textbf{The Role of Projects in PR Abandonment.}}} Our findings indicate that projects have a significant influence over PR abandonment. Specifically, we observed that the rate of abandoned PRs has significantly fluctuated throughout the history of projects, with some projects constantly decreasing the abandonment rate as they become mature, i.e., Ansible, Kibana, and Odoo (RQ\textsubscript{1}--RQ\textsubscript{2}). Our survey results suggest that projects typically undergo changes in their team, size, architecture, scope, policies, practices, or even ownership during their development lifecycle. Such changes bring with them both positive and negative aspects, which can fluctuate the rate of PR abandonment (\Cref{sec:perspectives}). Prior studies have also reported that project maturity has a mixed impact on the acceptance and review time of PRs \citep{tsay_influence_2014, yu_determinants_2016}. Our survey respondents recommend projects streamline their contribution process as much as possible to better accommodate new contributors.

\noindentparagraph{\emph{\textbf{The Role of PRs in PR Abandonment.}}} Our findings indicate that abandoned PRs are usually more complex than nonabandoned PRs. Specifically, we observed that complex PRs which have lengthier descriptions or more commits are more likely to get abandoned in most of our studied projects (RQ\textsubscript{1}--RQ\textsubscript{2}). A PR can be complex at submission time, when it contains too many commits or an abnormally lengthy description, or become more complex as its contributor submit additional commits (and thus makes more changed lines) during the review process to address the changes requested by the maintainers. Our survey results suggest that a complex PR is more likely to linger for a while before getting a first or even a follow-up review, especially if its contributor is not well-known to the maintainers (\Cref{sec:perspectives}). We also observed the lack of review from the maintainers as a frequent reason among abandoned PRs (RQ\textsubscript{3}). Prior studies have also reported that complex PRs negatively impact their acceptance and review time \citep{tsay_influence_2014, soares_acceptance_2015, yu_wait_2015, yu_determinants_2016, kononenko_studying_2018}. Our survey respondents recommend contributors make their PRs clear, concise, and focused as complex PRs are more difficult to review and require more interactions with the contributor to become ready (\Cref{sec:perspectives}). Also, maintainers expect PRs to have proper tests, formatting, documentation, and description of changes according to the project requirements (\Cref{sec:perspectives}).

\section{Related Work}
\label{sec:related_work}
\noindentparagraph{\emph{\textbf{PR Latency and Decision.}}} The literature has extensively studied how various technical, social, and personal factors influence the acceptance and review process of PRs. \citet{gousios_exploratory_2014} investigated how technical factors affect the merge decision and merge time of PRs. They found that the merge decision is mainly affected by whether the PR touches recently modified code. Also, they observed that the contributor's experience as well as the project's size, test coverage, and openness to external contributions influence the merge time. \citet{gousios_work_2015} found that the decision of maintainers to accept a PR is driven by its quality, especially conformance to the project style and architecture, code quality, and test coverage.

\citet{tsay_influence_2014} found that both technical and social factors influence PR acceptance, especially PRs with many comments are less likely to be accepted. Additionally, they observed that well-established projects are more conservative in accepting PRs. \citet{soares_acceptance_2015} found that the programming language, the number of commits, and the number of files added in a PR, as well as whether its contributor is an external developer and whether it is the contributor's first PR, influence the merge decision and merge time. \citet{yu_wait_2015} found that the PR size, the first response delay, and the availability of CI pipelines impact the review time of PRs. \citet{yu_determinants_2016} found that projects prefer PRs that are small, have less controversy, and are submitted by trusted contributors.

\citet{kononenko_studying_2018} found that the size of PRs, the number of participants in the review discussions, and the contributor's experience and affiliation influence both the review time and merge decision. Moreover, they reported that developers consider PR quality, type of change, and responsiveness of the contributor as important factors in the merge decision. Developers perceive the quality of a PR by its description, complexity, and revertability; and the quality of a review by its feedback, tests, and discussions. \citet{pinto_who_2018} found that in company-owned open-source projects, external contributors compared to the employees face significantly more rejections and have to wait longer to receive a decision on their contributions. \citet{zou_how_2019} found that PRs that violate the code style of projects are more likely to get rejected and take longer to get closed. \citet{lenarduzzi_does_2021} found that code quality does not affect the acceptance of PRs, and suggested that other factors such as the maintainer's reputation and the feature's importance might be more influential on PR acceptance.

Several studies have also investigated how demographic characteristics of contributors can influence the outcome of PRs. \citet{terrell_gender_2017} found that among external contributors, women whose gender is identifiable have lower acceptance rates. \citet{rastogi_biases_2016} and \citet{ rastogi_relationship_2018} also found that PRs are more likely to get accepted when both the contributors and the maintainers are from the same geographical location. Moreover, \citet{nadri_insights_2021} found evidence of bias against perceptible non-White races. Later, \citet{nadri_relationship_2021} found that contributions from perceptible White developers have a higher acceptance chance, and perceptible non-White contributors are more likely to get their PRs accepted if the maintainer is also from the same race. \citet{furtado_how_2021} also found that contributors from countries with low human development indexes submit fewer PRs but face the most rejections. Beside social and technical factors, \citet{iyer_effects_2021} also studied how personality traits \citep{deyoung_between_2007} influence PR acceptance. They found that personal and technical factors play a significant and comparable role in PR acceptance, but still not to the extent of social factors. Additionally, they observed that contributors who are more open and conscientious but less extroverted have a higher chance of getting their PRs accepted. Similarly, maintainers who are more conscientious, extroverted, and neurotic accept more PRs.

\noindentparagraph{\emph{\textbf{Duplicated PRs.}}} \citet{li_redundancy_2022} found that duplicate PRs waste human and computing resources and adversely impact OSS development. To facilitate studies on duplicated PRs, \citet{yu_dataset_2018} compiled a dataset of duplicated PRs from 26 popular GitHub projects. To identify duplicate PRs, \citet{li_detecting_2017} proposed an approach that uses textual information within PRs to automatically identify similar PRs. \citet{li_detecting_2021} extended the previous work by also considering the change information of PRs. \citet{ren_identifying_2019} proposed an approach to identify redundant code changes in forks as early as possible. \citet{wang_duplicate_2019} enhanced the performance of the previous approach by considering the time factor.

\noindentparagraph{\emph{\textbf{Abandoned PRs.}}} Recently, \citet{li_are_2021} manually examined 321 abandoned PRs from five GitHub projects (namely, Cocos2d-x, Kubernetes, Node.js, Rails, and Rust) to identify the reasons why PRs get abandoned by their contributors, the impact of abandoned PRs on the maintainers, and the strategies adopted by the projects to deal with abandoned PRs. Then, they quantified the frequency of the identified reasons, impacts, and strategies by surveying 710 developers of 100 popular GitHub projects. They found that the reasons why contributors abandon their PRs relate to the lack of maintainers' responsiveness and the lack of contributors' time and interest. While this study discussed the developers' perspective on PR abandonment, the influence of the factors related to PRs, contributors, review processes, and projects on the abandonment probability of PRs is still not known. To fill this knowledge gap, we curated a larger dataset consisting of 10 popular and mature GitHub projects and analyzed abandoned PRs from both a quantitative and qualitative perspective.

\section{Limitations}
\label{sec:limitations}
\noindentparagraph{\emph{\textbf{Threats to Internal Validity.}}} Threats to internal validity are concerned about the issues that might affect the validity of our findings. The first threat is related to our definition of abandoned PRs. We define abandoned PRs as those promising PRs that have been neither integrated nor rejected because their contributors have left the review process unfinished. While in our preliminary investigation, we rarely found cases where another developer continues an abandoned PR, but this can be systematically investigated in future studies. The second threat is related to the process of identifying abandoned PRs. Our heuristics may have missed some truly abandoned PRs and wrongly marked some PRs as abandoned. To mitigate this threat, we considered as many relevant keywords as possible by iteratively refining our keywords as we observed new patterns in the discussion comments of known abandoned PRs. Also, we assessed the quality of our dataset by manually investigating 100 abandoned PRs. The third threat is related to the process of identifying the reasons why PRs get abandoned by their contributors. We may have drawn wrong conclusions in card sorting because the coders may have had preconceptions. To minimize this bias, each PR was independently labeled by at least two authors, and then the three authors discussed and merged the labels. The fourth threat is related to the completeness of the abandonment reasons. To further minimize this risk, we coded all the remaining cards when saturation was reached in card sorting. We also performed a second pass over all cards to ensure that we did not miss any important information.

\noindentparagraph{\emph{\textbf{Threats to External Validity.}}} Threats to external validity are concerned with the generalizability of our findings across different projects. To conduct our study, we focused on ten popular GitHub projects with the richest historical PR data. Although the studied projects cover several different application domains and programming languages, they do not represent the entire open-source ecosystem. Therefore, our findings may not generalize beyond our studied projects, especially since we observe conflicting patterns across different projects due to their inherent differences. Future replication studies with a more diverse selection of projects both inside and outside the open-source ecosystem are required to obtain more widely applicable insights. Also, our survey findings are based on the responses from 16 participants. While these participants are all among the top core maintainers of the studied projects, different maintainers may have different perspectives, and thus our findings may not be generalized to other settings.

\section{Conclusion}
\label{sec:conclusion}
Abandoned PRs waste the time and effort of their contributors and their reviewers. To provide more comprehensive insights into the underlying dynamics of PR abandonment, we conducted a mixed-methods study on ten popular and mature GitHub projects. Using statistical techniques, we found that abandoned PRs tend to be more complex, their contributors tend to be less experienced, and their review processes tend to be lengthier than nonabandoned PRs. We then relied on machine learning techniques to determine the relative importance of the features and describe how each feature varies the predicted abandonment probability of PRs. We found that the features of review processes, contributors, and projects are more important for predicting PR abandonment than the features of PRs themselves. Specifically, PRs with more than three responses from either the participants or the contributors, and those submitted by novice contributors are more likely to get abandoned. Also, the abandonment probability changes as projects evolve, with half the projects showing a decrease in abandonment in their mature stages and the other half showing an increase in abandonment. To identify the probable reasons why contributors abandon their PRs, we manually examined a random sample of abandoned PRs. We found that difficulty addressing the maintainers’ comments, lack of review from the maintainers, difficulty resolving the CI failures, and difficulty resolving the merge issues are the most common reasons why contributors abandon their PRs. Finally, we surveyed the top core maintainers of the studied projects to gain additional insights on how they deal with or suggest dealing with abandoned PRs and their perspectives on our findings. Combining the findings from our research questions and survey responses, we discussed the role of PRs, contributors, review processes, and projects in PR abandonment.

\bibliographystyle{ACM-Reference-Format}
\bibliography{references}


\begin{thebibliography}{82}


\ifx \showCODEN    \undefined \def \showCODEN     #1{\unskip}     \fi
\ifx \showDOI      \undefined \def \showDOI       #1{#1}\fi
\ifx \showISBNx    \undefined \def \showISBNx     #1{\unskip}     \fi
\ifx \showISBNxiii \undefined \def \showISBNxiii  #1{\unskip}     \fi
\ifx \showISSN     \undefined \def \showISSN      #1{\unskip}     \fi
\ifx \showLCCN     \undefined \def \showLCCN      #1{\unskip}     \fi
\ifx \shownote     \undefined \def \shownote      #1{#1}          \fi
\ifx \showarticletitle \undefined \def \showarticletitle #1{#1}   \fi
\ifx \showURL      \undefined \def \showURL       {\relax}        \fi
\providecommand\bibfield[2]{#2}
\providecommand\bibinfo[2]{#2}
\providecommand\natexlab[1]{#1}
\providecommand\showeprint[2][]{arXiv:#2}

\bibitem[Apley and Zhu(2020)]%
        {apley_visualizing_2020}
\bibfield{author}{\bibinfo{person}{Daniel~W. Apley} {and}
  \bibinfo{person}{Jingyu Zhu}.} \bibinfo{year}{2020}\natexlab{}.
\newblock \showarticletitle{Visualizing the effects of predictor variables in
  black box supervised learning models}.
\newblock \bibinfo{journal}{\emph{Journal of the Royal Statistical Society:
  Series B (Statistical Methodology)}} \bibinfo{volume}{82},
  \bibinfo{number}{4} (\bibinfo{year}{2020}), \bibinfo{pages}{1059--1086}.
\newblock
\urldef\tempurl%
\url{https://doi.org/10.1111/rssb.12377}
\showDOI{\tempurl}


\bibitem[Ben-Shachar et~al\mbox{.}(2020)]%
        {ben-shachar_effectsize_2020}
\bibfield{author}{\bibinfo{person}{Mattan~S. Ben-Shachar},
  \bibinfo{person}{Daniel Lüdecke}, {and} \bibinfo{person}{Dominique
  Makowski}.} \bibinfo{year}{2020}\natexlab{}.
\newblock \showarticletitle{effectsize: {Estimation} of effect size indices and
  standardized parameters}.
\newblock \bibinfo{journal}{\emph{Journal of Open Source Software}}
  \bibinfo{volume}{5}, \bibinfo{number}{56} (\bibinfo{year}{2020}),
  \bibinfo{pages}{2815}.
\newblock
\urldef\tempurl%
\url{https://doi.org/10.21105/joss.02815}
\showDOI{\tempurl}


\bibitem[Bischl et~al\mbox{.}(2016)]%
        {bischl_mlr_2016}
\bibfield{author}{\bibinfo{person}{Bernd Bischl}, \bibinfo{person}{Michel
  Lang}, \bibinfo{person}{Lars Kotthoff}, \bibinfo{person}{Julia Schiffner},
  \bibinfo{person}{Jakob Richter}, \bibinfo{person}{Erich Studerus},
  \bibinfo{person}{Giuseppe Casalicchio}, {and} \bibinfo{person}{Zachary~M.
  Jones}.} \bibinfo{year}{2016}\natexlab{}.
\newblock \showarticletitle{mlr: {Machine} learning in {R}}.
\newblock \bibinfo{journal}{\emph{Journal of Machine Learning Research}}
  \bibinfo{volume}{17}, \bibinfo{number}{170} (\bibinfo{year}{2016}),
  \bibinfo{pages}{1--5}.
\newblock
\urldef\tempurl%
\url{https://jmlr.org/papers/v17/15-066.html}
\showURL{%
\tempurl}


\bibitem[Borges et~al\mbox{.}(2016)]%
        {borges_understanding_2016}
\bibfield{author}{\bibinfo{person}{Hudson Borges}, \bibinfo{person}{Andre
  Hora}, {and} \bibinfo{person}{Marco~Tulio Valente}.}
  \bibinfo{year}{2016}\natexlab{}.
\newblock \showarticletitle{Understanding the factors that impact the
  popularity of {GitHub} repositories}. In
  \bibinfo{booktitle}{\emph{Proceedings of the 32nd {International}
  {Conference} on {Software} {Maintenance} and {Evolution} ({ICSME})}}.
  \bibinfo{pages}{334--344}.
\newblock
\urldef\tempurl%
\url{https://doi.org/10.1109/ICSME.2016.31}
\showDOI{\tempurl}


\bibitem[Borges and Valente(2018)]%
        {borges_whats_2018}
\bibfield{author}{\bibinfo{person}{Hudson Borges} {and}
  \bibinfo{person}{Marco~Tulio Valente}.} \bibinfo{year}{2018}\natexlab{}.
\newblock \showarticletitle{What's in a {GitHub} star? {Understanding}
  repository starring practices in a social coding platform}.
\newblock \bibinfo{journal}{\emph{Journal of Systems and Software}}
  \bibinfo{volume}{146} (\bibinfo{year}{2018}), \bibinfo{pages}{112--129}.
\newblock
\urldef\tempurl%
\url{https://doi.org/10.1016/j.jss.2018.09.016}
\showDOI{\tempurl}


\bibitem[Bradley(1997)]%
        {bradley_use_1997}
\bibfield{author}{\bibinfo{person}{Andrew~P. Bradley}.}
  \bibinfo{year}{1997}\natexlab{}.
\newblock \showarticletitle{The use of the area under the {ROC} curve in the
  evaluation of machine learning algorithms}.
\newblock \bibinfo{journal}{\emph{Pattern Recognition}} \bibinfo{volume}{30},
  \bibinfo{number}{7} (\bibinfo{year}{1997}), \bibinfo{pages}{1145--1159}.
\newblock
\urldef\tempurl%
\url{https://doi.org/10.1016/S0031-3203(96)00142-2}
\showDOI{\tempurl}


\bibitem[Breiman(2001)]%
        {breiman_random_2001}
\bibfield{author}{\bibinfo{person}{Leo Breiman}.}
  \bibinfo{year}{2001}\natexlab{}.
\newblock \showarticletitle{Random forests}.
\newblock \bibinfo{journal}{\emph{Machine Learning}} \bibinfo{volume}{45},
  \bibinfo{number}{1} (\bibinfo{year}{2001}), \bibinfo{pages}{5--32}.
\newblock
\urldef\tempurl%
\url{https://doi.org/10.1023/A:1010933404324}
\showDOI{\tempurl}


\bibitem[Cliff(1993)]%
        {cliff_dominance_1993}
\bibfield{author}{\bibinfo{person}{Norman Cliff}.}
  \bibinfo{year}{1993}\natexlab{}.
\newblock \showarticletitle{Dominance statistics: {Ordinal} analyses to answer
  ordinal questions}.
\newblock \bibinfo{journal}{\emph{Psychological Bulletin}}
  \bibinfo{volume}{114}, \bibinfo{number}{3} (\bibinfo{year}{1993}),
  \bibinfo{pages}{494--509}.
\newblock
\urldef\tempurl%
\url{https://doi.org/10.1037/0033-2909.114.3.494}
\showDOI{\tempurl}


\bibitem[Cohen(1960)]%
        {cohen_coefficient_1960}
\bibfield{author}{\bibinfo{person}{Jacob Cohen}.}
  \bibinfo{year}{1960}\natexlab{}.
\newblock \showarticletitle{A coefficient of agreement for nominal scales}.
\newblock \bibinfo{journal}{\emph{Educational and Psychological Measurement}}
  \bibinfo{volume}{20}, \bibinfo{number}{1} (\bibinfo{year}{1960}),
  \bibinfo{pages}{37--46}.
\newblock
\urldef\tempurl%
\url{https://doi.org/10.1177/001316446002000104}
\showDOI{\tempurl}


\bibitem[Creswell and Creswell(2017)]%
        {creswell_research_2017}
\bibfield{author}{\bibinfo{person}{John~W. Creswell} {and}
  \bibinfo{person}{J.~David Creswell}.} \bibinfo{year}{2017}\natexlab{}.
\newblock \bibinfo{booktitle}{\emph{Research {Design}: {Qualitative},
  {Quantitative}, and {Mixed} {Methods} {Approaches}} (\bibinfo{edition}{5th}
  ed.)}.
\newblock \bibinfo{publisher}{SAGE Publications, Inc}.
\newblock
\urldef\tempurl%
\url{https://us.sagepub.com/en-us/nam/research-design/book255675}
\showURL{%
\tempurl}


\bibitem[Davis(2018)]%
        {davis_8_2018}
\bibfield{author}{\bibinfo{person}{Noah Davis}.}
  \bibinfo{year}{2018}\natexlab{}.
\newblock \bibinfo{title}{8\% of pull requests are doomed}.
\newblock
\newblock
\urldef\tempurl%
\url{https://codeclimate.com/blog/abandoned-pull-requests}
\showURL{%
\tempurl}


\bibitem[DeYoung et~al\mbox{.}(2007)]%
        {deyoung_between_2007}
\bibfield{author}{\bibinfo{person}{Colin~G. DeYoung}, \bibinfo{person}{Lena~C.
  Quilty}, {and} \bibinfo{person}{Jordan~B. Peterson}.}
  \bibinfo{year}{2007}\natexlab{}.
\newblock \showarticletitle{Between facets and domains: 10 aspects of the {Big}
  {Five}}.
\newblock \bibinfo{journal}{\emph{Journal of Personality and Social
  Psychology}} \bibinfo{volume}{93}, \bibinfo{number}{5}
  (\bibinfo{year}{2007}), \bibinfo{pages}{880--896}.
\newblock
\urldef\tempurl%
\url{https://doi.org/10.1037/0022-3514.93.5.880}
\showDOI{\tempurl}


\bibitem[Dormann et~al\mbox{.}(2013)]%
        {dormann_collinearity_2013}
\bibfield{author}{\bibinfo{person}{Carsten~F. Dormann}, \bibinfo{person}{Jane
  Elith}, \bibinfo{person}{Sven Bacher}, \bibinfo{person}{Carsten Buchmann},
  \bibinfo{person}{Gudrun Carl}, \bibinfo{person}{Gabriel Carré},
  \bibinfo{person}{Jaime~R. García~Marquéz}, \bibinfo{person}{Bernd Gruber},
  \bibinfo{person}{Bruno Lafourcade}, \bibinfo{person}{Pedro~J. Leitão},
  \bibinfo{person}{Tamara Münkemüller}, \bibinfo{person}{Colin McClean},
  \bibinfo{person}{Patrick~E. Osborne}, \bibinfo{person}{Björn Reineking},
  \bibinfo{person}{Boris Schröder}, \bibinfo{person}{Andrew~K. Skidmore},
  \bibinfo{person}{Damaris Zurell}, {and} \bibinfo{person}{Sven Lautenbach}.}
  \bibinfo{year}{2013}\natexlab{}.
\newblock \showarticletitle{Collinearity: {A} review of methods to deal with it
  and a simulation study evaluating their performance}.
\newblock \bibinfo{journal}{\emph{Ecography}} \bibinfo{volume}{36},
  \bibinfo{number}{1} (\bibinfo{year}{2013}), \bibinfo{pages}{27--46}.
\newblock
\urldef\tempurl%
\url{https://doi.org/10.1111/j.1600-0587.2012.07348.x}
\showDOI{\tempurl}


\bibitem[Evans(1996)]%
        {evans_straightforward_1996}
\bibfield{author}{\bibinfo{person}{James~D. Evans}.}
  \bibinfo{year}{1996}\natexlab{}.
\newblock \bibinfo{booktitle}{\emph{Straightforward {Statistics} for the
  {Behavioral} {Sciences}}}.
\newblock \bibinfo{publisher}{Thomson Brooks/Cole Publishing Co.}
\newblock
\urldef\tempurl%
\url{https://psycnet.apa.org/record/1995-98499-000}
\showURL{%
\tempurl}


\bibitem[Fawcett(2006)]%
        {fawcett_introduction_2006}
\bibfield{author}{\bibinfo{person}{Tom Fawcett}.}
  \bibinfo{year}{2006}\natexlab{}.
\newblock \showarticletitle{An introduction to {ROC} analysis}.
\newblock \bibinfo{journal}{\emph{Pattern Recognition Letters}}
  \bibinfo{volume}{27}, \bibinfo{number}{8} (\bibinfo{year}{2006}),
  \bibinfo{pages}{861--874}.
\newblock
\urldef\tempurl%
\url{https://doi.org/10.1016/j.patrec.2005.10.010}
\showDOI{\tempurl}


\bibitem[Fisher et~al\mbox{.}(2019)]%
        {fisher_all_2019}
\bibfield{author}{\bibinfo{person}{Aaron Fisher}, \bibinfo{person}{Cynthia
  Rudin}, {and} \bibinfo{person}{Francesca Dominici}.}
  \bibinfo{year}{2019}\natexlab{}.
\newblock \showarticletitle{All models are wrong, but many are useful:
  {Learning} a variable's importance by studying an entire class of prediction
  models simultaneously}.
\newblock \bibinfo{journal}{\emph{Journal of Machine Learning Research}}
  \bibinfo{volume}{20}, \bibinfo{number}{177} (\bibinfo{year}{2019}),
  \bibinfo{pages}{1--81}.
\newblock
\urldef\tempurl%
\url{https://jmlr.org/papers/v20/18-760.html}
\showURL{%
\tempurl}


\bibitem[Furtado et~al\mbox{.}(2021)]%
        {furtado_how_2021}
\bibfield{author}{\bibinfo{person}{Leonardo~B. Furtado}, \bibinfo{person}{Bruno
  Cartaxo}, \bibinfo{person}{Christoph Treude}, {and} \bibinfo{person}{Gustavo
  Pinto}.} \bibinfo{year}{2021}\natexlab{}.
\newblock \showarticletitle{How successful are open source contributions from
  countries with different levels of human development?}
\newblock \bibinfo{journal}{\emph{IEEE Software}} \bibinfo{volume}{38},
  \bibinfo{number}{2} (\bibinfo{year}{2021}), \bibinfo{pages}{58--63}.
\newblock
\urldef\tempurl%
\url{https://doi.org/10.1109/MS.2020.3044020}
\showDOI{\tempurl}


\bibitem[{GitHub}(2020)]%
        {github_state_2020}
\bibfield{author}{\bibinfo{person}{{GitHub}}.} \bibinfo{year}{2020}\natexlab{}.
\newblock \bibinfo{title}{The state of the {Octoverse}}.
\newblock
\newblock
\urldef\tempurl%
\url{https://octoverse.github.com}
\showURL{%
\tempurl}


\bibitem[{GitHub}(2021a)]%
        {github_enums_2021}
\bibfield{author}{\bibinfo{person}{{GitHub}}.}
  \bibinfo{year}{2021}\natexlab{a}.
\newblock \bibinfo{title}{Enums}.
\newblock
\newblock
\urldef\tempurl%
\url{https://docs.github.com/en/graphql/reference/enums}
\showURL{%
\tempurl}


\bibitem[{GitHub}(2021b)]%
        {github_issues_2021}
\bibfield{author}{\bibinfo{person}{{GitHub}}.}
  \bibinfo{year}{2021}\natexlab{b}.
\newblock \bibinfo{title}{Issues}.
\newblock
\newblock
\urldef\tempurl%
\url{https://docs.github.com/en/rest/reference/issues}
\showURL{%
\tempurl}


\bibitem[{GitHub}(2021c)]%
        {github_pulls_2021}
\bibfield{author}{\bibinfo{person}{{GitHub}}.}
  \bibinfo{year}{2021}\natexlab{c}.
\newblock \bibinfo{title}{Pulls}.
\newblock
\newblock
\urldef\tempurl%
\url{https://docs.github.com/en/rest/reference/pulls}
\showURL{%
\tempurl}


\bibitem[{GitHub}(2021d)]%
        {github_rest_2021}
\bibfield{author}{\bibinfo{person}{{GitHub}}.}
  \bibinfo{year}{2021}\natexlab{d}.
\newblock \bibinfo{title}{{REST} {API}}.
\newblock
\newblock
\urldef\tempurl%
\url{https://docs.github.com/en/rest}
\showURL{%
\tempurl}


\bibitem[{GitHub}(2021e)]%
        {github_stale_2021}
\bibfield{author}{\bibinfo{person}{{GitHub}}.}
  \bibinfo{year}{2021}\natexlab{e}.
\newblock \bibinfo{title}{Stale}.
\newblock
\newblock
\urldef\tempurl%
\url{https://github.com/marketplace/stale}
\showURL{%
\tempurl}


\bibitem[Gousios(2013)]%
        {gousios_ghtorent_2013}
\bibfield{author}{\bibinfo{person}{Georgios Gousios}.}
  \bibinfo{year}{2013}\natexlab{}.
\newblock \showarticletitle{The {GHTorent} dataset and tool suite}. In
  \bibinfo{booktitle}{\emph{Proceedings of the 10th {Working} {Conference} on
  {Mining} {Software} {Repositories} ({MSR})}}. \bibinfo{pages}{233--236}.
\newblock
\urldef\tempurl%
\url{https://doi.org/10.1109/MSR.2013.6624034}
\showDOI{\tempurl}


\bibitem[Gousios et~al\mbox{.}(2014)]%
        {gousios_exploratory_2014}
\bibfield{author}{\bibinfo{person}{Georgios Gousios}, \bibinfo{person}{Martin
  Pinzger}, {and} \bibinfo{person}{Arie van Deursen}.}
  \bibinfo{year}{2014}\natexlab{}.
\newblock \showarticletitle{An exploratory study of the pull-based software
  development model}. In \bibinfo{booktitle}{\emph{Proceedings of the 36th
  {International} {Conference} on {Software} {Engineering} ({ICSE})}}.
  \bibinfo{pages}{345--355}.
\newblock
\urldef\tempurl%
\url{https://doi.org/10.1145/2568225.2568260}
\showDOI{\tempurl}


\bibitem[Gousios and Zaidman(2014)]%
        {gousios_dataset_2014}
\bibfield{author}{\bibinfo{person}{Georgios Gousios} {and}
  \bibinfo{person}{Andy Zaidman}.} \bibinfo{year}{2014}\natexlab{}.
\newblock \showarticletitle{A dataset for pull-based development research}. In
  \bibinfo{booktitle}{\emph{Proceedings of the 11th {Working} {Conference} on
  {Mining} {Software} {Repositories} ({MSR})}}. \bibinfo{pages}{368--371}.
\newblock
\urldef\tempurl%
\url{https://doi.org/10.1145/2597073.2597122}
\showDOI{\tempurl}


\bibitem[Gousios et~al\mbox{.}(2015)]%
        {gousios_work_2015}
\bibfield{author}{\bibinfo{person}{Georgios Gousios}, \bibinfo{person}{Andy
  Zaidman}, \bibinfo{person}{Margaret-Anne Storey}, {and} \bibinfo{person}{Arie
  van Deursen}.} \bibinfo{year}{2015}\natexlab{}.
\newblock \showarticletitle{Work practices and challenges in pull-based
  development: {The} integrator's perspective}. In
  \bibinfo{booktitle}{\emph{Proceedings of the 37th {International}
  {Conference} on {Software} {Engineering} ({ICSE})}}.
  \bibinfo{pages}{358--368}.
\newblock
\urldef\tempurl%
\url{https://doi.org/10.1109/ICSE.2015.55}
\showDOI{\tempurl}


\bibitem[Grigorik(2021)]%
        {grigorik_gh_2021}
\bibfield{author}{\bibinfo{person}{Ilya Grigorik}.}
  \bibinfo{year}{2021}\natexlab{}.
\newblock \bibinfo{title}{{GH} {Archive}: {A} project to record the public
  {GitHub} timeline, archive it, and make it easily accessible for further
  analysis}.
\newblock
\newblock
\urldef\tempurl%
\url{https://www.gharchive.org}
\showURL{%
\tempurl}


\bibitem[{Haibo He} and Garcia(2009)]%
        {haibo_he_learning_2009}
\bibfield{author}{\bibinfo{person}{{Haibo He}} {and}
  \bibinfo{person}{Edwardo~A. Garcia}.} \bibinfo{year}{2009}\natexlab{}.
\newblock \showarticletitle{Learning from imbalanced data}.
\newblock \bibinfo{journal}{\emph{IEEE Transactions on Knowledge and Data
  Engineering}} \bibinfo{volume}{21}, \bibinfo{number}{9}
  (\bibinfo{year}{2009}), \bibinfo{pages}{1263--1284}.
\newblock
\urldef\tempurl%
\url{https://doi.org/10.1109/TKDE.2008.239}
\showDOI{\tempurl}


\bibitem[Harrell(2021)]%
        {harrell_hmisc_2021}
\bibfield{author}{\bibinfo{person}{Frank~E. Harrell}.}
  \bibinfo{year}{2021}\natexlab{}.
\newblock \bibinfo{title}{Hmisc: {Harrell} {Miscellaneous}}.
\newblock
\newblock
\urldef\tempurl%
\url{https://CRAN.R-project.org/package=Hmisc}
\showURL{%
\tempurl}


\bibitem[Hess and Kromrey(2004)]%
        {hess_robust_2004}
\bibfield{author}{\bibinfo{person}{Melinda~R. Hess} {and}
  \bibinfo{person}{Jeffrey~D. Kromrey}.} \bibinfo{year}{2004}\natexlab{}.
\newblock \showarticletitle{Robust confidence intervals for effect sizes: {A}
  comparative study of {Cohen}'s d and {Cliff}'s delta under non-normality and
  heterogeneous variances}. In \bibinfo{booktitle}{\emph{Presented at the
  {Annual} {Meeting} of the {American} {Educational} {Research} {Association}
  ({AERA})}}.
\newblock
\urldef\tempurl%
\url{https://citeseerx.ist.psu.edu/viewdoc/summary?doi=10.1.1.487.8299}
\showURL{%
\tempurl}


\bibitem[Hintze and Nelson(1998)]%
        {hintze_violin_1998}
\bibfield{author}{\bibinfo{person}{Jerry~L. Hintze} {and}
  \bibinfo{person}{Ray~D. Nelson}.} \bibinfo{year}{1998}\natexlab{}.
\newblock \showarticletitle{Violin plots: {A} box plot-density trace
  synergism}.
\newblock \bibinfo{journal}{\emph{The American Statistician}}
  \bibinfo{volume}{52}, \bibinfo{number}{2} (\bibinfo{year}{1998}),
  \bibinfo{pages}{181--184}.
\newblock
\urldef\tempurl%
\url{https://doi.org/10.1080/00031305.1998.10480559}
\showDOI{\tempurl}


\bibitem[Iyer et~al\mbox{.}(2021)]%
        {iyer_effects_2021}
\bibfield{author}{\bibinfo{person}{Rahul~N. Iyer}, \bibinfo{person}{S.~Alex
  Yun}, \bibinfo{person}{Meiyappan Nagappan}, {and} \bibinfo{person}{Jesse
  Hoey}.} \bibinfo{year}{2021}\natexlab{}.
\newblock \showarticletitle{Effects of personality traits on pull request
  acceptance}.
\newblock \bibinfo{journal}{\emph{IEEE Transactions on Software Engineering}}
  \bibinfo{volume}{47}, \bibinfo{number}{11} (\bibinfo{year}{2021}),
  \bibinfo{pages}{2632--2643}.
\newblock
\urldef\tempurl%
\url{https://doi.org/10.1109/TSE.2019.2960357}
\showDOI{\tempurl}


\bibitem[Jacques(2021)]%
        {jacques_pygithub_2021}
\bibfield{author}{\bibinfo{person}{Vincent Jacques}.}
  \bibinfo{year}{2021}\natexlab{}.
\newblock \bibinfo{title}{{PyGithub}: {Typed} interactions with the {GitHub}
  {API} v3}.
\newblock
\newblock
\urldef\tempurl%
\url{https://github.com/PyGithub/PyGithub}
\showURL{%
\tempurl}


\bibitem[Jones et~al\mbox{.}(1998)]%
        {jones_efficient_1998}
\bibfield{author}{\bibinfo{person}{Donald~R. Jones}, \bibinfo{person}{Matthias
  Schonlau}, {and} \bibinfo{person}{William~J. Welch}.}
  \bibinfo{year}{1998}\natexlab{}.
\newblock \showarticletitle{Efficient global optimization of expensive
  black-box functions}.
\newblock \bibinfo{journal}{\emph{Journal of Global Optimization}}
  \bibinfo{volume}{13}, \bibinfo{number}{4} (\bibinfo{year}{1998}),
  \bibinfo{pages}{455--492}.
\newblock
\urldef\tempurl%
\url{https://doi.org/10.1023/A:1008306431147}
\showDOI{\tempurl}


\bibitem[Kalliamvakou et~al\mbox{.}(2016)]%
        {kalliamvakou_-depth_2016}
\bibfield{author}{\bibinfo{person}{Eirini Kalliamvakou},
  \bibinfo{person}{Georgios Gousios}, \bibinfo{person}{Kelly Blincoe},
  \bibinfo{person}{Leif Singer}, \bibinfo{person}{Daniel~M. German}, {and}
  \bibinfo{person}{Daniela Damian}.} \bibinfo{year}{2016}\natexlab{}.
\newblock \showarticletitle{An in-depth study of the promises and perils of
  mining {GitHub}}.
\newblock \bibinfo{journal}{\emph{Empirical Software Engineering}}
  \bibinfo{volume}{21}, \bibinfo{number}{5} (\bibinfo{year}{2016}),
  \bibinfo{pages}{2035--2071}.
\newblock
\urldef\tempurl%
\url{https://doi.org/10.1007/s10664-015-9393-5}
\showDOI{\tempurl}


\bibitem[Khatoonabadi et~al\mbox{.}(2021)]%
        {khatoonabadi_gap2wss_2021}
\bibfield{author}{\bibinfo{person}{SayedHassan Khatoonabadi},
  \bibinfo{person}{Shahriar Lotfi}, {and} \bibinfo{person}{Ayaz Isazadeh}.}
  \bibinfo{year}{2021}\natexlab{}.
\newblock \showarticletitle{{GAP2WSS}: {A} genetic algorithm based on the
  {Pareto} principle for web service selection}.
\newblock \bibinfo{journal}{\emph{arXiv:2109.10430 [cs.NE]}}
  (\bibinfo{year}{2021}).
\newblock
\urldef\tempurl%
\url{https://arxiv.org/abs/2109.10430}
\showURL{%
\tempurl}


\bibitem[Kirk(1996)]%
        {kirk_practical_1996}
\bibfield{author}{\bibinfo{person}{Roger~E. Kirk}.}
  \bibinfo{year}{1996}\natexlab{}.
\newblock \showarticletitle{Practical significance: {A} concept whose time has
  come}.
\newblock \bibinfo{journal}{\emph{Educational and Psychological Measurement}}
  \bibinfo{volume}{56}, \bibinfo{number}{5} (\bibinfo{year}{1996}),
  \bibinfo{pages}{746--759}.
\newblock
\urldef\tempurl%
\url{https://doi.org/10.1177/0013164496056005002}
\showDOI{\tempurl}


\bibitem[Kononenko et~al\mbox{.}(2018)]%
        {kononenko_studying_2018}
\bibfield{author}{\bibinfo{person}{Oleksii Kononenko}, \bibinfo{person}{Tresa
  Rose}, \bibinfo{person}{Olga Baysal}, \bibinfo{person}{Michael Godfrey},
  \bibinfo{person}{Dennis Theisen}, {and} \bibinfo{person}{Bart de Water}.}
  \bibinfo{year}{2018}\natexlab{}.
\newblock \showarticletitle{Studying pull request merges: {A} case study of
  {Shopify}'s {Active} {Merchant}}. In \bibinfo{booktitle}{\emph{Proceedings of
  the 40th {International} {Conference} on {Software} {Engineering}: {Software}
  {Engineering} in {Practice} ({ICSE}-{SEIP})}}. \bibinfo{pages}{124--133}.
\newblock
\urldef\tempurl%
\url{https://doi.org/10.1145/3183519.3183542}
\showDOI{\tempurl}


\bibitem[{Kubernetes}(2021)]%
        {kubernetes_fejta-bot_2021}
\bibfield{author}{\bibinfo{person}{{Kubernetes}}.}
  \bibinfo{year}{2021}\natexlab{}.
\newblock \bibinfo{title}{fejta-bot}.
\newblock
\newblock
\urldef\tempurl%
\url{https://github.com/fejta-bot}
\showURL{%
\tempurl}


\bibitem[Landis and Koch(1977)]%
        {landis_measurement_1977}
\bibfield{author}{\bibinfo{person}{J.~Richard Landis} {and}
  \bibinfo{person}{Gary~G. Koch}.} \bibinfo{year}{1977}\natexlab{}.
\newblock \showarticletitle{The measurement of observer agreement for
  categorical data}.
\newblock \bibinfo{journal}{\emph{Biometrics}} \bibinfo{volume}{33},
  \bibinfo{number}{1} (\bibinfo{year}{1977}), \bibinfo{pages}{159--174}.
\newblock
\urldef\tempurl%
\url{https://doi.org/10.2307/2529310}
\showDOI{\tempurl}


\bibitem[Lenarduzzi et~al\mbox{.}(2021)]%
        {lenarduzzi_does_2021}
\bibfield{author}{\bibinfo{person}{Valentina Lenarduzzi}, \bibinfo{person}{Vili
  Nikkola}, \bibinfo{person}{Nyyti Saarimäki}, {and} \bibinfo{person}{Davide
  Taibi}.} \bibinfo{year}{2021}\natexlab{}.
\newblock \showarticletitle{Does code quality affect pull request acceptance?
  {An} empirical study}.
\newblock \bibinfo{journal}{\emph{Journal of Systems and Software}}
  \bibinfo{volume}{171} (\bibinfo{year}{2021}), \bibinfo{pages}{110806}.
\newblock
\urldef\tempurl%
\url{https://doi.org/10.1016/j.jss.2020.110806}
\showDOI{\tempurl}


\bibitem[Lethbridge et~al\mbox{.}(2005)]%
        {lethbridge_studying_2005}
\bibfield{author}{\bibinfo{person}{Timothy~C. Lethbridge},
  \bibinfo{person}{Susan~Elliott Sim}, {and} \bibinfo{person}{Janice Singer}.}
  \bibinfo{year}{2005}\natexlab{}.
\newblock \showarticletitle{Studying software engineers: {Data} collection
  techniques for software field studies}.
\newblock \bibinfo{journal}{\emph{Empirical Software Engineering}}
  \bibinfo{volume}{10}, \bibinfo{number}{3} (\bibinfo{year}{2005}),
  \bibinfo{pages}{311--341}.
\newblock
\urldef\tempurl%
\url{https://doi.org/10.1007/s10664-005-1290-x}
\showDOI{\tempurl}


\bibitem[Li et~al\mbox{.}(2017)]%
        {li_detecting_2017}
\bibfield{author}{\bibinfo{person}{Zhixing Li}, \bibinfo{person}{Gang Yin},
  \bibinfo{person}{Yue Yu}, \bibinfo{person}{Tao Wang}, {and}
  \bibinfo{person}{Huaimin Wang}.} \bibinfo{year}{2017}\natexlab{}.
\newblock \showarticletitle{Detecting duplicate pull-requests in {GitHub}}. In
  \bibinfo{booktitle}{\emph{Proceedings of the 9th {Asia}-{Pacific} {Symposium}
  on {Internetware} ({Internetware})}}. \bibinfo{pages}{1--6}.
\newblock
\urldef\tempurl%
\url{https://doi.org/10.1145/3131704.3131725}
\showDOI{\tempurl}


\bibitem[Li et~al\mbox{.}(2021a)]%
        {li_are_2021}
\bibfield{author}{\bibinfo{person}{Zhixing Li}, \bibinfo{person}{Yue Yu},
  \bibinfo{person}{Tao Wang}, \bibinfo{person}{Gang Yin},
  \bibinfo{person}{Shanshan Li}, {and} \bibinfo{person}{Huaimin Wang}.}
  \bibinfo{year}{2021}\natexlab{a}.
\newblock \showarticletitle{Are you still working on this? {An} empirical study
  on pull request abandonment}.
\newblock \bibinfo{journal}{\emph{IEEE Transactions on Software Engineering}}
  \bibinfo{volume}{early access} (\bibinfo{year}{2021}).
\newblock
\urldef\tempurl%
\url{https://doi.org/10.1109/TSE.2021.3053403}
\showDOI{\tempurl}


\bibitem[Li et~al\mbox{.}(2022)]%
        {li_redundancy_2022}
\bibfield{author}{\bibinfo{person}{Zhixing Li}, \bibinfo{person}{Yue Yu},
  \bibinfo{person}{Minghui Zhou}, \bibinfo{person}{Tao Wang},
  \bibinfo{person}{Gang Yin}, \bibinfo{person}{Long Lan}, {and}
  \bibinfo{person}{Huaimin Wang}.} \bibinfo{year}{2022}\natexlab{}.
\newblock \showarticletitle{Redundancy, context, and preference: {An} empirical
  study of duplicate pull requests in {OSS} projects}.
\newblock \bibinfo{journal}{\emph{IEEE Transactions on Software Engineering}}
  \bibinfo{volume}{48}, \bibinfo{number}{4} (\bibinfo{year}{2022}),
  \bibinfo{pages}{1309--1335}.
\newblock
\urldef\tempurl%
\url{https://doi.org/10.1109/TSE.2020.3018726}
\showDOI{\tempurl}


\bibitem[Li et~al\mbox{.}(2021b)]%
        {li_detecting_2021}
\bibfield{author}{\bibinfo{person}{Zhi-Xing Li}, \bibinfo{person}{Yue Yu},
  \bibinfo{person}{Tao Wang}, \bibinfo{person}{Gang Yin},
  \bibinfo{person}{Xin-Jun Mao}, {and} \bibinfo{person}{Huai-Min Wang}.}
  \bibinfo{year}{2021}\natexlab{b}.
\newblock \showarticletitle{Detecting duplicate contributions in pull-based
  model combining textual and change similarities}.
\newblock \bibinfo{journal}{\emph{Journal of Computer Science and Technology}}
  \bibinfo{volume}{36}, \bibinfo{number}{1} (\bibinfo{year}{2021}),
  \bibinfo{pages}{191--206}.
\newblock
\urldef\tempurl%
\url{https://doi.org/10.1007/s11390-020-9935-1}
\showDOI{\tempurl}


\bibitem[Macrae(2014)]%
        {macrae_added_2014}
\bibfield{author}{\bibinfo{person}{Callum Macrae}.}
  \bibinfo{year}{2014}\natexlab{}.
\newblock \bibinfo{title}{Added pixelmator. \#4781}.
\newblock
\newblock
\urldef\tempurl%
\url{https://github.com/Homebrew/homebrew-cask/pull/4781}
\showURL{%
\tempurl}


\bibitem[Mann and Whitney(1947)]%
        {mann_test_1947}
\bibfield{author}{\bibinfo{person}{Henry~B. Mann} {and}
  \bibinfo{person}{Donald~R. Whitney}.} \bibinfo{year}{1947}\natexlab{}.
\newblock \showarticletitle{On a test of whether one of two random variables is
  stochastically larger than the other}.
\newblock \bibinfo{journal}{\emph{Annals of Mathematical Statistics}}
  \bibinfo{volume}{18}, \bibinfo{number}{1} (\bibinfo{year}{1947}),
  \bibinfo{pages}{50--60}.
\newblock
\urldef\tempurl%
\url{https://doi.org/10.1214/aoms/1177730491}
\showDOI{\tempurl}


\bibitem[Molnar(2022)]%
        {molnar_interpretable_2022}
\bibfield{author}{\bibinfo{person}{Christoph Molnar}.}
  \bibinfo{year}{2022}\natexlab{}.
\newblock \bibinfo{booktitle}{\emph{Interpretable {Machine} {Learning}: {A}
  {Guide} for {Making} {Black} {Box} {Models} {Explainable}}
  (\bibinfo{edition}{2nd} ed.)}.
\newblock
\urldef\tempurl%
\url{https://christophm.github.io/interpretable-ml-book}
\showURL{%
\tempurl}


\bibitem[Molnar et~al\mbox{.}(2018)]%
        {molnar_iml_2018}
\bibfield{author}{\bibinfo{person}{Christoph Molnar}, \bibinfo{person}{Giuseppe
  Casalicchio}, {and} \bibinfo{person}{Bernd Bischl}.}
  \bibinfo{year}{2018}\natexlab{}.
\newblock \showarticletitle{iml: {An} {R} package for interpretable machine
  learning}.
\newblock \bibinfo{journal}{\emph{Journal of Open Source Software}}
  \bibinfo{volume}{3}, \bibinfo{number}{27} (\bibinfo{year}{2018}),
  \bibinfo{pages}{786}.
\newblock
\urldef\tempurl%
\url{https://doi.org/10.21105/joss.00786}
\showDOI{\tempurl}


\bibitem[Nadri et~al\mbox{.}(2021a)]%
        {nadri_insights_2021}
\bibfield{author}{\bibinfo{person}{Reza Nadri}, \bibinfo{person}{Gema
  Rodriguez-Perez}, {and} \bibinfo{person}{Meiyappan Nagappan}.}
  \bibinfo{year}{2021}\natexlab{a}.
\newblock \showarticletitle{Insights into nonmerged pull requests in {GitHub}:
  {Is} there evidence of bias based on perceptible race?}
\newblock \bibinfo{journal}{\emph{IEEE Software}} \bibinfo{volume}{38},
  \bibinfo{number}{2} (\bibinfo{year}{2021}), \bibinfo{pages}{51--57}.
\newblock
\urldef\tempurl%
\url{https://doi.org/10.1109/MS.2020.3036758}
\showDOI{\tempurl}


\bibitem[Nadri et~al\mbox{.}(2021b)]%
        {nadri_relationship_2021}
\bibfield{author}{\bibinfo{person}{Reza Nadri}, \bibinfo{person}{Gema
  Rodriguezperez}, {and} \bibinfo{person}{Meiyappan Nagappan}.}
  \bibinfo{year}{2021}\natexlab{b}.
\newblock \showarticletitle{On the relationship between the developer's
  perceptible race and ethnicity and the evaluation of contributions in {OSS}}.
\newblock \bibinfo{journal}{\emph{IEEE Transactions on Software Engineering}}
  \bibinfo{volume}{early access} (\bibinfo{year}{2021}).
\newblock
\urldef\tempurl%
\url{https://doi.org/10.1109/TSE.2021.3073773}
\showDOI{\tempurl}


\bibitem[Patil(2021)]%
        {patil_visualizations_2021}
\bibfield{author}{\bibinfo{person}{Indrajeet Patil}.}
  \bibinfo{year}{2021}\natexlab{}.
\newblock \showarticletitle{Visualizations with statistical details: {The}
  'ggstatsplot' approach}.
\newblock \bibinfo{journal}{\emph{Journal of Open Source Software}}
  \bibinfo{volume}{6}, \bibinfo{number}{61} (\bibinfo{year}{2021}),
  \bibinfo{pages}{3167}.
\newblock
\urldef\tempurl%
\url{https://doi.org/10.21105/joss.03167}
\showDOI{\tempurl}


\bibitem[Pedregosa et~al\mbox{.}(2011)]%
        {pedregosa_scikit-learn_2011}
\bibfield{author}{\bibinfo{person}{Fabian Pedregosa}, \bibinfo{person}{Gaël
  Varoquaux}, \bibinfo{person}{Alexandre Gramfort}, \bibinfo{person}{Vincent
  Michel}, \bibinfo{person}{Bertrand Thirion}, \bibinfo{person}{Olivier
  Grisel}, \bibinfo{person}{Mathieu Blondel}, \bibinfo{person}{Peter
  Prettenhofer}, \bibinfo{person}{Ron Weiss}, \bibinfo{person}{Vincent
  Dubourg}, \bibinfo{person}{Jake Vanderplas}, \bibinfo{person}{Alexandre
  Passos}, \bibinfo{person}{David Cournapeau}, \bibinfo{person}{Matthieu
  Brucher}, \bibinfo{person}{Matthieu Perrot}, {and} \bibinfo{person}{Édouard
  Duchesnay}.} \bibinfo{year}{2011}\natexlab{}.
\newblock \showarticletitle{Scikit-learn: {Machine} learning in {Python}}.
\newblock \bibinfo{journal}{\emph{Journal of Machine Learning Research}}
  \bibinfo{volume}{12}, \bibinfo{number}{85} (\bibinfo{year}{2011}),
  \bibinfo{pages}{2825--2830}.
\newblock
\urldef\tempurl%
\url{https://jmlr.org/papers/v12/pedregosa11a.html}
\showURL{%
\tempurl}


\bibitem[Pinto et~al\mbox{.}(2018)]%
        {pinto_who_2018}
\bibfield{author}{\bibinfo{person}{Gustavo Pinto}, \bibinfo{person}{Luiz~Felipe
  Dias}, {and} \bibinfo{person}{Igor Steinmacher}.}
  \bibinfo{year}{2018}\natexlab{}.
\newblock \showarticletitle{Who gets a patch accepted first? {Comparing} the
  contributions of employees and volunteers}. In
  \bibinfo{booktitle}{\emph{Proceedings of the 11th {International} {Workshop}
  on {Cooperative} and {Human} {Aspects} of {Software} {Engineering}
  ({CHASE})}}. \bibinfo{pages}{110--113}.
\newblock
\urldef\tempurl%
\url{https://doi.org/10.1145/3195836.3195858}
\showDOI{\tempurl}


\bibitem[Pinto et~al\mbox{.}(2016)]%
        {pinto_more_2016}
\bibfield{author}{\bibinfo{person}{Gustavo Pinto}, \bibinfo{person}{Igor
  Steinmacher}, {and} \bibinfo{person}{Marco~Aurélio Gerosa}.}
  \bibinfo{year}{2016}\natexlab{}.
\newblock \showarticletitle{More common than you think: {An} in-depth study of
  casual contributors}. In \bibinfo{booktitle}{\emph{Proceedings of the 23rd
  {International} {Conference} on {Software} {Analysis}, {Evolution}, and
  {Reengineering} ({SANER})}}. \bibinfo{pages}{112--123}.
\newblock
\urldef\tempurl%
\url{https://doi.org/10.1109/SANER.2016.68}
\showDOI{\tempurl}


\bibitem[Probst et~al\mbox{.}(2019)]%
        {probst_hyperparameters_2019}
\bibfield{author}{\bibinfo{person}{Philipp Probst}, \bibinfo{person}{Marvin~N.
  Wright}, {and} \bibinfo{person}{Anne‐Laure Boulesteix}.}
  \bibinfo{year}{2019}\natexlab{}.
\newblock \showarticletitle{Hyperparameters and tuning strategies for random
  forest}.
\newblock \bibinfo{journal}{\emph{WIREs Data Mining and Knowledge Discovery}}
  \bibinfo{volume}{9}, \bibinfo{number}{3} (\bibinfo{year}{2019}),
  \bibinfo{pages}{e1301}.
\newblock
\urldef\tempurl%
\url{https://doi.org/10.1002/widm.1301}
\showDOI{\tempurl}


\bibitem[{R Core Team}(2021)]%
        {r_core_team_r_2021}
\bibfield{author}{\bibinfo{person}{{R Core Team}}.}
  \bibinfo{year}{2021}\natexlab{}.
\newblock \bibinfo{title}{R: {A} language and environment for statistical
  computing}.
\newblock
\newblock
\urldef\tempurl%
\url{https://www.R-project.org}
\showURL{%
\tempurl}


\bibitem[Rastogi(2016)]%
        {rastogi_biases_2016}
\bibfield{author}{\bibinfo{person}{Ayushi Rastogi}.}
  \bibinfo{year}{2016}\natexlab{}.
\newblock \showarticletitle{Do biases related to geographical location
  influence work-related decisions in {GitHub}?}. In
  \bibinfo{booktitle}{\emph{Proceedings of the 38th {International}
  {Conference} on {Software} {Engineering} {Companion} ({ICSE}-{C})}}.
  \bibinfo{pages}{665--667}.
\newblock
\urldef\tempurl%
\url{https://doi.org/10.1145/2889160.2891035}
\showDOI{\tempurl}


\bibitem[Rastogi et~al\mbox{.}(2018)]%
        {rastogi_relationship_2018}
\bibfield{author}{\bibinfo{person}{Ayushi Rastogi}, \bibinfo{person}{Nachiappan
  Nagappan}, \bibinfo{person}{Georgios Gousios}, {and} \bibinfo{person}{André
  van~der Hoek}.} \bibinfo{year}{2018}\natexlab{}.
\newblock \showarticletitle{Relationship between geographical location and
  evaluation of developer contributions in {GitHub}}. In
  \bibinfo{booktitle}{\emph{Proceedings of the 12th {International} {Symposium}
  on {Empirical} {Software} {Engineering} and {Measurement} ({ESEM})}}.
  \bibinfo{pages}{1--8}.
\newblock
\urldef\tempurl%
\url{https://doi.org/10.1145/3239235.3240504}
\showDOI{\tempurl}


\bibitem[Ren et~al\mbox{.}(2019)]%
        {ren_identifying_2019}
\bibfield{author}{\bibinfo{person}{Luyao Ren}, \bibinfo{person}{Shurui Zhou},
  \bibinfo{person}{Christian Kästner}, {and} \bibinfo{person}{Andrzej
  Wąsowski}.} \bibinfo{year}{2019}\natexlab{}.
\newblock \showarticletitle{Identifying redundancies in fork-based
  development}. In \bibinfo{booktitle}{\emph{Proceedings of the 26th
  {International} {Conference} on {Software} {Analysis}, {Evolution} and
  {Reengineering} ({SANER})}}. \bibinfo{pages}{230--241}.
\newblock
\urldef\tempurl%
\url{https://doi.org/10.1109/SANER.2019.8668023}
\showDOI{\tempurl}


\bibitem[Seaman(1999)]%
        {seaman_qualitative_1999}
\bibfield{author}{\bibinfo{person}{Carolyn~B. Seaman}.}
  \bibinfo{year}{1999}\natexlab{}.
\newblock \showarticletitle{Qualitative methods in empirical studies of
  software engineering}.
\newblock \bibinfo{journal}{\emph{IEEE Transactions on Software Engineering}}
  \bibinfo{volume}{25}, \bibinfo{number}{4} (\bibinfo{year}{1999}),
  \bibinfo{pages}{557--572}.
\newblock
\urldef\tempurl%
\url{https://doi.org/10.1109/32.799955}
\showDOI{\tempurl}


\bibitem[Soares et~al\mbox{.}(2015)]%
        {soares_acceptance_2015}
\bibfield{author}{\bibinfo{person}{Daricélio~Moreira Soares},
  \bibinfo{person}{Manoel~Limeira de Lima~Júnior}, {and}
  \bibinfo{person}{Leonardo Murta}.} \bibinfo{year}{2015}\natexlab{}.
\newblock \showarticletitle{Acceptance factors of pull requests in open-source
  projects}. In \bibinfo{booktitle}{\emph{Proceedings of the 30th {Annual}
  {Symposium} on {Applied} {Computing} ({SAC})}},
  \bibfield{editor}{\bibinfo{person}{Alexandre Plastino}} (Ed.).
  \bibinfo{pages}{1541--1546}.
\newblock
\urldef\tempurl%
\url{https://doi.org/10.1145/2695664.2695856}
\showDOI{\tempurl}


\bibitem[Spearman(2010)]%
        {spearman_proof_2010}
\bibfield{author}{\bibinfo{person}{Charles Spearman}.}
  \bibinfo{year}{2010}\natexlab{}.
\newblock \showarticletitle{The proof and measurement of association between
  two things}.
\newblock \bibinfo{journal}{\emph{International Journal of Epidemiology}}
  \bibinfo{volume}{39}, \bibinfo{number}{5} (\bibinfo{year}{2010}),
  \bibinfo{pages}{1137--1150}.
\newblock
\urldef\tempurl%
\url{https://doi.org/10.1093/ije/dyq191}
\showDOI{\tempurl}


\bibitem[Steinmacher et~al\mbox{.}(2014)]%
        {steinmacher_preliminary_2014}
\bibfield{author}{\bibinfo{person}{Igor Steinmacher},
  \bibinfo{person}{Ana~Paula Chaves}, \bibinfo{person}{Tayana~Uchoa Conte},
  {and} \bibinfo{person}{Marco~Aurélio Gerosa}.}
  \bibinfo{year}{2014}\natexlab{}.
\newblock \showarticletitle{Preliminary empirical identification of barriers
  faced by newcomers to open source software projects}. In
  \bibinfo{booktitle}{\emph{Proceedings of the 28th {Brazilian} {Symposium} on
  {Software} {Engineering} ({SBES})}}. \bibinfo{pages}{51--60}.
\newblock
\urldef\tempurl%
\url{https://doi.org/10.1109/SBES.2014.9}
\showDOI{\tempurl}


\bibitem[Steinmacher et~al\mbox{.}(2018)]%
        {steinmacher_almost_2018}
\bibfield{author}{\bibinfo{person}{Igor Steinmacher}, \bibinfo{person}{Gustavo
  Pinto}, \bibinfo{person}{Igor~Scaliante Wiese}, {and}
  \bibinfo{person}{Marco~Aurélio Gerosa}.} \bibinfo{year}{2018}\natexlab{}.
\newblock \showarticletitle{Almost there: {A} study on quasi-contributors in
  open source software projects}. In \bibinfo{booktitle}{\emph{Proceedings of
  the 40th {International} {Conference} on {Software} {Engineering} ({ICSE})}}.
  \bibinfo{pages}{256--266}.
\newblock
\urldef\tempurl%
\url{https://doi.org/10.1145/3180155.3180208}
\showDOI{\tempurl}


\bibitem[Steinmacher et~al\mbox{.}(2013)]%
        {steinmacher_why_2013}
\bibfield{author}{\bibinfo{person}{Igor Steinmacher}, \bibinfo{person}{Igor
  Wiese}, \bibinfo{person}{Ana~Paula Chaves}, {and}
  \bibinfo{person}{Marco~Aurélio Gerosa}.} \bibinfo{year}{2013}\natexlab{}.
\newblock \showarticletitle{Why do newcomers abandon open source software
  projects?}. In \bibinfo{booktitle}{\emph{Proceedings of the 6th
  {International} {Workshop} on {Cooperative} and {Human} {Aspects} of
  {Software} {Engineering} ({CHASE})}}. \bibinfo{pages}{25--32}.
\newblock
\urldef\tempurl%
\url{https://doi.org/10.1109/CHASE.2013.6614728}
\showDOI{\tempurl}


\bibitem[Terrell et~al\mbox{.}(2017)]%
        {terrell_gender_2017}
\bibfield{author}{\bibinfo{person}{Josh Terrell}, \bibinfo{person}{Andrew
  Kofink}, \bibinfo{person}{Justin Middleton}, \bibinfo{person}{Clarissa
  Rainear}, \bibinfo{person}{Emerson Murphy-Hill}, \bibinfo{person}{Chris
  Parnin}, {and} \bibinfo{person}{Jon Stallings}.}
  \bibinfo{year}{2017}\natexlab{}.
\newblock \showarticletitle{Gender differences and bias in open source: {Pull}
  request acceptance of women versus men}.
\newblock \bibinfo{journal}{\emph{PeerJ Computer Science}}  \bibinfo{volume}{3}
  (\bibinfo{year}{2017}), \bibinfo{pages}{e111}.
\newblock
\urldef\tempurl%
\url{https://doi.org/10.7717/peerj-cs.111}
\showDOI{\tempurl}


\bibitem[Therox(2020)]%
        {therox_changes_2020}
\bibfield{author}{\bibinfo{person}{Orta Therox}.}
  \bibinfo{year}{2020}\natexlab{}.
\newblock \bibinfo{title}{Changes to how we manage {DefinitelyTyped}}.
\newblock
\newblock
\urldef\tempurl%
\url{https://devblogs.microsoft.com/typescript/changes-to-how-we-manage-definitelytyped}
\showURL{%
\tempurl}


\bibitem[Tsay et~al\mbox{.}(2014)]%
        {tsay_influence_2014}
\bibfield{author}{\bibinfo{person}{Jason Tsay}, \bibinfo{person}{Laura
  Dabbish}, {and} \bibinfo{person}{James Herbsleb}.}
  \bibinfo{year}{2014}\natexlab{}.
\newblock \showarticletitle{Influence of social and technical factors for
  evaluating contribution in {GitHub}}. In
  \bibinfo{booktitle}{\emph{Proceedings of the 36th {International}
  {Conference} on {Software} {Engineering} ({ICSE})}}.
  \bibinfo{pages}{356--366}.
\newblock
\urldef\tempurl%
\url{https://doi.org/10.1145/2568225.2568315}
\showDOI{\tempurl}


\bibitem[Wang et~al\mbox{.}(2019)]%
        {wang_duplicate_2019}
\bibfield{author}{\bibinfo{person}{Qingye Wang}, \bibinfo{person}{Bowen Xu},
  \bibinfo{person}{Xin Xia}, \bibinfo{person}{Ting Wang}, {and}
  \bibinfo{person}{Shanping Li}.} \bibinfo{year}{2019}\natexlab{}.
\newblock \showarticletitle{Duplicate pull request detection: {When} time
  matters}. In \bibinfo{booktitle}{\emph{Proceedings of the 11th
  {Asia}-{Pacific} {Symposium} on {Internetware} ({Internetware})}}.
  \bibinfo{pages}{1--10}.
\newblock
\urldef\tempurl%
\url{https://doi.org/10.1145/3361242.3361254}
\showDOI{\tempurl}


\bibitem[Wessel et~al\mbox{.}(2019)]%
        {wessel_should_2019}
\bibfield{author}{\bibinfo{person}{Mairieli Wessel}, \bibinfo{person}{Igor
  Steinmacher}, \bibinfo{person}{Igor Wiese}, {and}
  \bibinfo{person}{Marco~Aurélio Gerosa}.} \bibinfo{year}{2019}\natexlab{}.
\newblock \showarticletitle{Should {I} stale or should {I} close? {An} analysis
  of a bot that closes abandoned issues and pull requests}. In
  \bibinfo{booktitle}{\emph{Proceedings of the 1st {International} {Workshop}
  on {Bots} in {Software} {Engineering} ({BotSE})}}. \bibinfo{pages}{38--42}.
\newblock
\urldef\tempurl%
\url{https://doi.org/10.1109/BotSE.2019.00018}
\showDOI{\tempurl}


\bibitem[Wright and Ziegler(2017)]%
        {wright_ranger_2017}
\bibfield{author}{\bibinfo{person}{Marvin~N. Wright} {and}
  \bibinfo{person}{Andreas Ziegler}.} \bibinfo{year}{2017}\natexlab{}.
\newblock \showarticletitle{ranger: {A} fast implementation of random forests
  for high dimensional data in {C}++ and {R}}.
\newblock \bibinfo{journal}{\emph{Journal of Statistical Software}}
  \bibinfo{volume}{77}, \bibinfo{number}{1} (\bibinfo{year}{2017}),
  \bibinfo{pages}{1--17}.
\newblock
\urldef\tempurl%
\url{https://doi.org/10.18637/jss.v077.i01}
\showDOI{\tempurl}


\bibitem[Yu et~al\mbox{.}(2018)]%
        {yu_dataset_2018}
\bibfield{author}{\bibinfo{person}{Yue Yu}, \bibinfo{person}{Zhixing Li},
  \bibinfo{person}{Gang Yin}, \bibinfo{person}{Tao Wang}, {and}
  \bibinfo{person}{Huaimin Wang}.} \bibinfo{year}{2018}\natexlab{}.
\newblock \showarticletitle{A dataset of duplicate pull-requests in {GitHub}}.
  In \bibinfo{booktitle}{\emph{Proceedings of the 15th {International}
  {Conference} on {Mining} {Software} {Repositories} ({MSR})}}.
  \bibinfo{pages}{22--25}.
\newblock
\urldef\tempurl%
\url{https://doi.org/10.1145/3196398.3196455}
\showDOI{\tempurl}


\bibitem[Yu et~al\mbox{.}(2015)]%
        {yu_wait_2015}
\bibfield{author}{\bibinfo{person}{Yue Yu}, \bibinfo{person}{Huaimin Wang},
  \bibinfo{person}{Vladimir Filkov}, \bibinfo{person}{Premkumar Devanbu}, {and}
  \bibinfo{person}{Bogdan Vasilescu}.} \bibinfo{year}{2015}\natexlab{}.
\newblock \showarticletitle{Wait for it: {Determinants} of pull request
  evaluation latency on {GitHub}}. In \bibinfo{booktitle}{\emph{Proceedings of
  the 12th {Working} {Conference} on {Mining} {Software} {Repositories}
  ({MSR})}}. \bibinfo{pages}{367--371}.
\newblock
\urldef\tempurl%
\url{https://doi.org/10.1109/MSR.2015.42}
\showDOI{\tempurl}


\bibitem[Yu et~al\mbox{.}(2016)]%
        {yu_determinants_2016}
\bibfield{author}{\bibinfo{person}{Yue Yu}, \bibinfo{person}{Gang Yin},
  \bibinfo{person}{Tao Wang}, \bibinfo{person}{Cheng Yang}, {and}
  \bibinfo{person}{Huaimin Wang}.} \bibinfo{year}{2016}\natexlab{}.
\newblock \showarticletitle{Determinants of pull-based development in the
  context of continuous integration}.
\newblock \bibinfo{journal}{\emph{Science China Information Sciences}}
  \bibinfo{volume}{59}, \bibinfo{number}{8} (\bibinfo{year}{2016}),
  \bibinfo{pages}{080104}.
\newblock
\urldef\tempurl%
\url{https://doi.org/10.1007/s11432-016-5595-8}
\showDOI{\tempurl}


\bibitem[Zhang et~al\mbox{.}(2020)]%
        {zhang_shoulders_2020}
\bibfield{author}{\bibinfo{person}{Xunhui Zhang}, \bibinfo{person}{Ayushi
  Rastogi}, {and} \bibinfo{person}{Yue Yu}.} \bibinfo{year}{2020}\natexlab{}.
\newblock \showarticletitle{On the shoulders of giants: {A} new dataset for
  pull-based development research}. In \bibinfo{booktitle}{\emph{Proceedings of
  the 17th {International} {Conference} on {Mining} {Software} {Repositories}
  ({MSR})}}. \bibinfo{pages}{543--547}.
\newblock
\urldef\tempurl%
\url{https://doi.org/10.1145/3379597.3387489}
\showDOI{\tempurl}


\bibitem[Zhang et~al\mbox{.}(2022)]%
        {zhang_pull_2022}
\bibfield{author}{\bibinfo{person}{Xunhui Zhang}, \bibinfo{person}{Yue Yu},
  \bibinfo{person}{Georgios Gousios}, {and} \bibinfo{person}{Ayushi Rastogi}.}
  \bibinfo{year}{2022}\natexlab{}.
\newblock \showarticletitle{Pull request decisions explained: {An} empirical
  overview}.
\newblock \bibinfo{journal}{\emph{IEEE Transactions on Software Engineering}}
  \bibinfo{volume}{early access} (\bibinfo{year}{2022}).
\newblock
\urldef\tempurl%
\url{https://doi.org/10.1109/TSE.2022.3165056}
\showDOI{\tempurl}


\bibitem[Zhang et~al\mbox{.}(2021)]%
        {zhang_pull_2021}
\bibfield{author}{\bibinfo{person}{Xunhui Zhang}, \bibinfo{person}{Yue Yu},
  \bibinfo{person}{Tao Wang}, \bibinfo{person}{Ayushi Rastogi}, {and}
  \bibinfo{person}{Huaimin Wang}.} \bibinfo{year}{2021}\natexlab{}.
\newblock \showarticletitle{Pull request latency explained: {An} empirical
  overview}.
\newblock \bibinfo{journal}{\emph{arXiv:2108.09946 [cs.SE]}}
  (\bibinfo{year}{2021}).
\newblock
\urldef\tempurl%
\url{https://arxiv.org/abs/2108.09946}
\showURL{%
\tempurl}


\bibitem[Zhu et~al\mbox{.}(2016)]%
        {zhu_effectiveness_2016}
\bibfield{author}{\bibinfo{person}{Jiaxin Zhu}, \bibinfo{person}{Minghui Zhou},
  {and} \bibinfo{person}{Audris Mockus}.} \bibinfo{year}{2016}\natexlab{}.
\newblock \showarticletitle{Effectiveness of code contribution: {From}
  patch-based to pull-request-based tools}. In
  \bibinfo{booktitle}{\emph{Proceedings of the 24th {International} {Symposium}
  on {Foundations} of {Software} {Engineering} ({FSE})}}.
  \bibinfo{pages}{871--882}.
\newblock
\urldef\tempurl%
\url{https://doi.org/10.1145/2950290.2950364}
\showDOI{\tempurl}


\bibitem[Zou et~al\mbox{.}(2019)]%
        {zou_how_2019}
\bibfield{author}{\bibinfo{person}{Weiqin Zou}, \bibinfo{person}{Jifeng Xuan},
  \bibinfo{person}{Xiaoyuan Xie}, \bibinfo{person}{Zhenyu Chen}, {and}
  \bibinfo{person}{Baowen Xu}.} \bibinfo{year}{2019}\natexlab{}.
\newblock \showarticletitle{How does code style inconsistency affect pull
  request integration? {An} exploratory study on 117 {GitHub} projects}.
\newblock \bibinfo{journal}{\emph{Empirical Software Engineering}}
  \bibinfo{volume}{24}, \bibinfo{number}{6} (\bibinfo{year}{2019}),
  \bibinfo{pages}{3871--3903}.
\newblock
\urldef\tempurl%
\url{https://doi.org/10.1007/s10664-019-09720-x}
\showDOI{\tempurl}


\end{thebibliography}

\section*{Appendix}
\appendix
\section{Comparison of Abandoned and Nonabandoned PRs}
\label{appendix:stats}

\subsection{PR Features:}

\begin{figure}[H]
    \includegraphics[width=\textwidth]{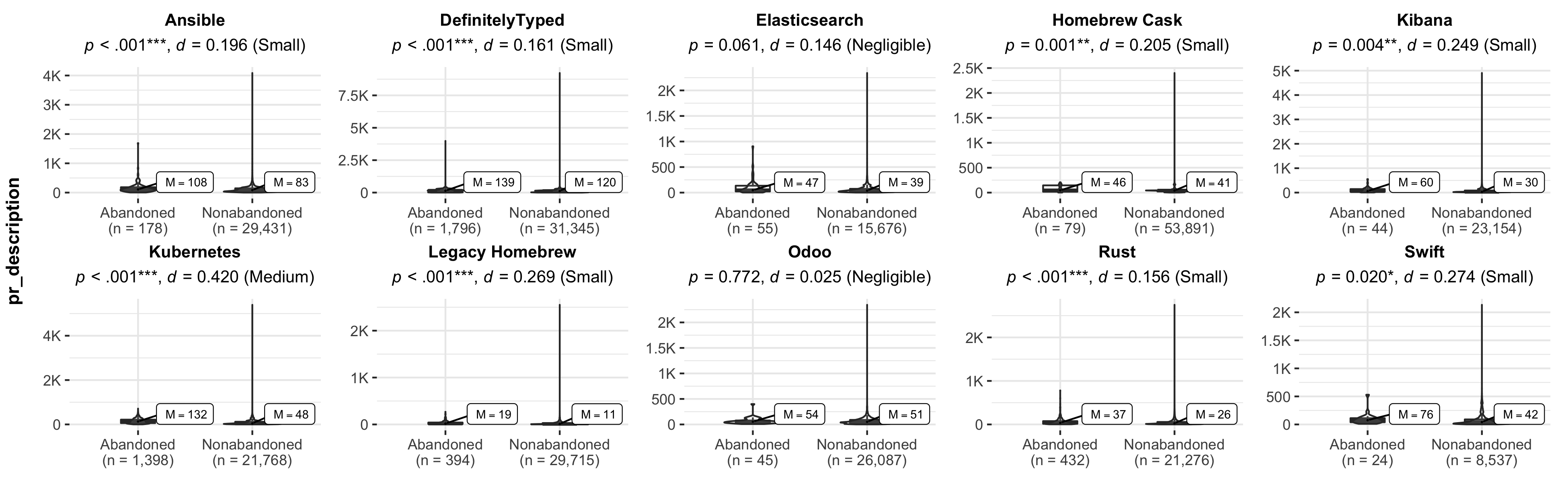}
    \caption{Comparison of abandoned and nonabandoned PRs wrt \textit{pr\_description} across the studied projects.}
\end{figure}

\begin{figure}[H]
    \includegraphics[width=\textwidth]{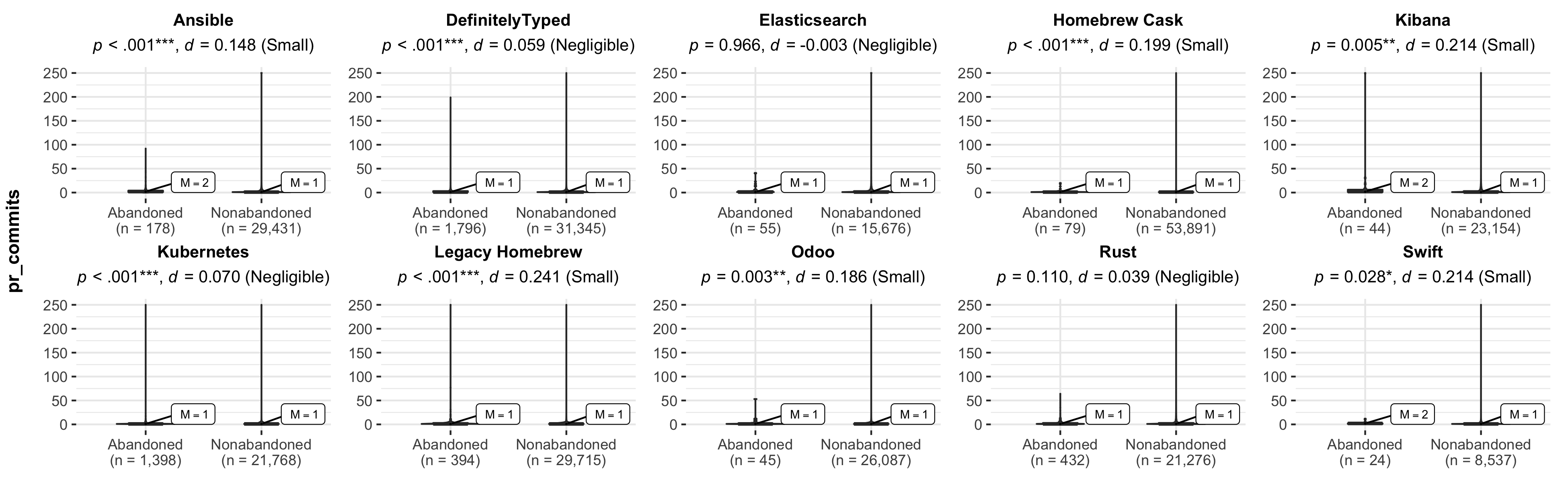}
    \caption{Comparison of abandoned and nonabandoned PRs wrt \textit{pr\_commits} across the studied projects.}
\end{figure}

\begin{figure}[H]
    \includegraphics[width=\textwidth]{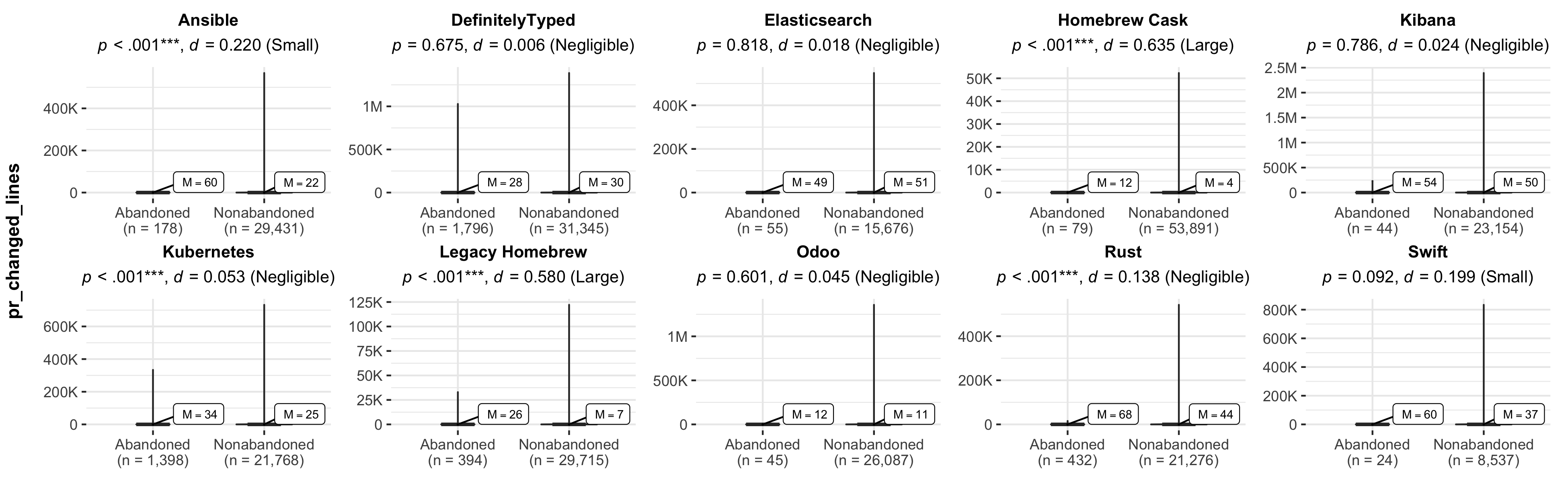}
    \caption{Comparison of abandoned and nonabandoned PRs wrt \textit{pr\_changed\_lines} across the studied projects.}
\end{figure}

\begin{figure}[H]
    \includegraphics[width=\textwidth]{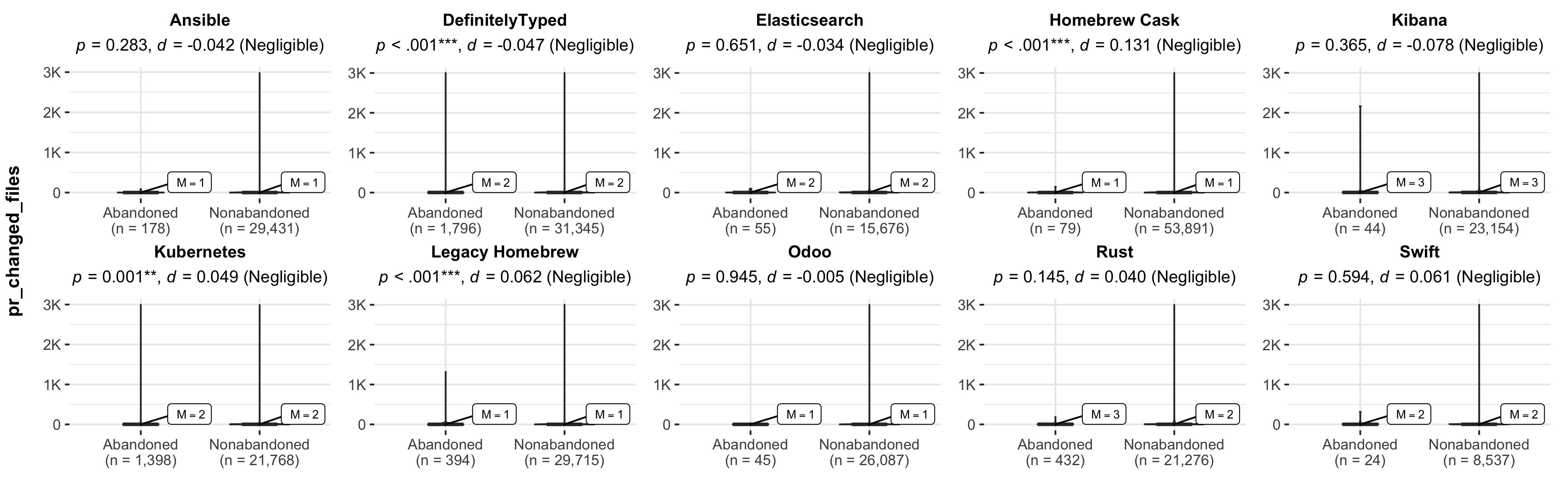}
    \caption{Comparison of abandoned and nonabandoned PRs wrt \textit{pr\_changed\_files} across the studied projects.}
\end{figure}

\subsection{Contributor Features:}

\begin{figure}[H]
    \includegraphics[width=\textwidth]{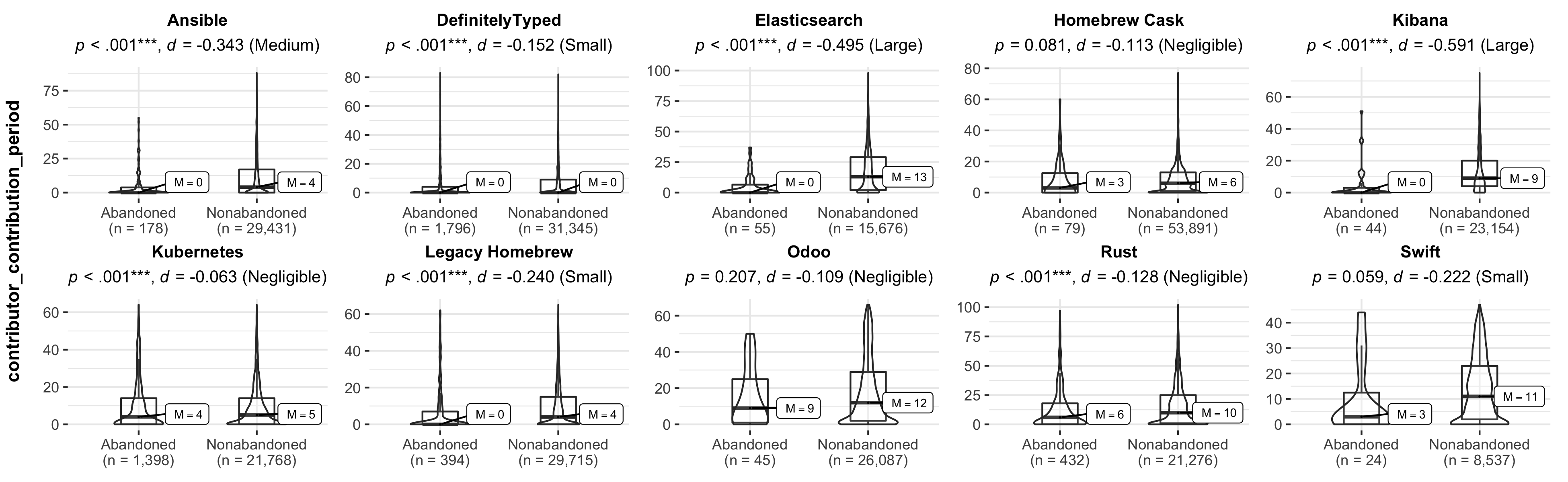}
    \caption{Comparison of abandoned and nonabandoned PRs wrt \textit{contributor\_contribution\_period} across the studied projects.}
\end{figure}

\begin{figure}[H]
    \includegraphics[width=\textwidth]{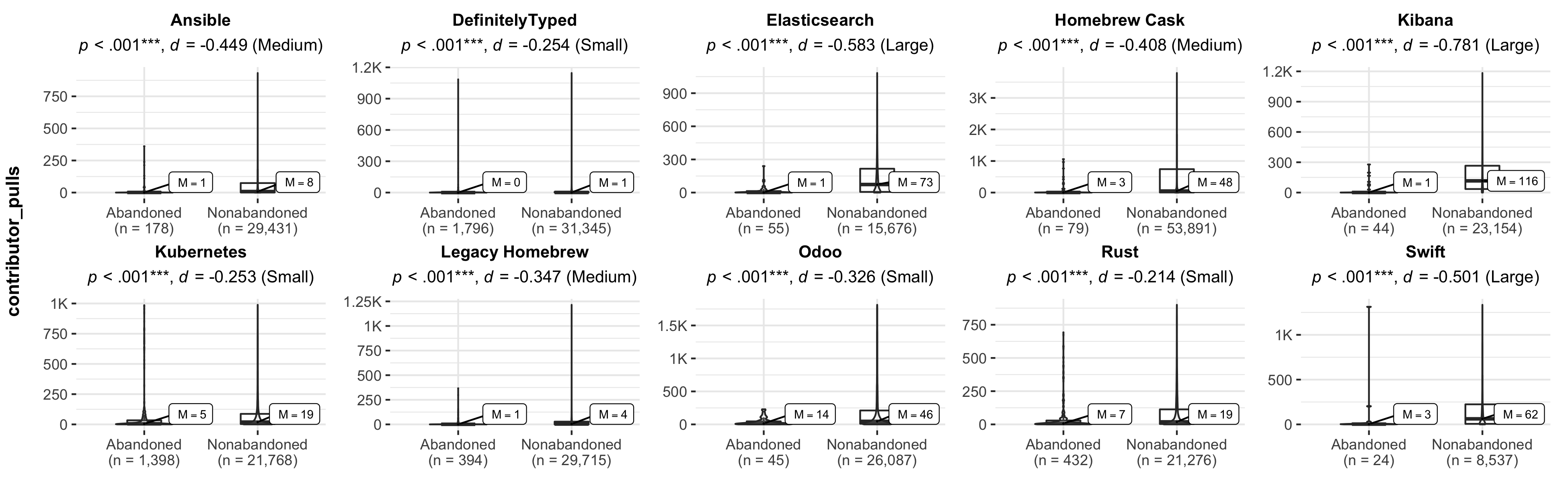}
    \caption{Comparison of abandoned and nonabandoned PRs wrt \textit{contributor\_pulls} across the studied projects.}
\end{figure}

\begin{figure}[H]
    \includegraphics[width=\textwidth]{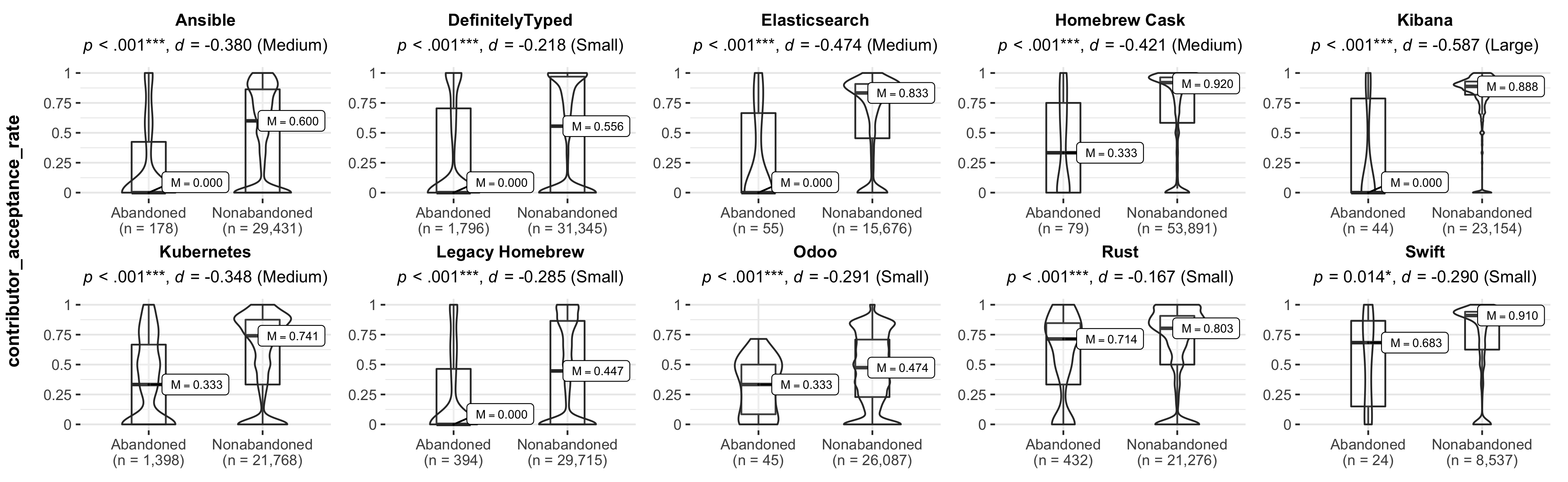}
    \caption{Comparison of abandoned and nonabandoned PRs wrt \textit{contributor\_acceptance\_rate} across the studied projects.}
\end{figure}

\begin{figure}[H]
    \includegraphics[width=\textwidth]{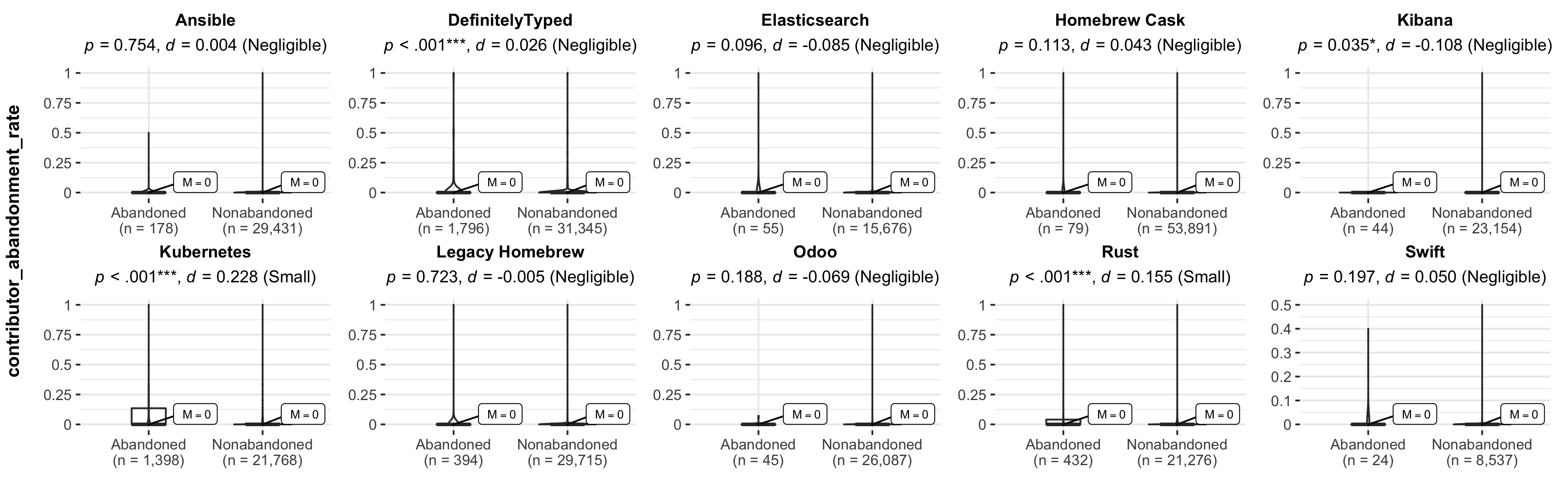}
    \caption{Comparison of abandoned and nonabandoned PRs wrt \textit{contributor\_abandonment\_rate} across the studied projects.}
\end{figure}

\subsection{Review Process Features:}

\begin{figure}[H]
    \includegraphics[width=\textwidth]{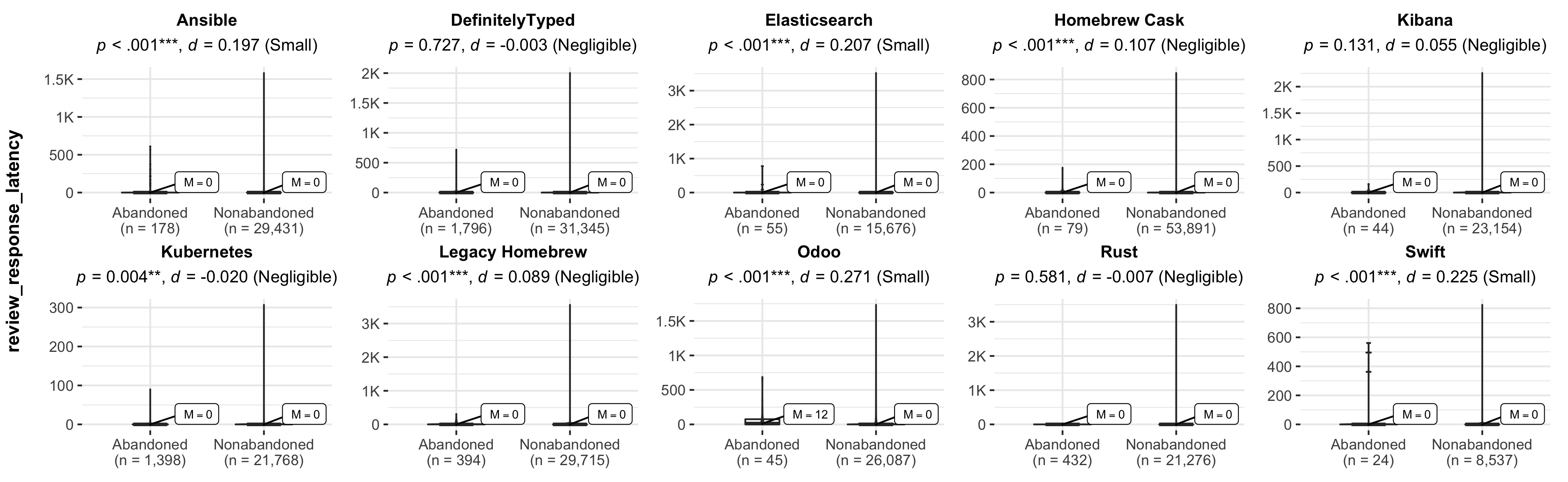}
    \caption{Comparison of abandoned and nonabandoned PRs wrt \textit{review\_response\_latency} across the studied projects.}
\end{figure}

\begin{figure}[H]
    \includegraphics[width=\textwidth]{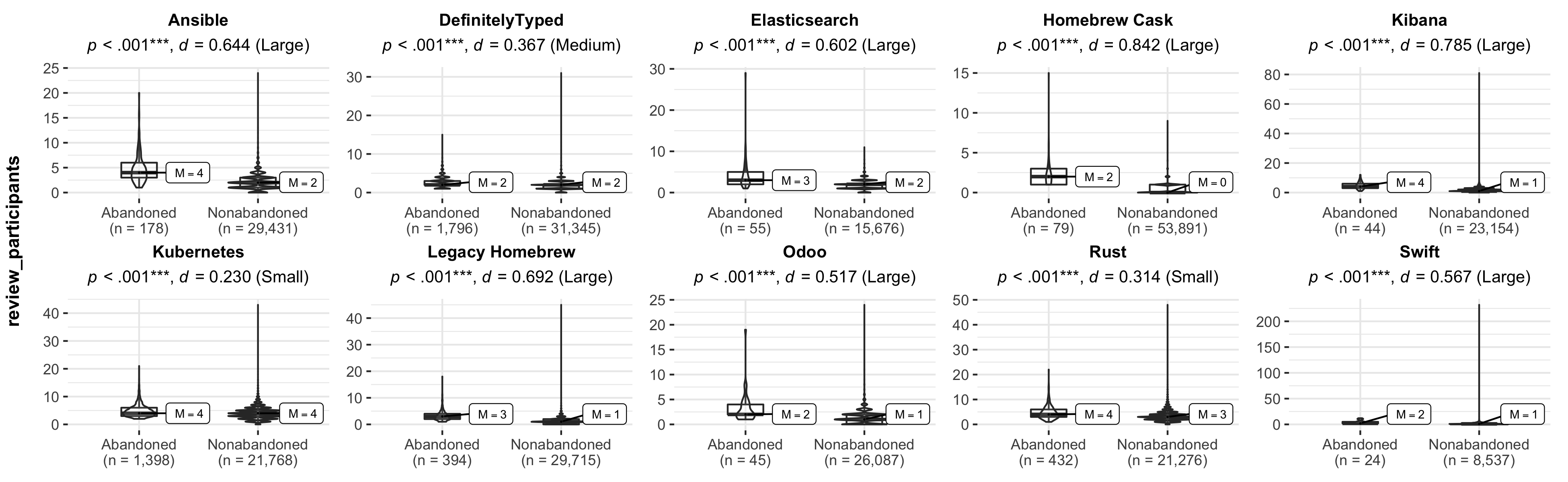}
    \caption{Comparison of abandoned and nonabandoned PRs wrt \textit{review\_participants} across the studied projects.}
\end{figure}

\begin{figure}[H]
    \includegraphics[width=\textwidth]{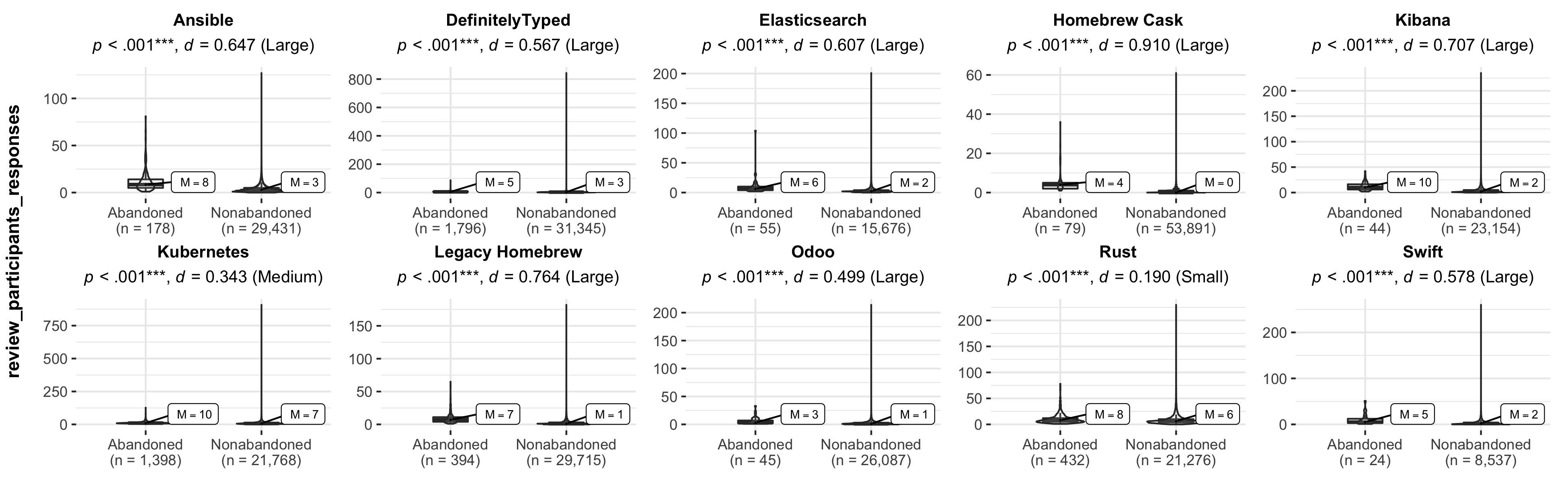}
    \caption{Comparison of abandoned and nonabandoned PRs wrt \textit{review\_participants\_responses} across the studied projects.}
\end{figure}

\begin{figure}[H]
    \includegraphics[width=\textwidth]{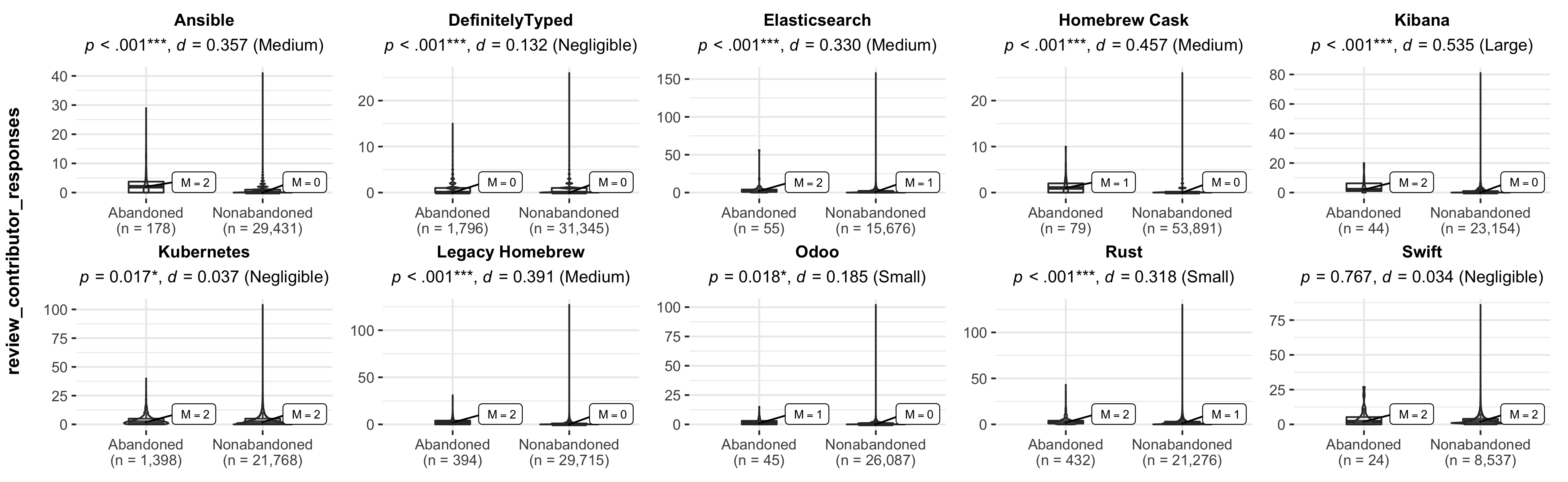}
    \caption{Comparison of abandoned and nonabandoned PRs wrt \textit{review\_contributor\_responses} across the studied projects.}
\end{figure}

\subsection{Project Features:}

\begin{figure}[H]
    \includegraphics[width=\textwidth]{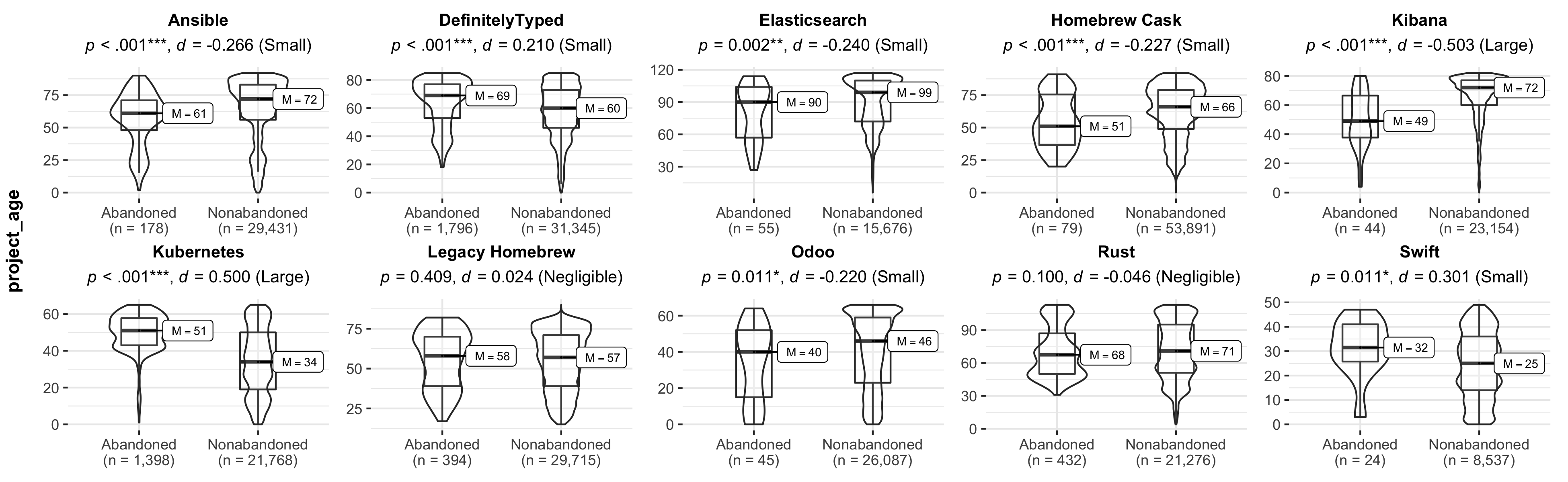}
    \caption{Comparison of abandoned and nonabandoned PRs wrt \textit{project\_age} across the studied projects.}
\end{figure}

\begin{figure}[H]
    \includegraphics[width=\textwidth]{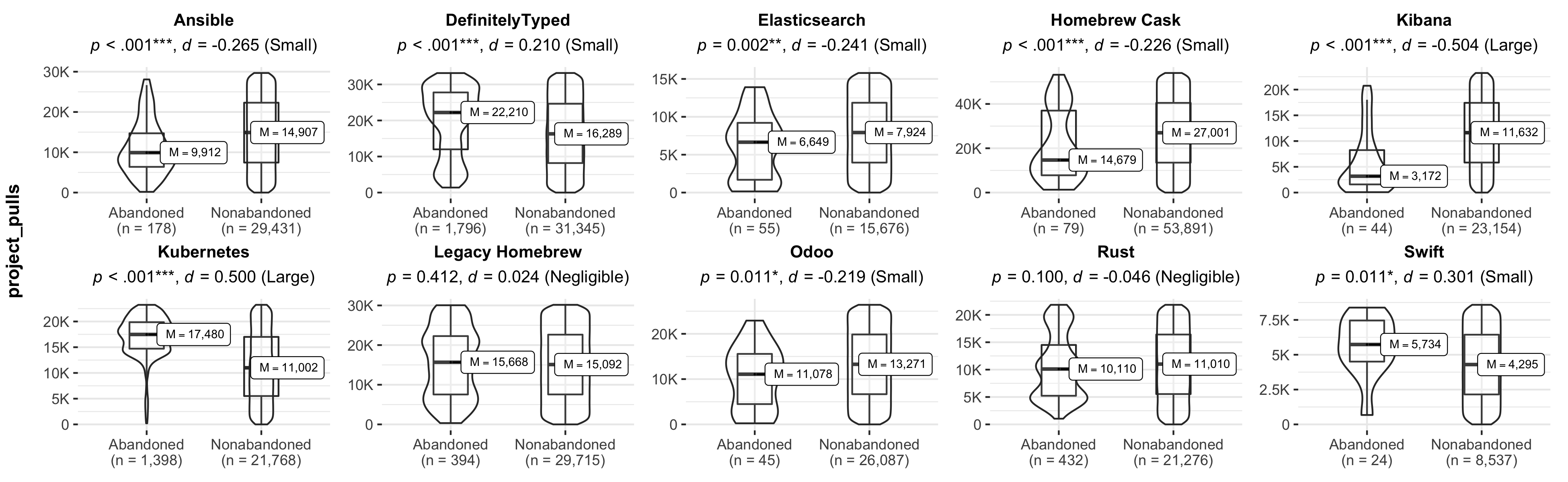}
    \caption{Comparison of abandoned and nonabandoned PRs wrt \textit{project\_pulls} across the studied projects.}
\end{figure}

\begin{figure}[H]
    \includegraphics[width=\textwidth]{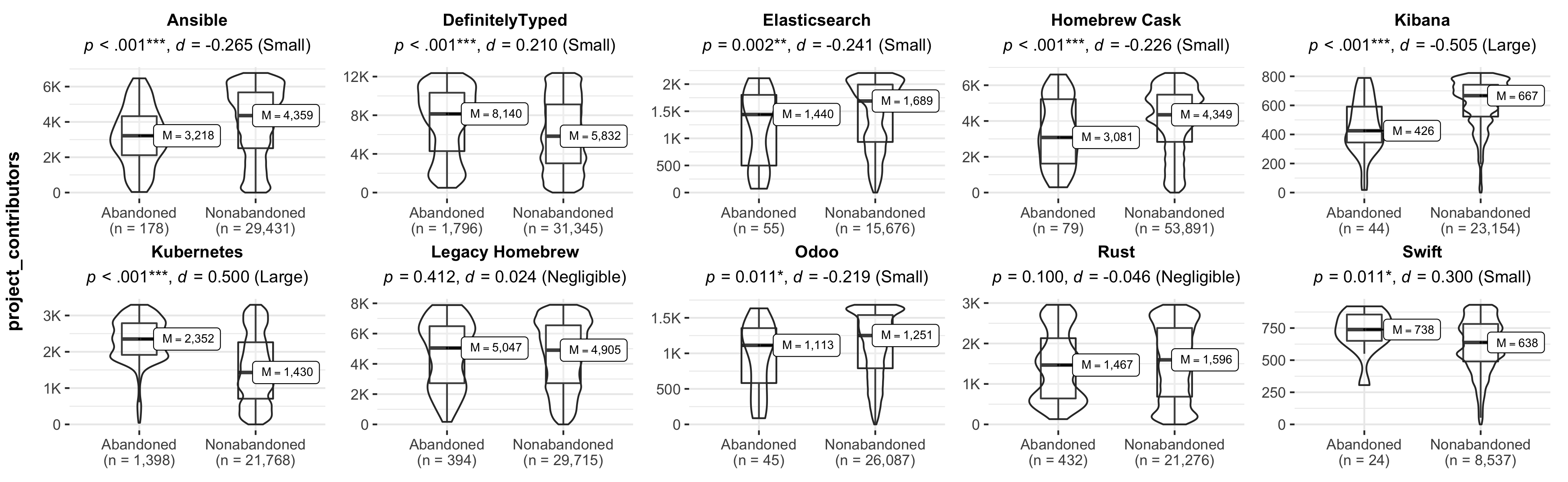}
    \caption{Comparison of abandoned and nonabandoned PRs wrt \textit{project\_contributors} across the studied projects.}
\end{figure}

\begin{figure}[H]
    \includegraphics[width=\textwidth]{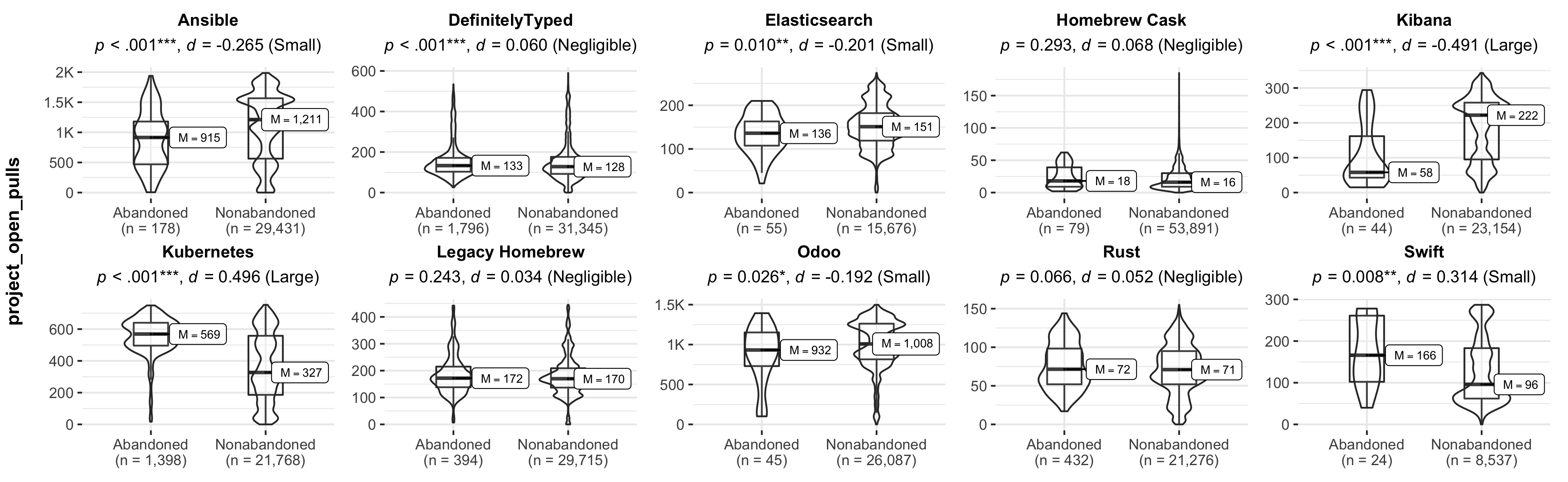}
    \caption{Comparison of abandoned and nonabandoned PRs wrt \textit{project\_open\_pulls} across the studied projects.}
\end{figure}

\section{ALE Plots for Different Features}
\label{appendix:ale}

\begin{figure}[H]
    \includegraphics[width=\textwidth]{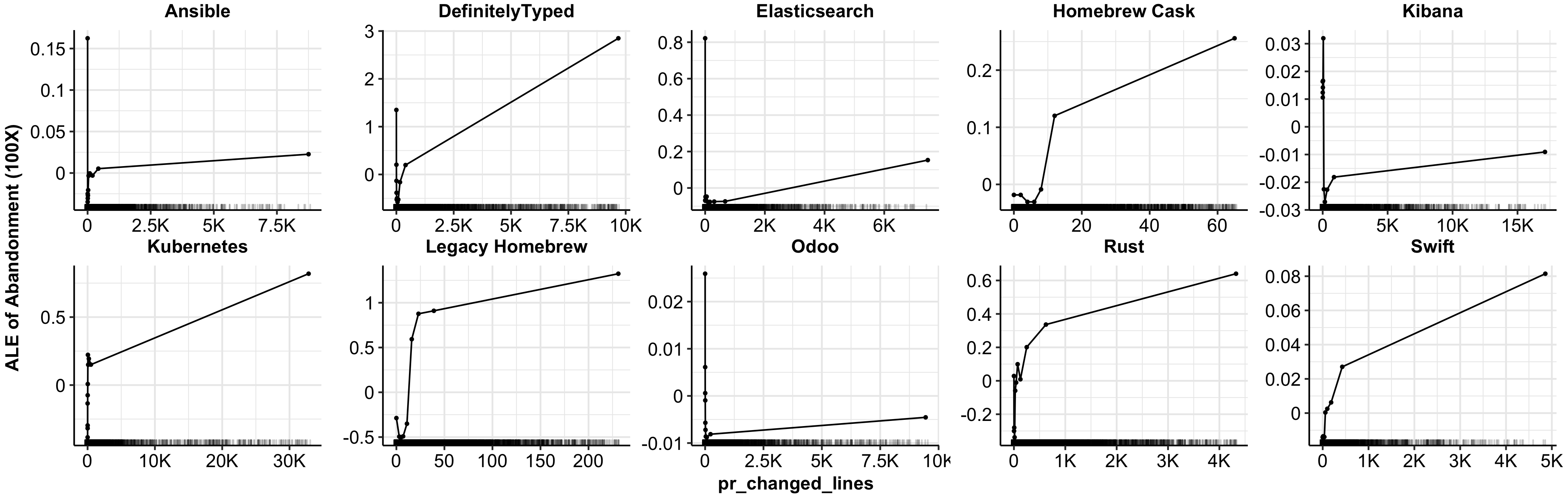}
    \caption{ALE plots showing how \textit{pr\_changed\_lines} varies the abandonment probability of PRs across the studied projects.}
\end{figure}

\begin{figure}[H]
    \includegraphics[width=\textwidth]{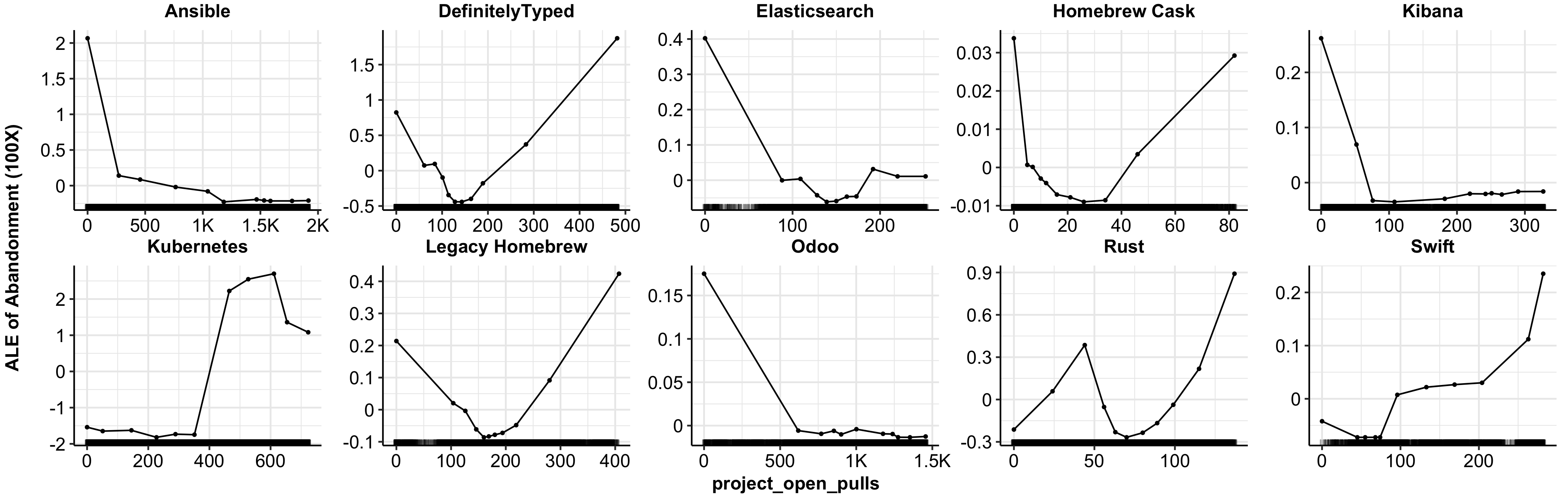}
    \caption{ALE plots showing how \textit{project\_open\_pulls} varies the abandonment probability of PRs across the studied projects.}
\end{figure}

\begin{figure}[H]
    \includegraphics[width=\textwidth]{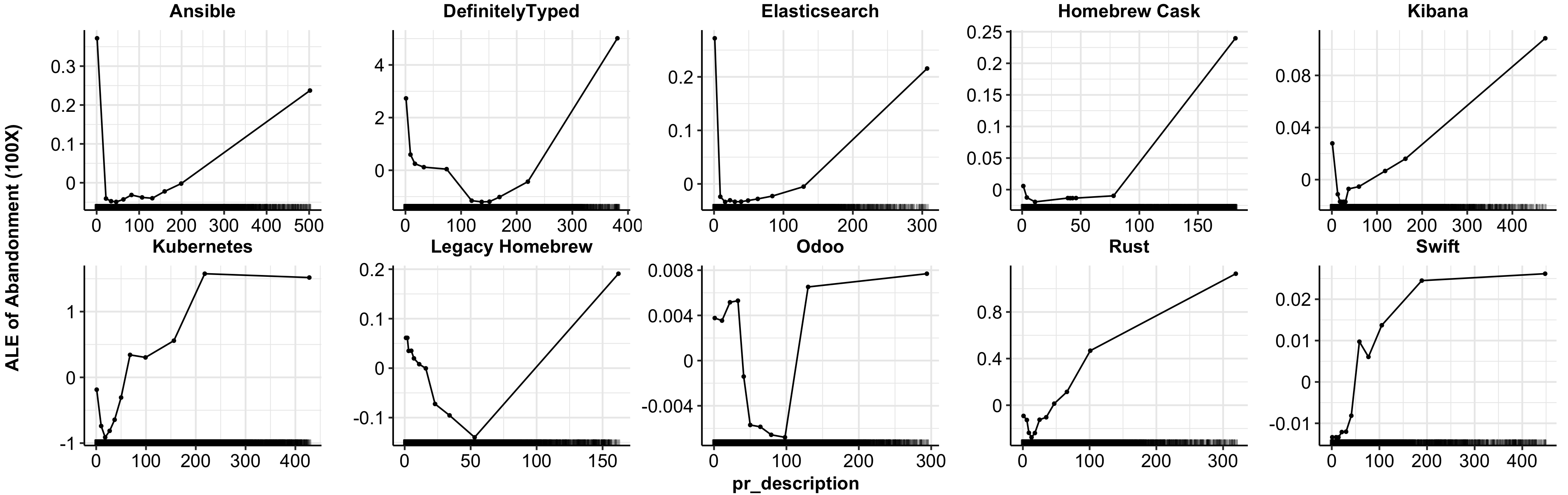}
    \caption{ALE plots showing how \textit{pr\_description} varies the abandonment probability of PRs across the studied projects.}
\end{figure}

\begin{figure}[H]
    \includegraphics[width=\textwidth]{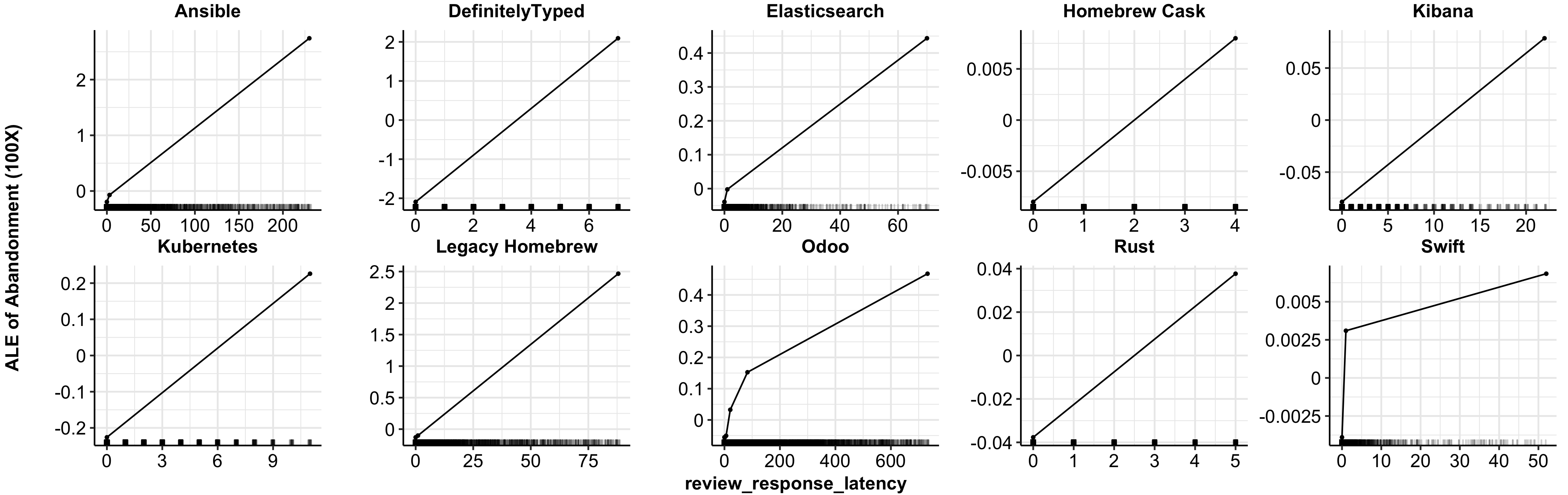}
    \caption{ALE plots showing how \textit{review\_response\_latency} varies the abandonment probability of PRs across the studied projects.}
\end{figure}

\begin{figure}[H]
    \includegraphics[width=\textwidth]{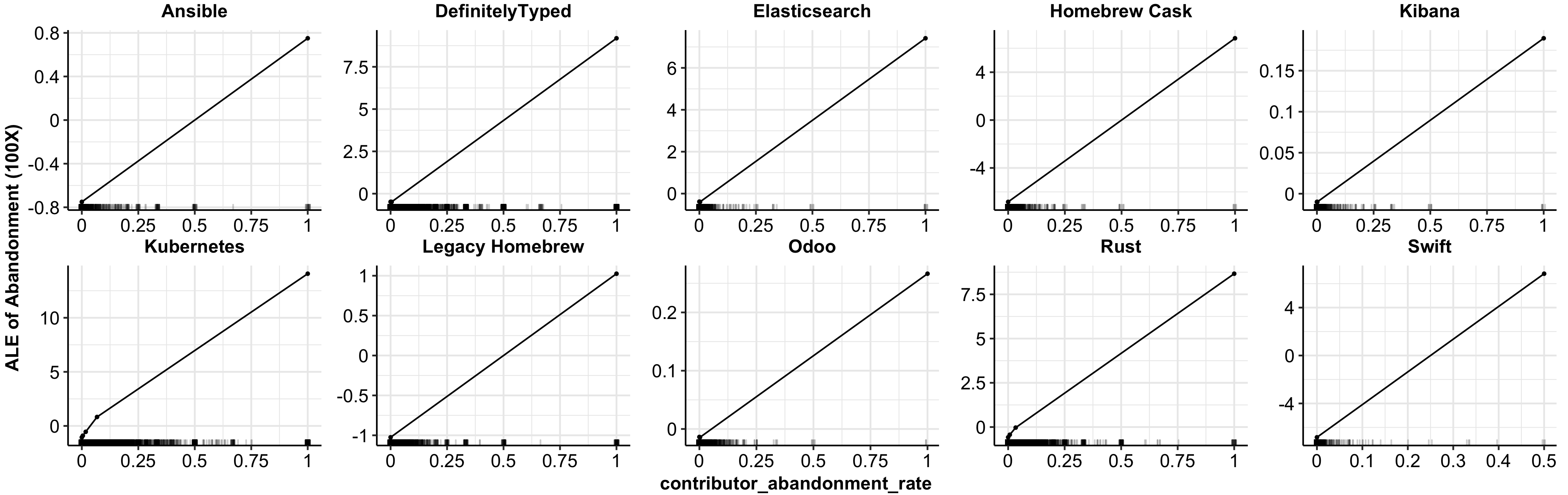}
    \caption{ALE plots showing how \textit{contributor\_abandonment\_rate} varies the abandonment probability of PRs across the studied projects.}
\end{figure}

\begin{figure}[H]
    \includegraphics[width=\textwidth]{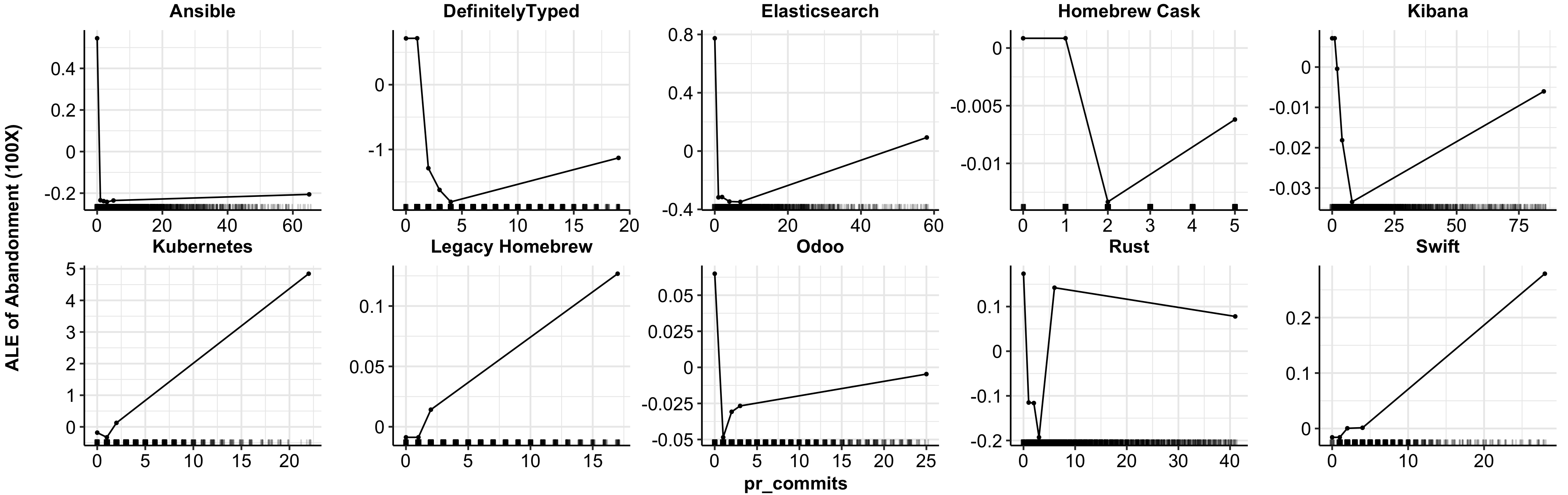}
    \caption{ALE plots showing how \textit{pr\_commits} varies the abandonment probability of PRs across the studied projects.}
\end{figure}

\end{document}